\documentclass[preprint2]{aastex}
\usepackage{graphicx}
\usepackage{longtable,lscape}

\usepackage{amsmath,amsfonts,amssymb}

\newcommand{\cN}{\mathcal{N}}

\newcommand{\wh}{\widehat}
\newcommand{\wt}{\widetilde}
\newcommand{\veps}{\varepsilon}

\def\build#1_#2^#3{\mathrel{\mathop{\kern 0pt#1}\limits_{#2}^{#3}}}

\shorttitle{New constraints on the dynamo mechanisms in K and M dwarfs}
\shortauthors{Houdebine, Mullan, Bercu, Paletou \& Gebran}

\begin{document}

\title{\small The Rotation-Activity Correlations in K and M dwarfs. II. New 
constraints on the dynamo mechanisms in late-K and M dwarfs before and at the 
transition to complete convection\thanks{Based on observations available at 
Observatoire de Haute Provence and the European Southern Observatory databases 
and on Hipparcos parallax measurements.}}

\author{\small E.R. Houdebine}
\affil{\scriptsize Armagh Observatory, College Hill, BT61 9DG Armagh, Northern 
Ireland}
\affil{\scriptsize Universit\'e de Toulouse, UPS-Observatoire 
Midi-Pyr\'en\'ees, IRAP, Toulouse, France}
\affil{\scriptsize CNRS, Institut de Recherche en Astrophysique et 
Plan\'etologie, 14 av. E. Belin, F--31400 Toulouse, France}
\email{\scriptsize eric\_houdebine@yahoo.fr}

\author{\small D.J. Mullan}
\affil{\scriptsize Department of Physics and Astronomy, University of Delaware,
 Newark, DE 19716, USA}
\email{\scriptsize mullan@udel.edu}

\author{\small B. Bercu}
\affil{\scriptsize Universit\'e de Bordeaux, Institut de Math\'ematiques,
UMR 5251, 351 cours de la lib\'eration, 33405 Talence cedex, France.}
\email{\scriptsize Bernard.Bercu@math.u-bordeaux1.fr}

\author{\small F. Paletou}
\affil{\scriptsize Universit\'e de Toulouse, UPS-Observatoire 
Midi-Pyr\'en\'ees, IRAP, Toulouse, France}
\affil{\tiny CNRS, Institut de Recherche en Astrophysique et Plan\'etologie, 
14 av. E. Belin, F--31400 Toulouse, France}

\author{\small M. Gebran}
\affil{\scriptsize Department of Physics \& Astronomy, Notre Dame 
University-Louaize, PO Box 72, Zouk Mika\"{e}l, Lebanon}


\newpage
\begin{abstract}
\small
We study the rotation-activity correlations (RACs) in a sample stars from 
spectral type dK4 to dM4. We study RACs using chromospheric data and coronal 
data. We study the Ca\,{\sc ii} line surface fluxes-$P/\sin i$ RACs. We fit 
the RACs with linear homoscedastic and heteroscedastic regression models. We 
find that these RACs differ substantially from one spectral sub-type to 
another. For dM3 and dM4 stars, we find that the RACs cannot be described by a 
simple model, but instead that there may exist two distinct RAC behaviors for 
the low activity and the high activity stellar sub-samples respectively. 
Although these results are preliminary and will need confirmation, the data  
suggest that these distinct RACs may be associated with different dynamo 
regimes.

We also study R'$_{HK}$ as a function of the Rossby number R$_{0}$. We find 
that: (i) For dK4 stars, we confirm R'$_{HK}$ as a function of R$_{0}$ agrees 
well with previous results for F-G-K stars. (ii) In dK6, dM2, dM3 and dM4 
stars, we find that, at a given R$_{0}$, the values of R'$_{HK}$  lie a factor 
of 3, 10, 20 and 90 respectively below the F-G-K RAC. Our results suggest a 
significant decrease in the efficiency of the dynamo mechanism(s) as regards 
chromospheric heating before and at dM3, i.e. before and at the TTCC. 

We also show that the ratio of coronal heating to chromospheric heating 
$L_{X}/L_{HK}$ increases by a factor of 100 between dK4 and dM4 stars. 
\end{abstract}
\normalsize


\section{Introduction}

In the present paper, we have two principal goals. First, we present data 
which have a bearing on the existence of correlations between rotation and 
activity (``RACs") in our sample of stars. Second, we ask if the empirical 
correlations can be interpreted in the context of stellar dynamo theories 
(which attempt to predict the properties of the magnetic fields which are 
generated in a star with a prescribed structure and rotation). The second goal 
is admittedly a challenging one: the basic mechanisms underlying stellar 
dynamo theory are being continuously improved as computer resources permit 
inclusion of more realistic physical effects. Rather than attempting to 
provide a comprehensive discussion of these complexities, we simplify our 
discussion by restricting attention to certain broad classes of dynamos.  We 
briefly outline the distinction between these classes in what follows. 
Whatever the source of the magnetic field, the observational consequences of 
such a field in a star is expected to be the same: dissipation of mechanical 
energy associated with magnetic processes leads to enhanced emission (relative 
to what is generated by the photosphere) from the chromosphere (in spectral 
lines such as Ca\,{\sc ii} H and K or $H_{\alpha}$) and from the corona (in 
X-ray continuum). These enhanced emissions arising from magnetic effects in 
the stellar atmosphere on the  are considered in this paper to be generic 
indicators of what we refer to as ``magnetic activity", or more briefly, 
``activity". In this paper, we seek to quantify how (an observational quantity 
that is associated with) ``activity" in low-mass dwarfs is correlated with (an 
observational quantity that is associated with) ``rotation". In order to set 
the stage for interpreting our result, we first need to define certain terms 
which are related to dynamo models. 

\subsection{Three classes of dynamo models}

Models of dynamos in the Sun and stars are typically based on ``mean-field 
electrodynamics" (MFE): the fluid flow and the magnetic field are separated 
into mean ($<u>$, $<B>$) and turbulent ($u'$, $B'$) components (e.g. Racine et 
al. 2011). Although the mean values of $u'$ and $B'$ are by definition zero, 
the mean value of their cross-product does {\it not} reduce to zero, but 
instead produces a non-zero turbulent MFE, $\it E$. The essence of MFE is to 
express the (vector) $\it E$ in terms of the large-scale (vector) magnetic 
field $<B>$ by means of a tensor expansion in the mean field plus its 
gradients. The leading order term in this expansion $\it E_i$ =  
$\alpha_{ij}$ $<B_j >$ is called the "$\alpha$-effect". Parker (1955) first 
suggested that the $\alpha$-effect might arise because cyclonic convective 
turbulence can systematically twist a large-scale magnetic field, and in the 
process, regenerate a large-scale poloidal field. All of the dynamos we 
consider here rely on the $\alpha$-effect, as well as on (at least) one more 
factor.

In order to construct a mean-field dynamo model, <B> is written as the sum of 
poloidal and toroidal components, and <u> is assumed to be directed purely in 
the azimuthal direction $\phi$, and to be axisymmetric, i.e. the angular 
velocity $\Omega$ can be a function of $r$ and $\theta$, but not of $\phi$. 
With these assumptions, the induction equation for the time-varying magnetic 
field can be separated into two equations, one for the poloidal field, the 
other for the toroidal field. Source terms ($S_p$, $S_t$) appear in both 
equations, and the dominant source terms determine the dynamo class. For the 
poloidal equation, $S_p$ is a single term, namely, the $\phi$ component of 
{\it E}, i.e. the $\alpha$-effect. For the toroidal equation, $S_t$ contains 
two terms: (i) includes a spatial gradient of $\Omega$; (ii) includes a 
spatial gradient of ${\it E}$. 

Three classes of mean-field dynamos are defined as follows: (a) 
$\alpha$-$\Omega$ dynamo, in which only term (i) is retained in the toroidal 
equation; (b) $\alpha ^2$ dynamo, in which only term (ii) is retained in the 
toroidal equation; (c) $\alpha ^2$-$\Omega$ dynamo, in which terms (i) and 
(ii) are both retained in the toroidal equation.   

\subsection{Stellar internal structure as it relates to dynamo activity}

Models of dwarf stars of spectral type F, G, K, and early M possess a 
radiative core and a convective envelope. In such stars, magnetic fields can 
be generated by an $\alpha$-$\Omega$ dynamo (also referred to as an interface 
dynamo, or shell dynamo, Parker 1975)\footnote{For a quantitative evaluation 
of such a dynamo in mid-K to early-M stars with known rotation periods, see 
Mullan et al. 2015}. The strongest toroidal magnetic fields in these stars 
expected to be produced in the vicinity of the interface (``tachocline") 
between the radiative core and the convective envelope where the differential 
rotation is the strongest. In stars which are massive enough to contain such 
an interface, and where therefore the possibility of an ``interface dynamo" 
(ID) exists, data pertaining to enhanced emission from such stars in 
chromospheric spectral lines and/or in coronal X-rays reveal clearly that 
there is a strong correlation between rotation and activity indicators (e.g. 
Pallavicini et al. 1981;  Wright et al. 2011), i.e. the RAC has a slope with 
a numerical value that is definitely non-zero.

However, since Limber (1958) derived models of cool main sequence stars with 
lower and lower masses, it has been widely believed by stellar evolution 
modelers that interfaces do not exist in all low-mass stars. An anonymous 
referee has pointed out that ``there has been no definitive evidence 
validating this theoretical prediction". In principle, asteroseismology of M 
dwarfs might eventually provide p-mode frequencies with enough precision that 
(by analogy with helioseismological data) signatures of the interface could be 
identified. But although theoretical models have been calculated for such 
stars (see Rodriguez-Lopez et al. 2014), no empirical data on reliable p-mode 
periods in M dwarfs are yet available. Despite the lack of definitive evidence 
at the present time, we shall adopt the widely held belief that stars on the 
main sequence undergo a Transition To Complete Convection (TTCC) at a certain 
spectral type. According to models, main sequence stars with masses that are 
less than a critical value of about 0.30-0.35~$M_{\odot}$ are completely 
convective. Such stars should have no tachocline whatsoever on the main 
sequence. And yet the empirical evidence shows that stars with masses less 
than 0.30-0.35~$M_{\odot}$ also show enhanced emission in chromospheric lines 
and in X-ray continuum. Since we have attributed such enhanced emission in 
warmer stars to magnetic activity, then it seems natural also to ascribe the 
enhanced emissions in completely convective stars to magnetic activity. Such 
activity must rely on a non-interface type of dynamo, possibly an 
$\alpha^{2}$ dynamo (distributive dynamo: DD), maintained by convective 
turbulence alone (e.g. Roberts \& Stix 1972, Rosner 1980, Dobler et al. 2006, 
Chabrier \& Kuker 2006). 

It is important to note that the existence of complete convection in a star 
does not mean that the star is necessarily completely convective at all stages 
of its main sequence phase of evolution. E.g., Feiden and Dotter (2013) report 
that a 0.3~$M_{\odot}$ model arrives on the main sequence with a radiative 
zone sandwiched between two convective zones: the models suggest that such a  
``sandwich" structure exists for several gigayears before the model becomes 
completely convective (see also Rodriguez-Lopez et al. 2014). However, at 
lower masses, 0.25~$M_{\odot}$, stellar models {\it are} completely convective 
at all ages (see Rodriguez-Lopez et al. 2014). Thus, when we use the label 
``completely convective" for stars of masses in the intermediate range 
(0.25-0.3)~$M_{\odot}$, the label involves not merely a question of the {\it 
mass} of the model, but also of the {\it age} of the model. This complication 
may cause some ambiguity when we attempt to interpret RACs in the vicinity of 
the TTCC.

For simplicity, it would be convenient if a given star could be assigned to 
having either an ID or a DD. But we recognize that such a simplified approach 
cannot be the whole story: in cool stars where an interface exists, the 
inevitable presence of a deep convective envelope gives rise to the possibility
 that ID and DD may both be operating (e.g. Brandenburg \& Subramanian 2005; 
Brown et al. 2010). 

In order to distinguish between the observational properties of ID and DD,
Durney, De Young \& Roxburgh (1993) pointed out some differences between ID 
and DD. First, the magnetic field created by an ID should depend strongly on 
rotation whereas the magnetic field created by a DD should not. In terms of 
the notation to be used in the present paper, this would lead one to expect 
that in an ID star, the RAC should have a slope which is definitely non-zero, 
whereas in a DD star, the RAC slope should be essentially zero. Second, 
although IDs can produce activity cycles in the large-scale field it was not 
clear (in 1993) how DD could ever give rise to a cycle.

Since 1993, the second feature has been called into question. E.g. Stefani \& 
Gerberth (2003) have demonstrated that cyclic behavior {\it is} possible in 
$\alpha ^2$ dynamo models provided that the following condition is satisfied: 
the $\alpha$-effect needs to have radial gradients which are sufficiently 
steep, including changes in the algebraic sign.

Moreover, recent modeling suggests that the third class of dynamo models 
($\alpha ^2$-$\Omega$) might be a better description to the solar dynamo 
(where there is certainly an interface) than an $\alpha$-$\Omega$ dynamo (e.g. 
Lawson et al 2015). From the perspective of a change from an $\alpha$-$\Omega$ 
dynamo to an $\alpha ^2$-$\Omega$ dynamo, perhaps in completely convective 
stars, we might also encounter a change from an $\alpha ^2$ dynamo to an 
$\alpha^2$-$\Omega$ dynamo. If this happens, then we might find that rather 
than seeing a zero slope for the RAC in a completely convective star, the 
inclusion of $\Omega$ in the dynamo model could lead to a non-zero RAC slope.

\subsection{Observational evidence for the TTCC?}

A long standing problem has been the search for an observational signature of
the putative transition between different types of dynamos at the TTCC. There 
has been several attempts to detect a change in magnetic activity diagnostics 
(e.g. Mullan \& MacDonald 2001) or magnetic field topologies (e.g. Donati et 
al. 2008, Morin et al. 2008, 2010, Phan-Bao et al. 2009, Stassun et al. 2011) 
in the vicinity of the TTCC. None of these studies have identified unambiguous 
and definitive  signatures of significant changes in magnetic activity at the 
TTCC. However, in a recent paper, West et al. (2015) have reported on a study 
of RACs in two groups of dwarfs: M1-M4, and M5-M8. West et al. (2015) found 
that the RAC in M1-M4 stars has a somewhat different slope than the RAC in 
M5-M8 stars (see Sect.~4 below for a more detailed discussion). One 
interpretation of this behavior is that something changes as regards the 
dynamo between M4 and M5. We will return to a discussion of this 
interpretation (as well as an alternative interpretation) in Section 4 below.

In order to address the TTCC-dynamo-transition topic meaningfully, we first 
need to identify at what spectral type the TTCC occurs. Limber (1958) states 
that ``in that part of the main sequence where the inner radiative region ... 
is becoming vanishingly small", the corresponding spectral type is M3V-M4V. 
Subsequently, Dorman et al. (1989) placed the TTCC at $M\sim 0.25 M_{\odot}$, 
which corresponds to the spectral subtype dM4. Even more recently, Chabrier \& 
Baraffe (1997) predict that the TTCC occurs at $M\sim 0.35 M_{\odot}$, which 
corresponds to the spectral subtype dM2. Therefore, it appears that the TTCC 
may lie somewhere in the range between subtypes dM2 and dM4, corresponding to 
masses of 0.25-0.4~$M_{\odot}$ (Stassun et al. 2011). The theoretical mass 
limit at the TTCC has been found to shift towards slightly smaller masses if 
different boundary conditions are used for the stellar models (Mullan et al. 
2015). (Even larger shifts of the TTCC towards lower masses were at one time 
proposed by Mullan and MacDonald 2001 if interior magnetic fields were to be 
as large as $10^{7-8}$ G. If fields as large as that were to exist inside 
stars, the definition of a ``fully convective star" could become more 
ambiguous: the onset of complete convection would then depend not only on the 
mass and age of a star, but would also depend on how strong its magnetic field 
is. However, we may not in fact need to worry about this ambiguity: Browning 
et al. (2016) have recently argued that such strong fields would be unstable. 

In standard main-sequence models, the radius of a model scales almost linearly 
with mass: therefore, the TTCC is expected to lie in the radius range of 
roughly 0.25-0.4~$R_{\odot}$. According to the radius-$T_{eff}$ calibration of 
Houdebine et al. (2016b) this radius range yields an effective temperature 
range of 3200-3500~K for the TTCC.

In this paper, we examine the topic of a possible dynamo transition at the 
TTCC using a more extensive and more fine-grained data set than has previously 
been available for study. 

\subsection{Dynamos: unsaturated and saturated}

Empirically, evidence for rotationally driven dynamos in cool stars first 
emerged  when researchers plotted the strength of chromospheric emission 
versus stellar rotation. E.g. See Kraft (1967), Vaughan et al. (1981), 
Soderblom (1982), Vogt et al. (1983), Noyes et al. (1984),  Marcy \& Chen 
(1992), Patten \& Simon (1996), Fekel (1997), Delfosse et al. (1998), Jeffries 
et al. (2000), Pizzolato et al. (2003), Mohanty \& Basri (2003), Browning et 
al. (2010), Wright et al. (2011), Rebassa-Mansergas et al. (2013), West et al. 
(2015).  

The key signature is the following: emission in chromospheric spectral lines 
is observed to be stronger in stars with faster rotation speed $v_r$ (or 
shorter rotation period P) (e.g. Vaughan et al 1981). This is defined as a 
``Rotation-Activity Correlation" (RAC). The existence of an RAC is consistent 
with the expectations that (i) the faster the rotation is, the stronger are 
the magnetic fields which can be generated (e.g. Mullan et al. 2015), and (ii) 
stronger fields are associated with stronger chromospheric heating (Skumanich 
et al. 1975) and with stronger coronal heating (e.g. Mullan 2009). Thus, if 
chromospheric emission intensity is plotted as a function of $P$, it is found 
that over a certain range of periods, the RAC has a clearly {\it negative} 
slope.  

However, as more data are accumulated, it emerges that the negative slope of 
the RAC does {\it not} extend indefinitely to shorter and shorter $P$. Instead,
 when $P$ becomes shorter than a certain value $P_{c}$, no further increase in 
emission occurs: for $P\leq P_{c}$, the RAC becomes flat (e.g. Vilhu 1984), 
with a slope of zero. By definition, in the flattened portion of the RAC, 
increasing rotation does {\it not} result in increased chromospheric/coronal 
emission. For solar type stars, the transition to a flat curve occurs for 
$P\approx$3 days. Vilhu suggested the term ``saturated" to refer to such 
conditions. Similarly, Wright et al. (2011) found that the 
saturation regime occurs at about $R_{0}\simeq 0.8$. Reiners et al. (2009) 
found that the sturation occurs at the critical Rossby number $R_{0}\simeq 
0.1$. We shall compare these values to those we obtain for our stellar 
samples below. Pizzolato et al. (2003) found that saturation occurs from a 
period of about $\approx 2$ days in solar type stars to about $\approx 10$ 
days in M dwarfs. Rebassa-Mansergas et al. (2013) found that stars with 
$v\sin i \geq 5\ hm\ s^{-1}$ are all in the saturated regime. This yields a 
rotation period of $\approx 3$ days for stars at the spectral type dM3 (see 
Paper~I). Therefore, we shall consider that saturation occurs for periods 
of about $\approx 3$ days or $R_{0}\simeq 0.5$. Since in our stellar samples 
we have stars with shorter rotation periods or $R_{0}$, we shall consider 
below that these stars likely lie in the saturated regime.

 Vilhu suggested that ``saturation" might be due to a complete 
coverage of the star's surface by magnetic fields. Another explanation of 
``saturation" was offered by Mullan (1984) in terms of the maximum possible 
flux of mechanical energy that can be generated by convection. In the present 
paper, we shall not attempt to identify the physical process which leads to 
saturation. Instead, we shall adopt a purely empirical approach, and we shall 
refer to the flat portion of an RAC as ``saturated". In the same vein, we 
shall refer to an RAC with a statistically significant negative slope as 
``unsaturated".  To the extent that an RAC owes its existence to the operation 
of a dynamo of some kind, we can say that the dynamo reveals itself in two 
regimes: saturated and unsaturated. The properties of the two different dynamo 
regimes we find below (see Section 3.8.1) for the low and high activity 
sub-samples respectively cannot be due to effects of the chromospheric 
response to non-thermal heating mechanisms. In fact, there does not seem to be 
any definitive evidence that the influence of a magnetic field in a stellar 
atmosphere ever attains a saturated level. In support of this claim, we note 
that the response of stellar chromospheres to non-thermal heating mechanisms 
is continuous and monotonically increasing from basal chromospheres to flaring 
chromospheres (e.g. Houdebine 1992, Houdebine \& Doyle 1994a, Houdebine \& 
Doyle 1994b, Houdebine et al. 1995, Houdebine \& Stempels 1997, Houdebine 
2009, Houdebine 2010). Therefore, the two different types of RACs we find 
below for the low and high activity stars respectively highlight the different 
properties of the dynamo mechanisms in these two types of stars.

It seems to us that in order to study the properties of stellar dynamos most 
profitably, it would be preferable to concentrate as much as possible on stars 
in the {\it unsaturated} regime. The reason for this claim is that, when 
conditions are saturated, there are extra factors which come into play which 
may obscure some physical properties that are directly associated with dynamo 
action. In view of Vilhu's identification of a critical period $P_{c}$, it 
seems that the {\it slowest} rotators have the best chance of being in the 
unsaturated regime. For that reason, we consider it worthwhile to push the 
spectroscopic measurements of stellar rotation towards the {\it smallest 
possible values} of $v\sin i$ which can be reliably measured.

In general, since slow rotation means less chromospheric heating, we expect 
that the slowest rotators in our dataset will be stars which are low-activity 
stars classified as dK and dM (i.e. those which by definition show no emission 
in the Balmer lines), while the fastest rotators will be stars which are 
highly active stars classified as dKe and dMe (i.e. those where by definition 
the Balmer lines have an emission core). In what follows, we shall be 
especially interested in determining the (negative) slope of the RAC for 
low-activity stars (e.g. see Fig.~13 below, lower panel).

\subsection{Extending our previous work}

One reason for the present study has emerged from recent work by Houdebine \& 
Mullan (2015: hereafter HM): they found that another diagnostic of magnetic 
fields,  namely, the efficiency 
of magnetic braking (which manifests itself in the rotational velocity), 
undergoes a detectable change at spectral sub-type dM3. Basing their analysis 
on a new data set of precise rotational velocities, HM found that the mean 
rotation period of M3 stars is abnormally large compared to those of the 
adjoining spectral types dM2 and dM4. This indicates that the dM3 stars have 
been slowed down more than the stars in the immediately adjacent sub-types dM2 
and dM4. This excess slowing at dM3 may be associated with a change in 
magnetic properties of the stars at dM3. Specifically, HM suggested that the 
change might be associated with an earlier report (Mullan et al 2006) that the 
lengths of flaring magnetic loops undergo a significant increase at spectral 
type dM3. 

In the present paper, we extend the work of HM in two distinct ways. First, we 
expand the database of precise rotational properties of K and  M dwarfs of 
various sub-types. Second, we expand and analyze a separate database which 
deals with the second physical parameter which enters into the RAC: the 
radiative properties which are associated with magnetic ``activity" in our 
sample of rotating stars. Our goal here is to use the activity data to 
construct RACs. Moreover, we quantify the RACs in two different parts of the 
stellar atmosphere: the chromosphere (using the Ca\,{\sc ii} lines) and the 
corona (using $L_X$). The present study is based on a larger sample of stars, 
and a finer grid of spectral sub-types, than have been used previously in 
constructing RACs for K and M stars. Our goal is to explore whether the 
unusual rotational signature reported by HM at dM3 is accompanied by unusual 
behavior in the activity indicators of either chromosphere or corona or both, 
either as regards the intensity of the radiation, or as regards the slopes of 
the RACs. 

A key physical factor which is known to be well correlated with chromospheric 
emission and coronal emission has to do with magnetic fields on the stellar 
surface (e.g. Skumanich et al 1975; Schrijver et al. 1989). The existence of 
RAC's may be interpreted as an indication that the surface magnetic field 
intensities are correlated with the stellar rotation rate.  This relationship 
is to be expected on the basis of standard dynamo theory (e.g. Parker 1979, 
Krause \& Radler 1980, Mullan et al. 2015). However, most of the observational 
investigations cited above suffer from two inadequacies: (i) they included 
only a few stars which are rotating slowly enough to be in the unsaturated 
regime, and (ii) the targets included stars which were spread out over a broad 
range of spectral types. In this paper, we attempt to remedy both of these 
inadequacies.

\subsection{Aspects of the data used in the present study}

In an effort to extend the RACs to slow rotators among late-type dwarfs, we 
have been reporting, over the past several years, improved spectroscopic 
measurements of rotational broadening $v\sin i$ in stars of spectral sub-types 
dK4 (Houdebine 2011a, Paper XVI thereafter), dK6 (Houdebine et al. 2016, Paper 
I), dM2 (Houdebine 2008, Paper VIII, Houdebine 2010a, Paper XIV), dM3 (HM) and 
dM4 (Houdebine 2012a, Paper XVII, Houdebine et al. 2016, Paper I). 

Combining our (previous) rotational measures with our measures of the 
Ca\,{\sc ii} line equivalent widths (EW), we have already reported RACs for 
dK4 (Houdebine 2011a, Paper XVI), dM2 ( Houdebine 2011b, Paper XV) and dM4 
stars (Houdebine 2012b, Paper XVIII), for slow and rapid rotators alike. In 
those earlier papers, we proposed empirical RACs which included large samples 
of slow rotators. We found that for later spectral types it is crucial to 
examine a fine grid of spectral sub-types. Specifically, we found that the 
RAC's vary significantly between dK4 to dM4 stars: the RAC's were found to 
have different gradients and different saturation levels (Papers XV, XVIII and 
the present study). The previous data, combined with the present study, now 
provide us with large enough data-sets in each spectral sub-type that we can 
investigate with improved confidence the differences (if any) between the 
RAC's in five different spectral sub-types. In view of the results reported in 
HM, it is notable that the present study enables us to study the RAC's in the 
vicinity of the TTCC.

When we consider {\it slow} rotators, our data confirm that the mean rotation 
periods of stars in the range dK4-dM4 in general decrease with decreasing 
effective temperature (see HM and Paper I). However, we also find that 
something unusual happens in the rotation rates between dM2 and dM4. The 
overall trend towards decreasing rotation period as we go from dK4 to dM4 is 
interrupted at spectral sub-type dM3: at that sub-type, the mean rotation 
period increases to a local peak, such that the mean rotational period at dM3 
is {\it longer} than the overall trend between dK4 and dM4 would have 
predicted (HM). But when we extend our investigation  to include {\it fast} 
rotators among the dK4-dM4 stars, there is found to be the following overall 
trend: the mean rotation period tends to increase slightly from dK4 to dM4. 
But once again, at dM3, an exception is found: the mean rotation period of 
fast rotators at dM3 is locally significantly {\it longer} than the overall 
trend would have predicted (HM). HM interpret these abnormally long rotation 
periods at sub-type dM3 as possibly being associated with the occurrence of 
increasing coronal loop lengths (previously reported by Mullan et al. 2006 in 
a study of flare stars). The mean rotation period of the slow rotators is an 
important constraint on the temporal history of the dynamo mechanisms and 
magnetic braking mechanisms (HM, Paper I). 

\subsection{Studying RACs in various formats}

An important aspect of the present paper is that we wish to investigate the 
RACs in various formats. In the first place, we construct RACs separately for 
the chromosphere and the corona. Moreover, we explore correlations between 
various observations of the ``activity" and various aspects of ``rotation". As 
an example, we will examine, for the H and K lines of Ca\,{\sc ii}, a plot of 
the quantity $R'_{HK}$ as a function of the Rossby number $R_{0}$. We also 
plot the Ca\,{\sc ii} surface flux as a function of $P/\sin i$. We use these 
different formats in an attempt to improve our chances of identifying changes 
(if any) in the dynamo regime in the vicinity of the TTCC. We shall find that 
changes near the TTCC are more readily detectable in some RACs than in 
others. This may explain why the empirical detection of the TTCC has been 
elusive in the past.

\section{Selection of spectroscopic data}

Stepie\'n (1989, 1993, 1994) has reported that RACs exhibit certain 
differences 
at different spectral types. As an extension of this finding, we have found, 
in previous studies (Paper XVIII, HM) and also in the present study, that it 
is important in constructing RACs to select samples of stars with $T_{eff}$ 
values which are confined within a narrow range. There are two principal 
reasons for this, one related to the choice of an optimal set of photospheric 
absorption lines, and the second related to our analysis of chromospheric 
emission lines. As regards the choice of photospheric lines, we have already 
discussed the first of the above reasons at length in Paper I in the context 
of optimizing the measurement of rotational velocities at each spectral 
subtype between dK4 and dM4. In the present paper, we turn now to the second 
reason.

In the context of {\it chromospheric} analysis, it is important to deal with 
stars with closely similar $T_{eff}$ when we are attempting to quantify with 
as much precision as possible the EW of the chromospheric lines (Ca\,{\sc ii} 
resonance doublet and H$_{\alpha}$). These lines inevitably include some 
contributions from the background photospheric continuum and from the 
temperature minimum region (e.g. Houdebine \& Doyle 1994, Cram \& Mullan 1979, 
Houdebine \& Stempels 1997, Paper XV). Initially, our samples of stars had 
been selected for the purpose of chromospheric modelling studies (e.g. 
Houdebine \& Stempels 1997, Houdebine 2009b Paper XII, Houdebine 2010b Paper 
IX), and in these studies the selection of stars with closely similar spectral 
types was essential in order to develop reliable grids of semi-empirical model 
chromospheres, each of which would be superposed on a particular photospheric 
model. 

Based on our previous papers, we have found that the most suitable initial 
selection parameter when we wish to identify a homogeneous sample of K or M 
dwarfs belonging to a specific sub-type is the (R-I) color: this color is 
sensitive to $T_{eff}$, but less so to metallicity (e.g. Leggett 1992, Ramirez 
\& Melendez 2005, Mann et al. 2015). Moreover, broad-band colors of high 
precision are widely available in the literature for many of the cool dwarfs 
which are of interest to us.

In the present paper, as an example of how we selected data for our RAC 
studies, we now describe how we gathered the relevant data for a sample of dK6 
stars. We selected a sample of 419 late K dwarfs on the basis of (R-I) 
measurements available in the literature. For example, our sample of dK6 stars 
(see Table~1 below) contains stars with (R-I)$_{C}$ 
(i.e. (R-I) color in the Cousins system) in the range [0.684;0.816] which also 
corresponds to (R-I)$_{K}$ ((R-I) in the Kron system) in the range 
[0.503;0.613] according to the transformation formulae of Leggett (1992) (see 
Leggett 1992 for more information on the Cousin's and Kron photometric 
systems). According to Kenyon and Hartmann (1995), this range of colors is 
centered on (R-I)$_{C}$=0.75, i.e. the spectral type dK7. However, when we 
compiled and derived effective temperatures (see Paper I) for this sample of 
late-K dwarfs, we found in average higher temperatures than what would be 
expected from the (R-I)$_{C}$-$T_{eff}$ tabulation of Kenyon and Hartmann 
(1995) (see Paper I). Our dK6 stellar sample contains stars that have similar 
(R-I)$_C$ colours and the same effective temperatures to within $\pm$110 K 
(see Paper I). We refer to HM for a discussion of corresponding data for our 
sample of dM3 stars. 

The literature provided us with a starting list of a large number (419) of 
late-K dwarfs. Searching through databases at the European Southern Observatory
 (ESO) and Observatoire de Haute Provence (OHP), we identified spectra of 112 
different stars which are suitable for our purposes. The final list of 105 
stars for which the available spectra would allow us to determine reliable
rotational data (i.e. $v\sin i$ values) in our sample of late-K dwarfs has 
already been provided in Paper I. For the present paper, we found spectra 
which would allow us to make reliable measurements of the EWs of Ca\,{\sc ii} 
and H$_{\alpha}$ for a sub-sample of only 89 late-K dwarfs. It is the 
combination of reliable rotations (Paper I) and reliable EWs of chromospheric 
lines which enable us to undertake, in the present paper, the study of the RAC 
in our sub-sample of dK6 dwarfs.
 
The spectra which we use for determining the Ca\,{\sc ii} and H$_{\alpha}$ 
equivalent widths in the present study of dK4-dM4 stars came from three 
different \'echelle spectrographs; HARPS (High Accuracy Radial velocity Planet 
Search, ESO), SOPHIE (OHP) and FEROS (The Fiber-fed Extended Range Optical 
Spectrograph). We included in our sample the SOPHIE observations obtained in 
the High Efficiency (HE) mode. The modes in which the spectra were obtained 
are indicated in Table~1 together with the corresponding Signal to Noise (S/N) 
ratio. For further details of the spectrographs, see HM. 

As a second example of how we selected pertinent data for our spectral 
sub-samples, we consider here briefly our sample of dM3 stars. As described in 
HM, westarted off with a list of 381 dM3 objects based on (R-I) data in the 
literature. Searching through the same databases as above, we found suitable 
observations which allowed us to determine reliable $v\sin i$ values for 86 
different dM3 stars. A subsequent search of the available spectra in the above 
databases allowed us to obtain reliable measurements of the EWs of 
Ca\,{\sc ii} and H$_{\alpha}$ for a sub-sample of only 59 M3 dwarfs for our 
chromospheric RAC (excluding probable spectroscopic binaries and low $\sin i$ 
stars).

\subsection{Biases in our stellar samples}

The stars in our samples include all stars from all observing programs 
which have been carried out with HARPS and SOPHIE for stars belonging to the 
following spectral sub-types: dK4, dK6, dM2, and dM3. For dM4 stars we 
compiled all measurements of $v\sin i$ available in the literature (see Paper 
I). For the dK6 and dM3 samples, we also supplemented our own measurements 
with measurements available in the literature, notably for active stars (see 
Paper I).

In the HARPS and SOPHIE databases, many of the spectra were obtained in 
connection with planet-search programs. In such programs, observers tend to 
avoid stars with high levels of magnetic activity. Therefore our spectral 
samples are likely to be biased in general towards low activity stars, i.e. 
stars which are in the unsaturated portions of the RACs. For reasons outlined 
above (see Section 1.1), we consider this bias to be an advantage in the 
present study. The biases in our samples may contribute somewhat to the 
density of the sampling at different parts of the RAC, but this is not 
expected to cause significant discrepancies as regards the overall RACs. Note 
that for the dM4 and dK6 samples, the sampling of the RACs should be more 
complete as regards the measures of $P/\sin i$ (because we included other 
measures of $v\sin i$ from the literature, see Paper~I).

\section{The rotation-activity correlations (RACs) in late-K and M dwarfs}

In this section, we first (Sub-section 3.1) evaluate the surface fluxes in the 
continuum in the vicinity of the Ca\,{\sc ii} resonance doublet by using our 
estimates of $T_{eff}$ and also by using the synthetic spectra of de Laverny 
et al. (2012).

Then, we construct the RACs of M and K dwarfs using a variety of approaches: 
some approaches may facilitate the extraction of information that is more 
difficult to extract by means of other approaches. In all of the approaches, 
we plot a quantity related to ``activity" as the ordinate, and a quantity 
related to ``rotation" as the abscissa.

Our first approach (Sub-Sections 3.2, 3.3, 3.4, 3.5, and 3.6) specifies the 
``activity" of a star in terms of the surface fluxes of chromospheric lines. 
Combining these surface flux results with our results for $P/\sin i$ (Paper I),
 we construct a set of RACs for M and K dwarfs belonging to various spectral 
sub-types (see Figs. 2, 3, 5, 8, and 12). We first analyze the RAC for K4 and 
K6 dwarfs. Then we re-visit the RAC in M2 dwarfs that was first investigated 
in Paper~XV. We then analyse the RAC in M3 dwarfs for which $v\sin i$ measures 
were reported in HM and for which we present new measurements of the 
chromospheric line equivalent widths here. We also re-visit the RAC in M4 
dwarfs that was studied in Paper XVIII with the new values of the stellar 
parameters determined in Paper I. Finally, we compare the RACs for five 
spectral K and M subtypes, dK4, dK6, dM2, dM3 and dM4, and draw conclusions 
about the differential variations of the RAC from mid-K dwarfs to M4 dwarfs.

We emphasize that our RACs are plotted as a function of the projected rotation 
period $P/\sin i$ and not the rotation period $P$. The average $\sin i$ is 
0.6. Then in order to recover the RACs as a function of $P$, one must multiply 
our RACs by a factor of 1.67. The scatter due to variable $\sin i$ is included 
in our correlations. Nevertheless, we show that one can obtain reasonably good 
empirical RACs in spite of these uncertainties and the uncertainties on our 
measures. 

In Sub-Sections 3.7 and 3.8, we present an overview of the systematic 
properties of the RACs in all of our sub-types, first as regards the {\it 
slopes} of the RACs (Sub-Section 3.7), and then as regards the absolute values 
of the chromospheric emission levels (Sub-Section 3.8). 

In Sub-Section 3.9, we switch to a different approach to constructing RACs. As 
regards rotation, we switch to the Rossby number. As regards chromospheric 
activity, we switch to a quantity which expresses the output power in the H 
and K lines of Ca\,{\sc ii} as a fraction of the star's output power 
($L_{bol}$) (Fig.~15).

Switching our attention to the corona (Sub-Section 3.10), we construct a 
different type of RAC for our targets, this time referring to conditions in 
the corona, rather than the chromosphere. The coronal RAC is obtained by 
plotting $L_X$/$L_{bol}$ as a function of the Rossby number (Fig.~16). 
Comparing ``activity" in chromosphere and corona (Sub-Section 3.11; Figs.~17, 
18) can provide information as to how any given M dwarf star partitions its 
deposition of mechanical energy between chromosphere and corona. To quantify 
this partition, in Sub-Section 3.11 we examine the correlations between $L_X$ 
and $L_{HK}$ (see Figs.~17, 18).   

In the final Sub-Section (3.12), we summarize the properties of the various 
RACs which our data have enabled us to construct.

\subsection{The mean fluxes in the vicinity of the Ca\,{\sc ii} resonance 
doublet.}

\begin{figure*} 
\vspace{-2.5cm}
\hspace{-2.5cm}
\includegraphics[width=12cm,angle=-90]{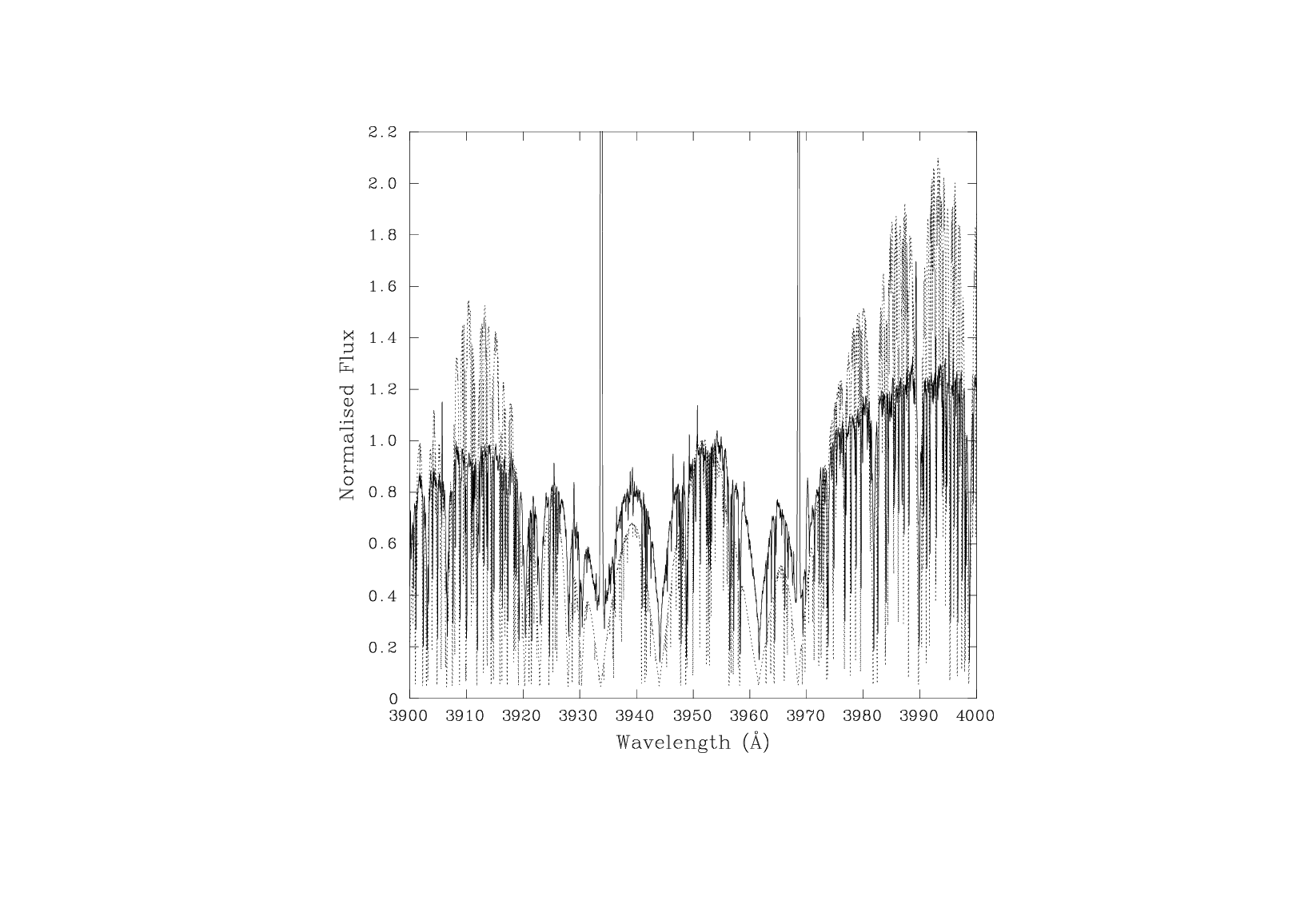}
\vspace{-1.5cm}
\hspace{-2.5cm}
\includegraphics[width=12cm,angle=-90]{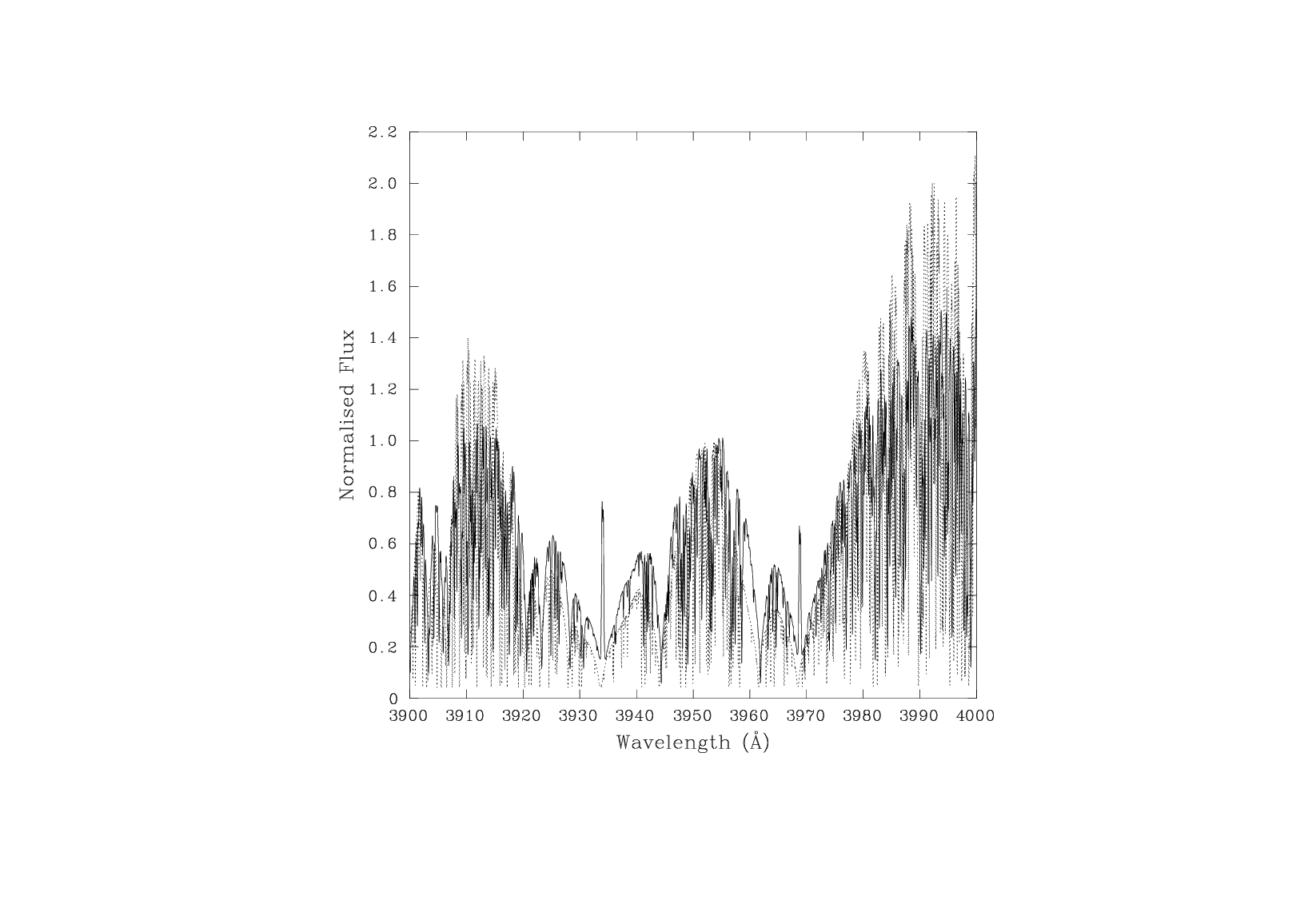}
\caption[]{Upper panel: the observed spectrum of Gl 205 (dM2, solid line) 
together with the theoretical spectrum of de Laverny et al. (2012) for an 
effective temperature of 3500~K (dotted line). Note substantial 
disagreements between the observations and the model. Lower panel: the 
observed spectrum of Gl 570A (dK4, solid line) together with the model for an 
effective temperature of 4500~K (dotted line). Here again there are 
significant differences between the model and observations. All the spectra 
are normalised for the continuum flux at about 3950~\AA, i.e. between the H 
\& K line centers.}
\end{figure*}

In order to evaluate the efficiency of the dynamo mechanism(s) as a function 
of spectral type in M and K dwarfs, it seems preferable to inter-compare the 
RAC's calibrated in terms of absolute energy fluxes, rather than confining our 
attention to the values of the equivalent width (EW). (Nevertheless, the EW 
have proven useful in terms of surface magnetic fields: see Mullan et al. 
2015.) In order to convert from EW to energy fluxes, we calculated the surface 
fluxes in the continuum in the vicinity of the Ca\,{\sc ii} lines from the 
theoretical model atmospheres of de Laverny et al. (2012) and Palacios et al. 
(2010)
\footnote{ http://npollux.lupm.univ-montp2.fr/ or ftp://ftp.oca.eu/pub/laverny/DEPOT/AMBRE$\_$Grid$\_$Flux/} for 
$log(g)=5.0$, [M/H]=0.0 and $\alpha$=0.0. We found that in M2, M3 
and M4 dwarfs, for our observations, the continuum at about 3950~\AA\ 
represents a good evaluation of the background continuum flux for the Ca\,{\sc 
ii} lines. But there are significant discrepancies between the observations 
and the models of de Laverny et al. (2012). We show a spectrum of a dM2 star 
(Gl 205) in Fig.~1 together with the model of de Laverny et al. (2012) for an 
effective temperature of 3500~K. We normalised these two spectra at 1 for the 
continuum flux at 3950~\AA. As one can see in this figure, there are large 
differences in the continuum fluxes at 3910~\AA\ and 4000~\AA\ between the 
model and the observation. We believe these differences are due to missing 
opacities in the models. We also show in the lower panel of Fig.~1 the 
spectrum of Gl 570A together with the model for an effective temperature of 
4500~K. Here again we note important differences: in the spectrum of Gl 570A, 
the continuum flux at 3950~\AA\ is not a good estimate of the continuum flux 
in the vicinity of the Ca\,{\sc ii} lines. In this case, one has to interpolate
 between the fluxes at 3910~\AA\ and 4000~\AA. Therefore, for our estimates of 
the theoretical surface fluxes in the vicinity of the Ca\,{\sc ii} lines for 
M2, M3 and M4 dwarfs, we took the average of the flux at 3950~\AA\ and the 
value interpolated between 3910~\AA\ and 4000~\AA. For K6 and K4 dwarfs, we 
used the value interpolated between 3910~\AA\ and 4000~\AA.

Using this approach, we found that the mean surface fluxes in the continuum 
in the vicinity of the Ca\,{\sc ii} lines are: $5.51\times 10^{5}$, 
$2.18\times 10^{5}$, $4.53\times 10^{4}$, $2.72\times 10^{4}$, and $1.74\times 
10^{4}\ ergs\ s^{-1}\ cm^{-2}\ \AA ^{-1}$ for dK4, dK6, dM2, dM3 and dM4 stars 
respectively. The strong decline (factor of $\sim$30) in surface flux in the 
violet with decreasing $T_{eff}$  from dK4 to dM4 is apparent in these numbers.
 Using these figures, we are now in a position to examine quantitatively how 
the surface fluxes in the Ca\,{\sc ii} lines behave as a function of 
$P/\sin i$. In the subsequent sections, the Ca\,{\sc ii} surface fluxes were 
computed for each star according to its effective temperature. 
We expect that the models give estimates of the continuum surface fluxes 
with a precision of the order of 40\%. As mentioned above, this figure is far 
below the factor of $\sim$30 in the continuum surface fluxes in the violet 
with decreasing $T_{eff}$ from dK4 to dM4. Therefore, the decline in the 
Ca\,{\sc ii} surface fluxes that we observe below (see Figure 14) in the RACs 
from dK4 to dM4 is highly significant. 

\subsection{The RAC in K4 dwarfs}

\begin{figure*} 
\vspace{-0.5cm}
\hspace{-2.5cm}
\includegraphics[width=14cm,angle=-90]{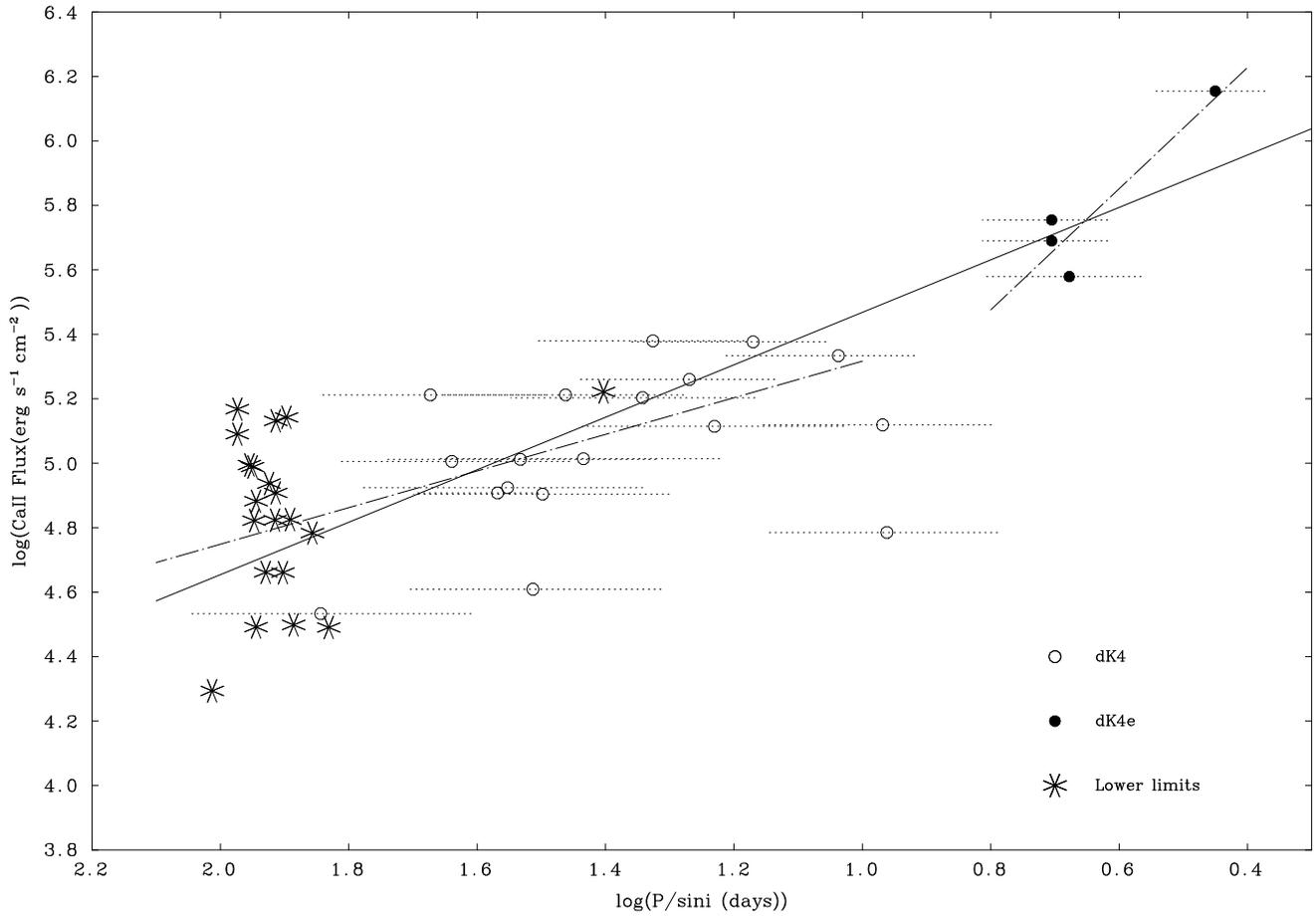}
\vspace{-0.5cm}
\caption[]{Correlation between the mean surface flux of the Ca\,{\sc ii} 
resonance doublet and log($P/\sin i$) for stars with spectral sub-type dK4. 
We overplot the heteroscedastic linear least square fit (LSF: solid line). We 
also plot the heteroscedastic linear LSF separately to the low activity and to 
the high activity stellar subsamples (see the two distinct straight dot-dashed 
lines). 
}
\end{figure*}

In this subsection, we re-investigate the RACs in our sample of dK4 stars, 
that was first investigated in Houdebine (2012b, Paper XVIII). We re-iterate 
that the RACs are of fundamental importance in order to constrain an essential 
parameter of the dynamo mechanism(s): the role played by rotation. In our 
previous studies (e.g. Paper VII, Paper XIV, Paper XVI, Paper XVII, HM), we 
reported on our results for $v\sin i$ and $P/\sin i$ for stars of low activity 
level (i.e. slow rotators) for the spectral sub-types dK4, dM2, dM3 and dM4. 
In Paper I we have reported similar rotational data for dK6 stars. Combining 
these rotational data with the surface fluxes of chromospheric lines in dK and 
dM starsstars provides a unique opportunity to investigate the RACs in a 
fine-grained sample of M and K dwarfs which are sub-divided across five 
closely-spaced, but distinct, spectral sub-types.

We show in Fig.~2 the RAC which we have obtained for our dK4 stars. In 
previously published studies of RACs in cool dwarfs, the majority of the 
results were reported in terms of  homoscedastic linear least square fits 
(LSF) in order to fit their observations. In the present study, we propose 
different approaches. First, we perform homoscedastic linear and quadratic 
least square fits to our samples of low+high activity stars (dK+dKe or dM+dMe).
 Second, we perform heteroscedastic linear and quadratic least square fits 
(LSF that account for measurement errors, see the Appendix) to our samples of 
low+high activity stars. Third, we also compute homoscedastic and 
heteroscedastic linear least square fits to two particular sub-sets of our 
data, namely, stars in which activity is at a low level (dK, dM) and at a high 
level (dKe, dMe) respectively. We inter-compare the results of these fits for 
various spectral sub-types in turn in this subsection and in the following 
subsections. As we shall see, these various fits allow us to account for the 
complexity of our data sets. In our least-square fits, we do not include in 
general the suspected low $\sin i$ stars, lower limits or spectroscopic 
binaries. However, in the present sub-section, for dK4 stars, it appears that 
some of the lower limit measurements do correlate with the other measurements 
of dK4 stars. Therefore, we included some of these measurements in our 
correlations (we did not include obvious outliers).

The heteroscedastic linear least-squares fit to the data in Fig.~2 is shown by 
the solid line. The equation for the solid line in Fig. 2 is as follows:

\begin{equation}
F_{CaII} = 1.916\pm  0.32\times 10^{6} \times (P/sini)^{-0.8140\pm 0.059}.
\end{equation}

The $\chi^{2}$ for the fit in Eq.~(1) is only 0.033 for our sample of 34 
dK4+dK4e stars. The statistical significance of this fit is 99.9\% (see 
Table~1). (Results of all least-squares fits for this and subsequent spectral 
sub-types can be found in Table 1.)

The homoscedastic linear least-squares fit yields similar results: 

\begin{equation}
F_{CaII} = 1.57\pm  0.34\times 10^{6} \times (P/sini)^{-0.756\pm 0.05}.
\end{equation}

The $\chi^{2}$ for this fit is 0.032 for 34 dK4+dK4e stars. The correlation 
coefficient for this fit is found to be 0.884 and the statistical significance 
of the correlation is at least 99.9\% (see Table~1). Therefore, these two fits 
are highly statistically significant. The reason for studying heteroscedastic 
fits in addition to the homoscedastic fits is that our $P/\sin i$ measurements 
have large errors and these errors vary as a function of the values of $P/\sin 
i$ (rapid rotators have smaller errors than slow rotators). The presence of 
these variable errors may yield heteroscedastic fits that are in some regards 
different from the homoscedastic fits. We shall see examples of this statement 
in some of the following subsections.

Two aspects of the fits in eqs. (1) and (2) will be referred to in the 
subsequent discussion. First, the exponent of $P/\sin i$  will be referred to 
as the ``RAC slope" in a plot of $log(F_{CaII})$ versus $log(P/\sin i)$. 
Systematic changes in the numerical value of the RAC slope will lead us to an 
important conclusion of the present paper. Second, the numerical coefficient 
closest to the equals sign is a measure of the absolute level of the 
chromospheric emission among the stars in the sample. Systematic changes in 
the numerical value of the coefficient will also be an important conclusion of 
this paper. 

So far in this sub-section, we have done an analysis of all of our dK4 stars, 
i.e. we have combined both the dK4 stars and the dK4e stars into a single 
sample. Now we split our sample up into two groups (dK4 in one, dK4e in the 
second), and analyze each group separately. Specifically, we now apply 
heteroscedastic and homoscedastic linear LSF to the sub-samples of only the 
low activity stars (dK4) and then repeat the exercise including only the high 
activity stars (dK4e) (see Table~1). The reason for 
performing these separate fits is that we found that the linear fits did not 
reproduce well both the low activity and high activity sub-samples: this 
result led us to wonder if these two samples might contain different dynamo 
modes. If it turns out that indeed different dynamo modes are at work in slow 
rotators and fast rotators, then separate analyses of the data sets is 
warranted. We find that the linear heteroscedastic LSF to the dK4 low activity 
stars yield:

\begin{equation}
F_{CaII} = 7.66\pm  1.89\times 10^{5}\times (P/sini)^{-0.568\pm 0.084}.
\end{equation}

The $\chi^{2}$ for this fit is 0.034 for our sample of 30 dK4 stars and the 
statistical significance of the correlation is at least 99.9\% (see Table~1). 
This fit is shown as the straight dot-dashed line in Fig.~2. For the 
homoscedastic fit, we obtain:

\begin{equation}
F_{CaII} = 9.33\pm  3.16\times 10^{5}\times (P/sini)^{-0.624\pm 0.11}.
\end{equation}

The $\chi^{2}$ for this fit is only 0.031 for 30 dK4 stars, the correlation 
coefficient is 0.744 and the statistical significance is at least 99.9\% (see 
Table~1). Therefore, both of these fits (homo and hetero) are highly 
statistically significant in spite of the scatter in the data. We note that 
the differences between the homoscedastic and the heteroscedastic fits fall 
within the uncertainties of the parameters of the fits.

The linear heteroscedastic LSF to the dK4e high activity stars yield:

\begin{equation}
F_{CaII} = 9.73\pm  5.07\times 10^{6}\times (P/sini)^{-1.877\pm 0.51}.
\end{equation}

The $\chi^{2}$ for this fit is 0.038 for 4 dK4e stars and the statistical 
significance is 99.7\% (see Table~1). This fit is shown as the upper dot-dashed
straight line in Fig.~2. We must admit that our dK4e sample contains only 4 
stars and that therefore the RAC for dK4e stars is not well constrained. 
However, as we shall see in the subsequent sub-sections, we find that this RAC 
is consistent with the trend of the high activity star RACs at other spectral 
types. 

An important feature of our results emerges when we compare Eqs. (1) and (3), 
and when we compare Eqs. (2), (4) and (5). Whether we consider homo- or 
hetero-scedastic results, we find that the gradient of the RAC for the low 
activity stars alone (-0.57, -0.64) is shallower in magnitude than that of the 
combined sample of low+high activity stars (-0.81, -0.76), and that the 
gradient of the RAC for the high activity stars alone (-1.88, -1.877) is 
larger in magnitude than that of the combined sample of low+high activity 
stars (-0.81, -0.76). We shall find that this occurs systematically between 
the low activity sub-samples and the full samples for all five of our spectral 
sub-types (see Table~1 and Sections 3.3-3.6 below). The effect is more 
pronounced as the spectral type increases (see also Sect.~3.7). For the high 
activity sub-samples, we shall find that the slope is steeper than that of the 
combined samples for dK4, dK6 and dM2 stars, but that it reverses for the dM3 
and dM4 stars (at the TTCC and beyond) and becomes shallower than that of the 
combined samples. As a consequence, the linear fits to the combined samples of 
low+high activity stars tend to overestimate the slope in the low activity 
sub-samples, and underestimate the slopes in the high activity stars 
sub-samples for dK4, dK6 and dM2 stars. Therefore, we also performed 
homoscedastic and heteroscedastic quadratic LSF to our samples of low+high 
activity stars for our five spectral sub-types. The quadratic fits allow us 
to reproduce both the shallower gradient of the linear LSF among the low 
activity stars as well as the higher fluxes among the high activity stars. 
This shows that the quadratic fit may give a better description of the data 
than the linear fit for the combined samples of low+high activity stars. 
However, according to an anonymous referee comments, the quadratic fits may 
not be significantly different from the results of the linear fits. To argue 
on this point, the referee performed a simulation on sub-samples of dM and dMe 
stars with random errors typical of those we find in this study. The referee 
found also a shallower slope among his sub-sample of slow rotators. Given the 
evidence at hand, the signatures identified as possible evidences supporting 
the case for a quadratic fits are most probably due to random errors working 
in combination with a fairly narrow $log(P/\sin i$) domain as compared to 
those errors. Therefore, although the slopes of the high activity sub-samples 
in dK4, dK6 and dM2 are in favor of a quadratic description of the data, we 
cannot yet conclude that quadratic fits represent definitely a better 
representation of the full datasets. We can also note that the $\chi^{2}$s are 
comparable between linear and quadratic fits. More data will be required to 
conclude.

As an anonymous referee rightly pointed to, if the data are localized 
to a domain that is only 2-3 times as large as a typical error bar, random 
measurement errors can easily randomize the data and weaken the slope of any 
true correlation. However, this point is not statistically correct. If the 
distribution of the errors is Gaussian (which we assume here), then the correct
 parameter to compare to the RAC domain $R$ is $\delta = 
\frac{<error>}{\sqrt{n}}$ where $n$ is the number of measures. The value of 
$\frac{R}{\delta}$ is 17.2 for our low activity dK4 star sample which is far 
larger the values of 2-3 mentionned above. It appears that $\frac{R}{\delta}$ 
lies in the range 17 to 40 for our low activity and high activity RACs. The 
levels of confidence of a given LSF depends not only on the mean error but 
{\em alo on the number of measures}. Therefore, our LSFs should be established 
to a fairly high level of confidence, which is confirmed by the high 
statistical significances that we obtained for our least square fits (see 
Table~1). Nevertheless, the simulation of the referee is of interest to us. We 
take this important point into account and we emphasize here the preliminary 
character of our results. Indoubtedly, these interesting results should be 
comfirmed with additional data, obtained preferably with a higher resolution 
spectrograph for the slow rotators such as ESPRESSO (ESO, $R=220,000$). 
Therefore, in the following sub-sections and Sect.~3.7, our results on the 
separate fits to the sub-samples of low and high activity stars should be 
considered with caution and are only preliminary. They should be confirmed 
with larger stellar samples.

\subsection{The RAC in K6 dwarfs}

\begin{figure*} 
\vspace{-0.5cm}
\hspace{-2.5cm}
\includegraphics[width=14cm,angle=-90]{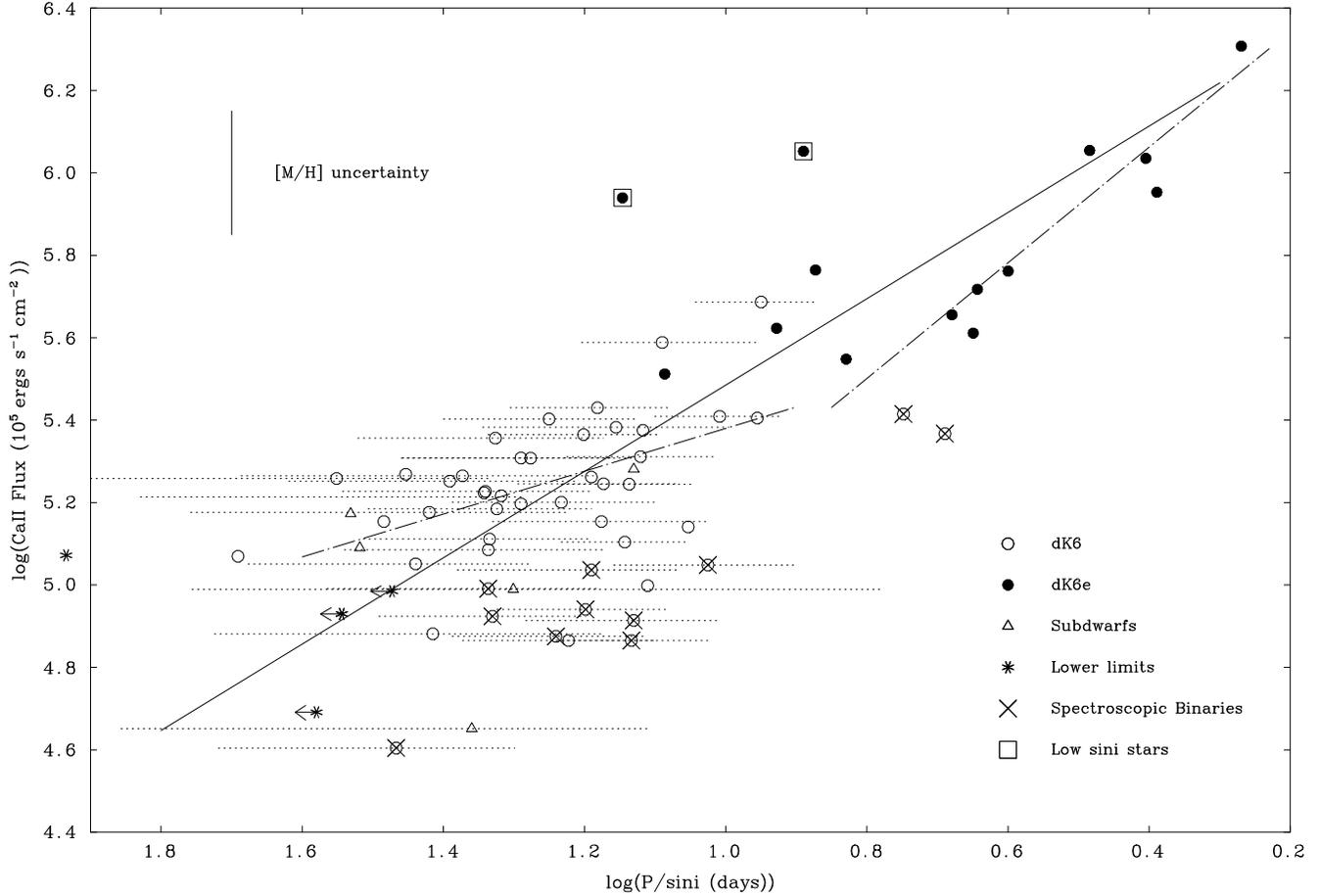}
\vspace{-0.5cm}
\caption[]{Correlation between the logarithm of the mean surface flux of the 
Ca\,{\sc ii} resonance doublet and log($P/\sin i$) for stars with spectral 
sub-type dK6. We overplot the heteroscedastic linear LSF (solid line). 
We also plot the heteroscedastic linear LSF to the low activity stellar 
subsample (dot-dashed straight line between $log(P/\sin i) = 1.6$  and 0.9) 
and the heteroscedastic linear LSF to the high activity stellar subsample (a 
separate dot-dashed straight line between $log(P/\sin i)=0.85$ and 0.25). We 
show in the top-left of the figure an estimate of the error on the Ca\,{\sc 
ii} mean fluxes due to metallicity effects. 
}
\end{figure*}

In this subsection, we discuss the results for our sample of dK6 stars. For 
the sake of consistency with previous measurements in this series of papers, 
and to avoid duplication in describing the method, we refer the reader to 
Papers VI and XV to see how we evaluate the EW for the Ca\,{\sc ii} resonance 
doublet and the H$_{\alpha}$ line. In Table 2, we list the EW we have 
obtained for the Ca\,{\sc ii} resonance doublet and for the H$_{\alpha}$ 
line for our dK6 stars.

We note that our dK6 stellar sample contains only one star with H$_\alpha$ 
definitely in emission (i.e. a dK6e star): Gl~517 (EQ Vir), with the fastest 
rotation ($v\sin i$ = $9.77\ km\ s^{-1}$, $P$/sin$i$ = 3.05 days, see Paper I).
 Our ``dK6" sample also includes one star of ``intermediate" activity with 
H$_{\alpha}$ neither in emission nor absorption (i.e. a dK6(e) star): Gl~208 
(Table~2). As regards EQ Vir, this star has been classified in the literature 
as a dK5 star. In fact, Gl 517 is a BY~Dra star with colours which vary with 
time (depending on spot coverage). 

Our observations contain many slow rotators but very few fast rotators (dKe 
stars). Therefore, in order to complete our sample and have a better defined 
RAC for dK6 stars, in Paper I we compiled $v\sin i$ and $P\sin i$ measures 
from the literature. We found 43 additional $v\sin i$ and $P\sin i$ 
measures which brings our total compilation of measures to 150. 

We also retrieved some FEROS spectra from the ESO Archive for Gl~142, 
Gl~885A, Gl~900 and HIP~113597. The measures of the Ca\,{\sc ii} EW for these 
stars are given in Table~2. We also searched the literature for other 
measurements of the Ca\,{\sc ii} EW. We found only a few measurements of the 
Ca\,{\sc ii} EW or flux for our dK6 stars (Marilli et al. (1986), Rutten 
(1987), Duncan et al. (1991), Browning et al. (2010)). These measures are 
also listed in Table~2.

We saw in Paper~XV (see also Sect.~3.4) that correction for metallicity effects
 in dM2 stars is essential in order to obtain a good correlation between the 
Ca\,{\sc ii} EW and $P/\sin i$. This is due to the fact that the Ca\,{\sc ii} 
line formation depends sensitively on the Ca abundance (Houdebine \& Panagi 
2016, in preparation). Here the same applies to dK6 stars and to the Ca\,{\sc 
ii} surface fluxes. In Paper I we compiled [M/H] measures for our dK6 stellar 
sample and here we computed the Ca\,{\sc ii} surface fluxes corrected for 
metallicity effects, assuming a proportionality between [M/H] and Ca\,{\sc ii} 
surface fluxes. 

We show in Fig.~3 the RAC which we have obtained for our dK6 stars. The 
heteroscedastic linear LSF to the data in Fig.~3 is shown 
by the solid line. One of the stars in our sample, Gl~208 (a dM6(e) star with 
$P$/sin$i$ =  8.94 days), is probably also a fast rotator: but its value of 
$v\sin i$ is too small (and its value of $P/\sin i$ is too large) to be 
entirely consistent with its observed activity level. It is possible that we 
are viewing this star close to its rotation axis, i.e. $\sin i$ may be
 atypically low. We also report on two relatively active stars with a rather 
slow rotation: Gl~455.1 and Gl~907.1. These stars depart noticeably from the 
main correlation. We believe that these stars also have low $\sin i$. Most of 
the other stars follow the solid-line correlation fairly well with a few 
exceptions: a group of stars that lie below the main correlation and are low 
activity-fast rotators. One of them is a subdwarf: their internal structure 
may cause their dynamos to operate differently from those in main sequence K 
stars. For the few other stars, similar discrepancies from the RAC have been 
found in dM2 stars (Paper XV and Sect.~3.4 below); they will also be noted 
among dM3 stars (see Sect.~3.5 below). Reiners et al. (2012) also observed a 
few stars which are discrepant compared to the global trend: Reiners et al. 
disregarded such objects as being due to measurement errors. However, the 
repetition with which they appear in our samples, and the fact that they are 
relatively rapidly rotating, lead us to believe that these {\em low-activity 
relatively fast rotators} may indeed exist as a possibly significant sub-set 
of late K and M dwarfs. The best explanation we have is that, in a star such 
as Gl~412.3AB, which exhibits a slight asymmetry when its photospheric 
spectral lines are subjected to rotational analysis (see Paper I), these stars 
may be unresolved spectroscopic binaries: as such, they could yield abnormally 
broadened cross-correlation profiles (see also Gl~186AB in Paper I). Therefore 
these stars are probably spectroscopic binaries that are unresolved at the 
time the observations were made. This is quite plausible since binarity is a 
common characteristic of late-type dwarfs. We label these stars as 
spectroscopic binaries in Fig.~3.

Including the known binaries in our dK6 sample, the number of probable 
spectroscopic binaries yield a proportion of 20\% binaries in our full 
sample. This is consistent with the recent finding of Ward-Duoung et al. (2015)
that binary stars constitute a percentage of about 25\% of their stellar 
sample for $M_{*}\sim 0.7M_{\odot}$. We shall see in the subsequent subsections
 that the proportion of binaries amounts to about 28\%, 37\% and 48\% for our 
dM2, dM3 and dM4 stellar samples respectively. Our dM2 and dM3 results are 
again compatible with the fraction of about 32-37\% derived by Ward-Duoung et 
al. (2015) for these stars. For our dM4 sample, our figure of 48\% is somewhat 
larger than the 37\% fraction expected at this spectral type. This 
disagreement may find a simple explanation in the bias in detecting parallaxes
for nearby M dwarfs. Indeed, faint single M4 dwarfs are much more difficult to 
detect than the companions to nearby brighter M and K dwarfs. Therefore, our 
dM4 sample, which is largely based on parallax surveys of nearby dwarfs is 
biased towards the detection of faint companions to nearby brighter M and K 
dwarfs. This should explain why we have a larger binary fraction for the M4 
dwarf sample.

We should like to emphasize four points regarding the plot in Fig.~3. (i) 
Metallicity differences from star to star contribute to the scatter about 
the dK6 RAC in Fig.~3. By analogy, in Paper~XV we reported that metallicity 
differences among dM2 stars are responsible for the greater part of the 
scatter in the RAC. (ii) Another contribution to the scatter is that our 
statistics on the Ca\,{\sc ii} line EW are poor; we have very few measures of 
the Ca\,{\sc ii} line EW for most of our dK6 stars compared to what we have 
for the stars in our dM2 sample, and the Ca\,{\sc ii} line EW is known to 
vary with rotation and with the phase of the activity cycle (e.g. Baliunas et 
al. 1995). However, we have little or no information as to which part of the 
cycle our observations happened to ``catch" any particular star. (iii) The 
value of $\sin i$ may vary from 1 to 0 and certainly contributes to the 
scatter in Fig.~3. (iv) The range of effective temperatures in our dK6 sample 
is somewhat larger than in our samples of dM2, dM3 and dM4 stars (e.g. only 
$\pm 70$~K in dM3 stars, where $T_{eff}$ ranges from 3210 to 3350~K). 

We would like to emphasize as well that a few of the active stars in our dK6 
stellar sample are candidate young stars that may not yet have contracted to 
the Main-Sequence (MS). A reliable way to identify the Pre-Main-Sequence (PMS) 
stars in our samples is the stellar radius: the fact is, PMS stars that have 
not yet contracted to the MS have abnormally large radii. We identified three 
such stars in our dK6 sample: GJ 1177A, GJ 182 and GJ 425B are possible PMS 
stars (see Paper I). However, Gl 425B is a rather low activity star. As such, 
the abnormally large radius for this star is probably due to binarity. 
Nevertheless, we find that these stars {\em do correlate well with the MS 
stars}. In our M2 sample (see next Section), we have identified also two PMS 
stars: GJ 1264 and GJ 803. But again, as we shall see, these PMS stars do 
correlate very well with the other MS stars. In our M3 sample, GJ 277A is a 
possible PMS star. In our M4 sample, GJ 2069A, GJ 3322, GJ 669A, GJ 695B, 
GJ 812A are possible PMS stars according to their radii. {\em We emphasize 
that all these stars do not rotate especially fast. There are many MS stars 
that rotate faster.} These PMS stars are expected actually to spin up as they 
contract to the MS, and young MS stars are expected to be the fastest rotators 
(e.g. Barnes 2003; for a theoretical model of this process, see e.g. Fig.~2 in 
MacDonald \& Mullan 2003). That is what we actually observe in our samples of 
stars: young MS stars are the fastest rotators (e.g. among our M4 sample; GJ 
3631, GJ 3789, GJ 4020B, GJ 4338B, GJ 431, GJ 630.1, GJ 791.2A) with $v\sin i$ 
in excess of 15 $km\ s^{-1}$ and up to 56 $km\ s^{-1}$.

All our RACs demonstrate that PMS stars do not stand out as significantly 
different from the main correlations of the MS stars (see subsequent Sections).
 Numerous previous studies also found similar results: e.g. Mamajek \& 
Hillebrand (2008), Browning et al. (2010), Christian et al. (2011), West et 
al. (2015). This finding is rather intriguing since PMS stars may have 
internal structures which differ from those of MS stars: in fact, some PMS 
stars may even be fully convective. More investigation is required to confirm 
this result, but so far, in our samples, we found no definitive evidence that 
PMS stars obey different RACs from those of MS stars.

The heteroscedastic linear LSF to the RAC in Fig.~3 gives the following:

\begin{equation}
F_{CaII} = 3.402\pm  0.021\times 10^{6} \times (P/sini)^{-1.047\pm 0.042}.
\end{equation}

The parameters of this LSF are given in Table~1. The $\chi^{2}$ is 0.093 and 
the statistical significance is 99.9\% for 55 dK6 stars. The homoscedastic 
linear LSF to the same data set gives:

\begin{equation}
F_{CaII} = 1.95\pm  0.34\times 10^{6} \times (P/sini)^{-0.81\pm 0.06}.
\end{equation}

The correlation coefficient for this fit is 0.876, the statistical significance
 is 99.9\% and the $\chi^{2}$ is only 0.021 for 55 dK6 stars. We find in this 
case that the heteroscedastic and the homoscedastic linear fits give slightly 
different results for the slope and for the flux amplitude: This is due to the 
fact that high activity stars (fast rotators) have smaller uncertainties in 
$P/\sin i$ compared to low activity stars, and this give them much higher 
weights in the least square fit. As a result, the heteroscedastic linear 
solution fits the high activity stars better, whereas the homoscedastic linear 
solution clearly underestimates the fluxes among high activity stars. 
Therefore, in this case the heteroscedastic fit gives more sensible results.

For future reference, we note that in Fig.~3, the correlation in Eq.~(6) spans 
the entire range of $P/\sin i$ values for which we have dK6 data. In 
particular, it is important to notice that the dK6 data exhibit no evidence 
for a flattening (or ``saturation") of the RAC at the shortest periods 
($P/\sin i$ = 1.8 days). This is consistent with previous results for the 
earlier spectral type (dK4) which also showed no signs of saturation (see 
Sect.~3.2). Neither is there evidence for saturation among dM2 stars (see 
Fig.~5 below).

Having analyzed the combined samples of slow and fast rotators, we now turn to 
performing heteroscedastic and homoscedastic linear LSF on the sub-samples of 
low activity stars (dK6) and high activity stars (dK6e) separately. The 
heteroscedastic linear LSF to the dK6 low activity stars yields:

\begin{equation}
F_{CaII} = 8.23\pm  2.23\times 10^{5}\times (P/sini)^{-0.531\pm 0.12}.
\end{equation}

The $\chi^{2}$ for this fit is 0.017 and the statistical significance is 99.9\%
 for 41 dK6 stars (see Table~1). This fit is shown as the straight dot-dashed 
line in the lower left portion of Fig.~3. The homoscedastic linear LSF to this 
sub-sample gives:

\begin{equation}
F_{CaII} = 1.29\pm  0.36\times 10^{6}\times (P/sini)^{-0.637\pm 0.11}.
\end{equation}

The correlation coefficient for this fit is 0.706, the statistical significance
 is 99.9\% and the $\chi^{2}$ is only 0.014 for 41 dK6 stars. Therefore, these 
two fits are highly statistically significant at a confidence level greater 
than 99.9\%.

We also fitted the high activity stellar sub-sample. The heteroscedastic 
linear LSF to the dK6 high activity stars yields:

\begin{equation}
F_{CaII} = 4.20\pm  3.52\times 10^{6}\times (P/sini)^{-1.402\pm 0.43}.
\end{equation}

The $\chi^{2}$ for this fit is 0.023 and the statistical significance is 99.9\%
 for 9 dK6e stars (see Table~1). This fit is shown as the straight dot-dashed 
line in the upper right portion of Fig.~3.

We again find that the gradient of the RAC for the low activity stars alone 
(-0.86, -0.64) seems shallower than that of the low+high activity star sample 
(-1.05, -0.81) and that the gradient of the RAC for the high activity stars 
alone (-1.402, -1.364) seems steeper than that of the low+high activity star 
sample (-1.05, -0.81). We find that this difference in RAC steepness occurs 
systematically between the low activity sub-samples and the full samples for 
all five of our spectral sub-types (see Table~1 and Sections 3.4-3.6 below). 
The parameter $\frac{R}{\delta}$ is 17.20 for the high activity star sub-sample
 and 39.68 for the low activity star sub-sample. Therefore, the parameter 
 $\delta =\frac{<error>}{\sqrt{n}}$ which is a normalized estimate of the mean 
error on the measurements is much smaller than the period domains of the RACs 
in both cases. Hence, these LSFs should be relatively well established, which 
corresponds to the high statistical significances we obtain (Table~1). However,
 we consider these results as still preliminary because the domains of the 
RACs are relatively small compared to the typical uncertainties on individual 
measures.

If this difference between the RACs of the low and high activity sub-samples 
is confirmed to be true, this may represent a discovery of interest: Indeed, 
to the extent that the slope of an RAC is related in some way to an underlying 
dynamo process, the difference which we found between the slopes of the low 
and high activity sub-samples suggests that there may exist two different 
dynamo regimes for these two sub-samples of stars. We find that this 
difference between low and high activity regimes persists for all our samples 
of stars with different spectral sub-types. However, we observe a difference 
for stars which are more massive than the TTCC and stars which are less 
massive than the TTCC. For the former, the slope among high activity stars is 
{\em steeper} than the slope among the low+high activity stars, whereas for 
the latter, we find that the slope among high activity stars is {\em shallower}
 than the slope among the low+high activity stars. This preliminary result 
could be of interest for the dynamo mechanisms. More data with a higher 
resolution spectrograph will be needed to confirm these results.

\subsection{The RAC in M2 dwarfs}

We have re-investigated the RAC in the sample of dM2 stars of Paper~XV. We 
have given the new stellar parameters and new $P/\sin i$ values in Paper~I. We 
show the raw data in Fig.~4. We can see in this figure that there is a large 
scatter among M2 dwarf low activity stars. Most of the subdwarfs also lie 
significantly apart from most of the M2 dwarfs. Most of this large scatter is 
due to metallicity effects on the Ca\,{\sc ii} line formation.

We used the empirical correlation found by Houdebine (2008, Paper VII) to 
determine the metallicity for each star as a function of its radius (see 
Paper~I). We use these values here to correct the Ca\,{\sc ii} surface fluxes 
for metallicity effects, assuming a proportionality between surface flux and 
metallicity (optically thin case). The Ca\,{\sc ii} line EW, surface fluxes 
and surface fluxes corrected for metallicity are all listed in Table~3.

\begin{figure*} 
\vspace{-0.5cm}
\hspace{-2.5cm}
\includegraphics[width=14cm,angle=-90]{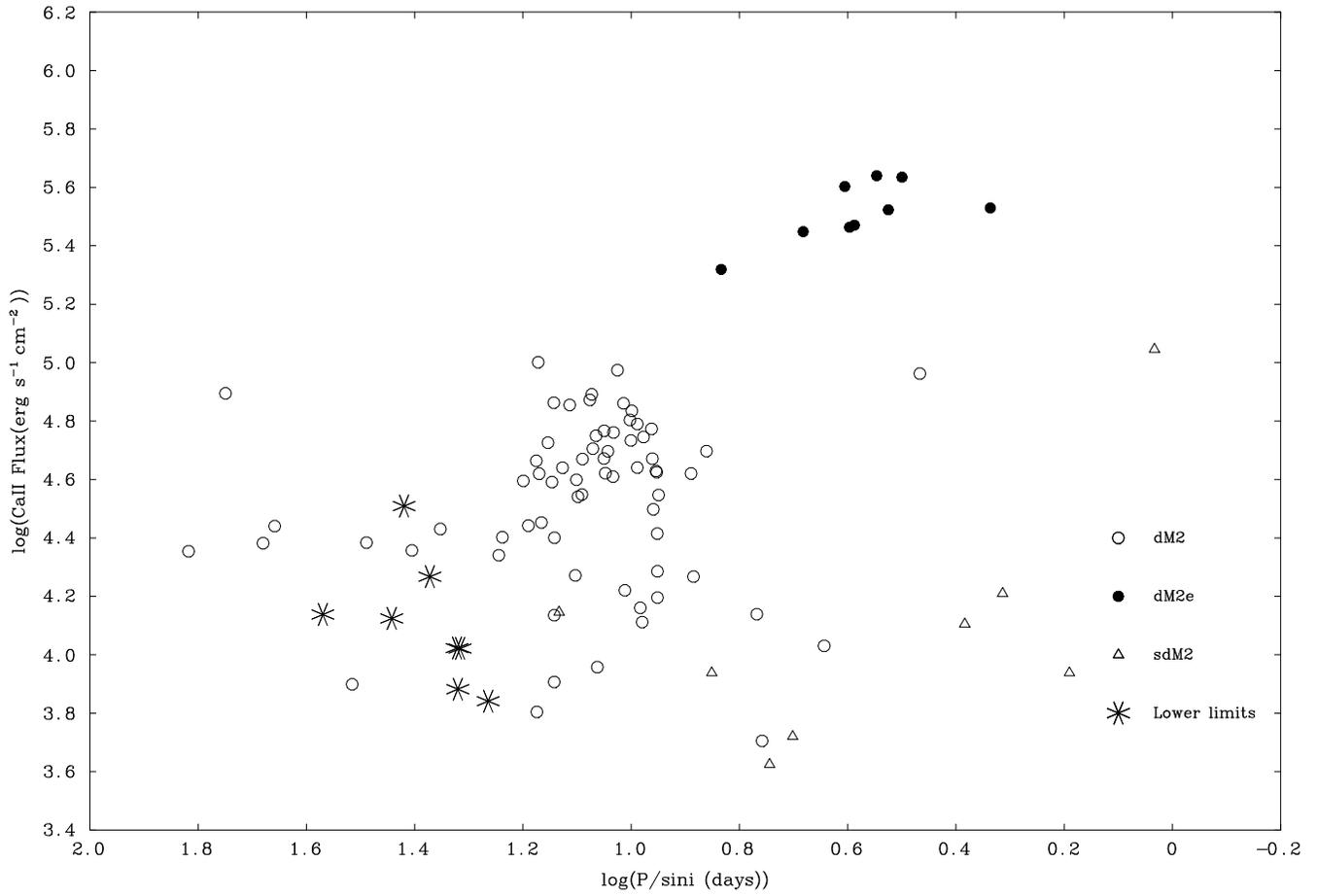}
\vspace{-0.5cm}
\caption[]{Correlation between the logarithm of the mean surface flux of the 
Ca\,{\sc ii} resonance doublet and log($P/\sin i$) for stars with spectral 
sub-type dM2. In this plot, no corrections have been made for metallicity 
differences.}
\end{figure*}

We show in Fig.~5 the Ca\,{\sc ii} surface fluxes (after the data have been 
corrected for metallicity effects) as a function of $P/\sin i$. The 
improvement of the correlation between this figure and Fig.~4 is striking: it 
shows how important it is to correct the RAC for metallicity effects in M 
dwarfs. In Fig.~5, the scatter has been much reduced among the slow rotators, 
and also among the fast rotators. Moreover, even subdwarfs correlate with 
normal metallicity dwarfs. 

In Fig. 5, we plot the uncertainties on $P/\sin i$ as dotted lines for each 
star. For these uncertainties, we assumed an uncertainty of $\pm 
0.14\ km\ s^{-1}$ for measures of $v\sin i$ below $1\ km\ s^{-1}$, $\pm 0.30\ 
km\ s^{-1}$ for $v\sin i$ between $1\ km\ s^{-1}$ and $6\ km\ s^{-1}$, and 
$\pm 0.50\ km\ s^{-1}$ for $v\sin i$ above $6\ km\ s^{-1}$ (see Paper I). 
Uncertainties in the values of $P/\sin i$ have already been included in Figs.~2
 and 3 above, following the same prescription as we describe here. Similar 
uncertainties will be included in Figs.~8 and 12 below in connection with dM3 
and dM4 stars respectively.

In Fig. 5, the straight solid line shows the heteroscedastic linear LSF to the 
data. There is little  difference between this fit and the linear LSF that was 
obtained in Paper~XV. There is a moderate shift between the two correlations, 
and the gradient is almost unchanged (see Table~2). We identified some low 
activity-relatively fast rotators in this M2 dwarf sample. As already mentioned
 in the previous section, these stars are probably unresolved spectroscopic 
binaries. So we have labelled these stars ``spectroscopic binaries" in Fig.~5. 
There are also a few stars that have too long rotation periods for their 
surface fluxes. We believe these stars are a sub-group of stars with low 
$\sin i$. These stars are shown as squares in Fig.~5.

\begin{figure*} 
\vspace{-0.5cm}
\hspace{-2.5cm}
\includegraphics[width=14cm,angle=-90]{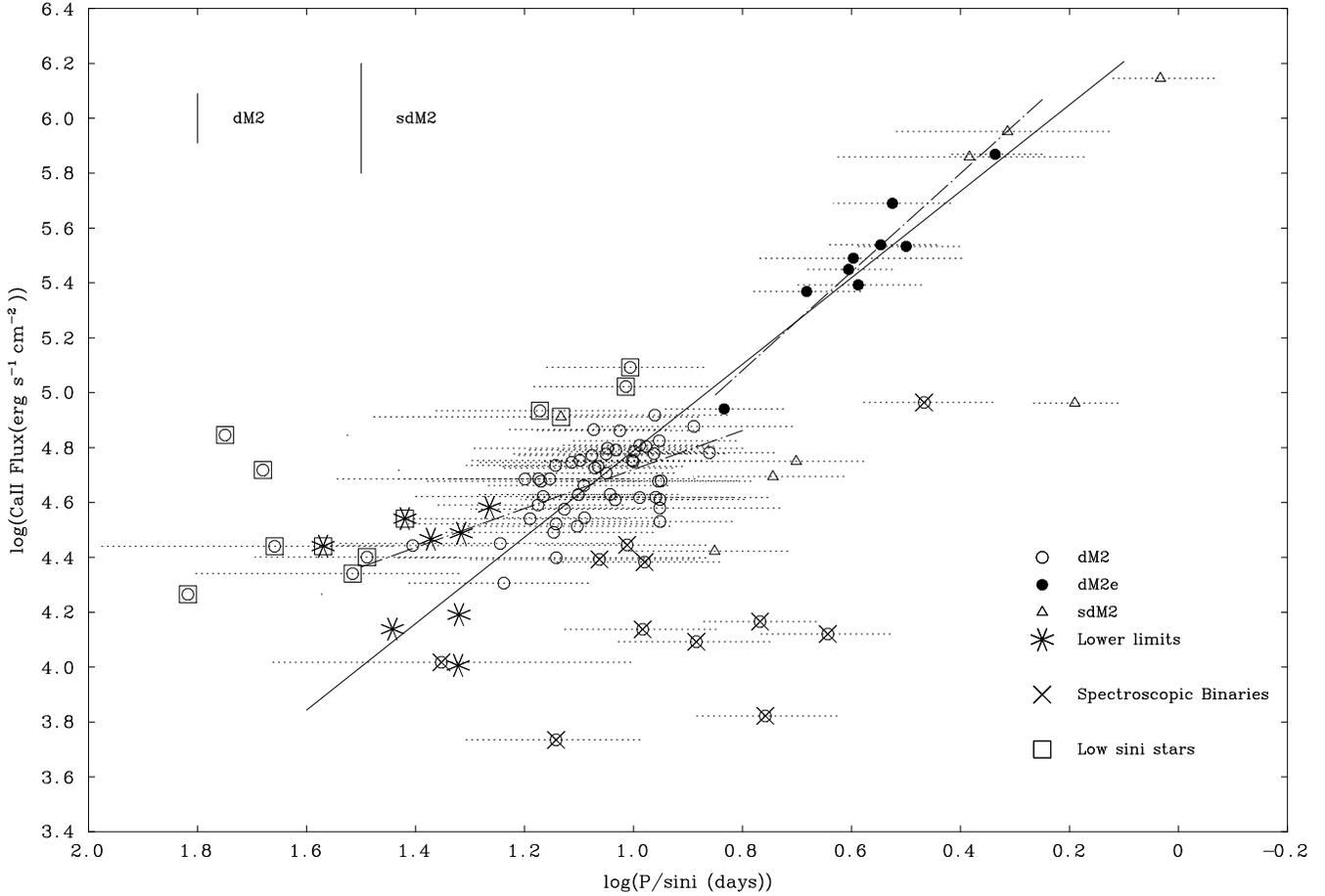}
\vspace{-0.5cm}
\caption[]{Correlation between the logarithm of the mean surface flux of the 
Ca\,{\sc ii} resonance doublet corrected for metallicity effects and 
log($P/\sin i$) for stars with spectral sub-type dM2. The comparison between 
this correlation and that of Fig~4 is striking: it shows how important it is 
to correct for metallicity effects the RAC in M dwarfs. In this figure, even 
subdwarfs now correlate with normal metallicity dwarfs. The straight solid 
line shows the heteroscedastic linear LSF to the data. The dot-dashed lines 
show the heteroscedastic linear LSF to the low activity star and high activity 
star sub-samples. We show in the upper-left corner of the figure estimates of 
the uncertainties on the Ca\,{\sc ii} mean surface fluxes due to the 
corrections for metallicity effects, for dwarfs and subdwarfs.}
\end{figure*}

The heteroscedastic linear LSF to the RAC in Fig.~5 gives the following:

\begin{equation}
F_{CaII} = 2.312\pm  0.276\times 10^{6} \times (P/sini)^{-1.575\pm 0.058}.
\end{equation}

The parameters of this LSF are given in Table~1. The $\chi^{2}$ is 0.028 for 
66 dM2 stars.

The homoscedastic linear LSF to the RAC in Fig.~5 gives the following:

\begin{equation}
F_{CaII} = 1.89\pm  0.85\times 10^{6} \times (P/sini)^{-1.481\pm 0.068}.
\end{equation}

The parameters of this LSF are given in Table~2. The correlation coefficient 
is 0.949 and the $\chi^{2}$ is 0.018 for 66 dM2 stars. The gradient of 
this correlation (-1.481) is very close to that of the correlation found in 
Paper~XV (-1.53). In fact there are few differences between the two results: 
the correlation has only been shifted slightly in $P/\sin i$. Both the 
homoscedastic and the heteroscedastic give highly significant correlations 
with a statistical significance better than 99.9\%. The heteroscedastic 
linear fit yields a somewhat slightly larger slope than the homoscedastic 
linear fit. This is because the errors on the dM2e stars are significantly 
smaller than those of the dM2 stars, and therefore the weights of the dM2e 
stars in the correlations are larger than those of dM2 stars in the 
heteroscedastic fit. As a consequence, the heteroscedastic fit goes through 
the sub-sample of dM2e stars whereas the homoscedastic fit underestimates 
clearly the higher fluxes of the dM2e stars. Therefore the heteroscedastic fit 
provides globally a better evaluation of the slope for the dM2+dM2e stars.

We also performed linear LSF to the sub-samples of only the dM2 low activity 
and the high activity dM2e stars (see Table~1). The heteroscedastic linear LSF 
to the dM2 low activity stars sub-sample yields:

\begin{equation}
F_{CaII} = 2.68\pm  0.77\times 10^{5} \times (P/sini)^{-0.709\pm 0.14}.
\end{equation}

Therefore, the gradient of the RAC for the low activity dM2 stars is again 
significantly shallower than that of the combined sample of low+high activity 
stars.

For the homoscedastic linear LSF to the dM2 low activity stars, we obtain:

\begin{equation}
F_{CaII} = 4.17\pm  1.15\times 10^{5} \times (P/sini)^{-0.891\pm 0.12}.
\end{equation}

Both the heteroscedastic and the homoscedastic fits show that the gradient of 
the RAC for the low activity dM2 stars is again (as for dK4 and dK6 stars) 
significantly shallower than those of the combined sample of low+high activity 
stars. The $\chi^{2}$ for these fits are 0.046 and 0.013 respectively. These 
fits are statistically significant at a confidence level better than 99.9\%. 
We note that the slope of the heteroscedastic fit is slightly shallower than 
that of the homoscedastic fit and is also shallower than the value found for 
the heteroscedastic fit for the dK6 sample. However, if we take into account 
errors (see Table~1), there are in fact no significant differences between 
these values.

The heteroscedastic linear LSF to the dM2e high activity star sub-sample 
yields:

\begin{equation}
F_{CaII} = 3.28\pm  0.54\times 10^{6} \times (P/sini)^{-1.793\pm 0.14}.
\end{equation}

Therefore, the slope of the RAC for the high activity dM2e stars is again 
steeper than that of the combined sample of low+high activity stars. We note 
that this slope gets closer to that of the combined sample as we move from dK4 
to dM2. The differences between these two slopes were larger for dK4 and dK6 
stars.

The parameter $\frac{R}{\delta}$ is 23.45 for the high activity star sub-sample
 and 34.77 for the low activity star sub-sample. Therefore, the parameter 
 $\delta =\frac{<error>}{\sqrt{n}}$ which is a normalized estimate of the mean 
error on the measurements is much smaller than the period domains of the RACs 
in both cases. Hence, these LSFs should be relatively well established, which 
corresponds to the high statistical significances we obtain (Table~1). However,
 we consider again these results as still preliminary because the domains of 
the RACs are relatively small compared to the typical uncertainties on 
individual measures.

\subsection{The RAC in M3 dwarfs}

In HM, we have already listed the results for the rotational parameters 
$v\sin i$ and $P/\sin i$ of our dM3 stellar sample. However, we re-computed 
the stellar parameters for our sample of dM3 stars in Paper~I according to the 
new results of Mann et al. (2015). The revised values of the parameters for 
the dM3 sample can be found in Paper~I. In the present paper, we report on 
chromospheric line data for dM3 stars: these data were not a part of HM. In 
Table~4, we list the EW we have obtained for the Ca\,{\sc ii} resonance 
doublet and for the H$_{\alpha}$ line for our dM3 stars.

In order to supplement our sample of $v\sin i$ and Ca\,{\sc ii} measurements, 
and in order to have an unbiased sample of measurements, we searched the 
literature for additional $v\sin i$ and Ca\,{\sc ii} measurements: We found 
several additional stars with $v\sin i$ measurements (see Paper~I). But we 
found only a few measurements of the Ca\,{\sc ii} lines for these stars 
(Stauffer \& Hartmann 1986, Rutten 1987, Giampapa et al. 1989, Rutten et al. 
1989). Instead, we found several measures of the $H_{\alpha}$ EW (Soderblom et 
al. 1991, Hawley et al. 1996, Kamper et al. 1997, Christian \& Mathioudakis 
2002, Gizis et al. 2002, Pace 2013). Fortunately, there is a relatively tight 
correlation between the mean Ca\,{\sc ii} line mean EW and $H_{\alpha}$ EW in 
dM3 and dM4 stars. Therefore, we decided to infer the Ca\,{\sc ii} line mean 
EW from the measures of the $H_{\alpha}$ EW. We show in Fig.~6 the 
relationship between the Ca\,{\sc ii} line mean EW and the $H_{\alpha}$ EW for 
our measurements of dM3 stars (see Table~4). As one can see the correlation 
between these two parameters is very good for all stars except Gl~644AB 
which lies slightly below. The homoscedastic linear least square fit to this 
data except Gl~644AB gives:

\begin{equation}
EW_{CaII} = 2.90\pm 0.05\times EW_{H_{\alpha}} + 1.14\pm 0.125
\end{equation}

The parameters of this fit are given in Table~2. The fit is very good with a 
correlation coefficient of 0.9994. We give the Ca\,{\sc ii} line mean EW 
computed from the $H_{\alpha}$ EW in Table~4.

We show the RAC for dM3 stars for the raw data in Fig.~7. One can see in 
this figure that the scatter is very large among both dM3 and dM3e stars. The 
same kind of scatter was observed among the raw data of dM2 stars (Fig.~4).  
Considering the good correction we had for the metallicity in the previous 
section for dM2 stars, we decided to compile all the metallicities published 
for our initial selection list of 381 dM3 stars, and try to obtain a 
metallicity-radius correlation for these stars, similarly to the dM2 stars in 
Paper~VII. We found metallicities from the literature for 147 dM3 stars. 
This data and the radius-[M/H] relationship were reported in Paper~I. Data for 
M3 stars corrected for metallicity can be found in Table 5 and Fig. 8.

\begin{figure*} 
\vspace{-0.5cm}
\hspace{-2.5cm}
\includegraphics[width=14cm,angle=-90]{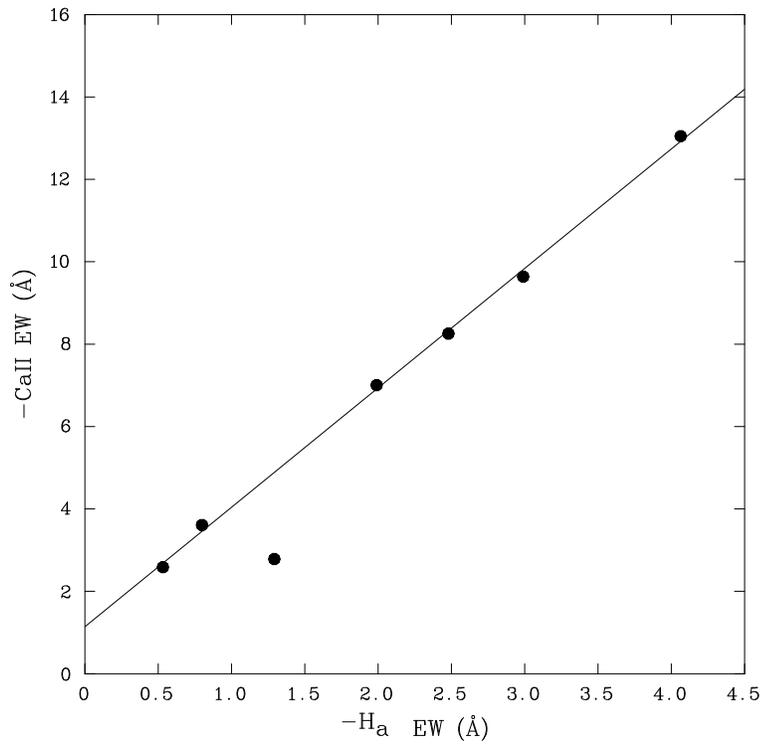}
\vspace{-0.5cm}
\caption[]{Correlation between the mean EW of the Ca\,{\sc ii} resonance 
doublet and the $H_{\alpha}$ EW for stars with spectral sub-type dM3e. The 
solid line is the least square fit to the data.}
\end{figure*}

\begin{figure*} 
\vspace{-0.5cm}
\hspace{-2.5cm}
\includegraphics[width=14cm,angle=-90]{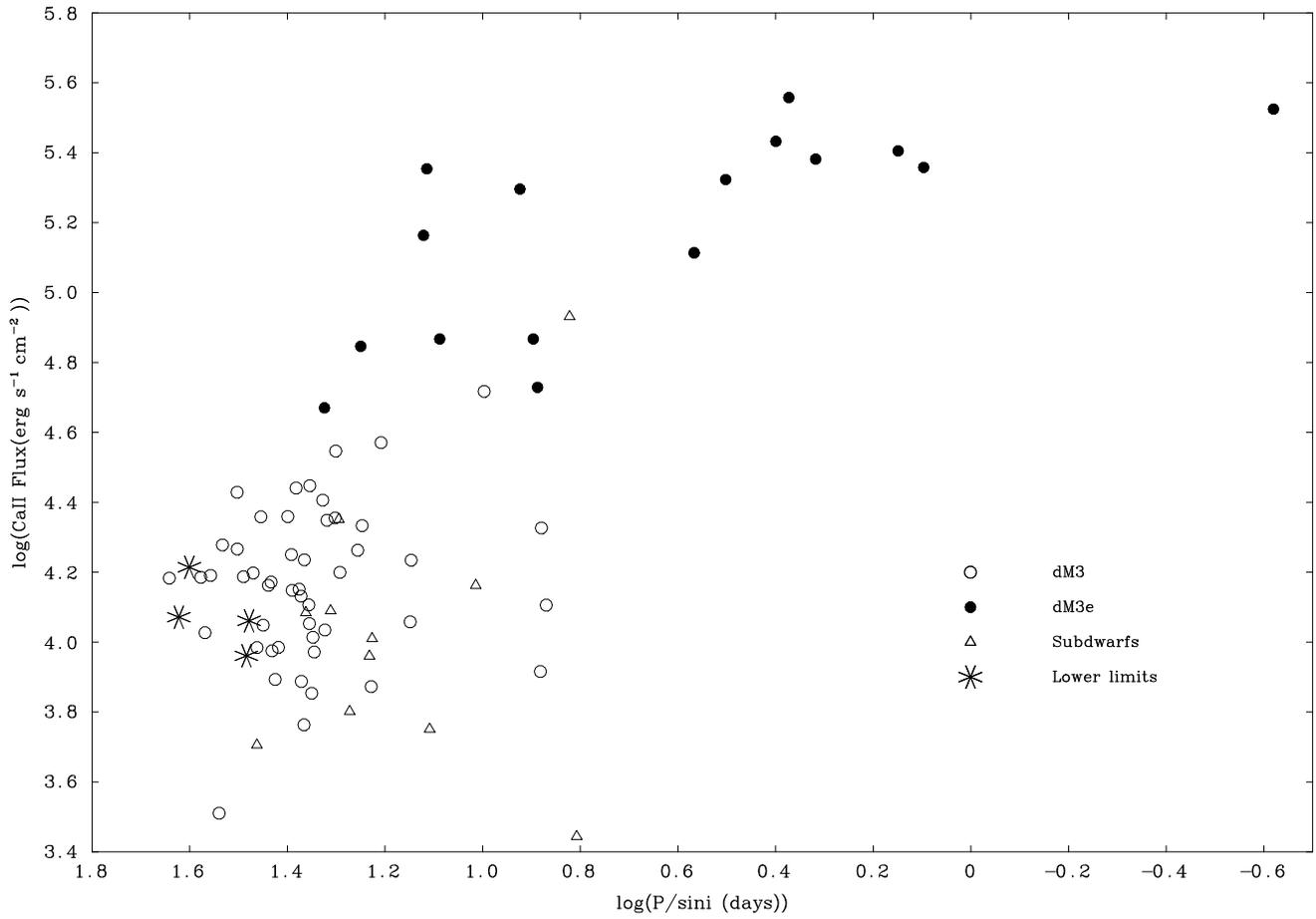}
\vspace{-0.5cm}
\caption[]{Correlation between the logarithm of the mean surface fluxes of the 
Ca\,{\sc ii} resonance doublet and log($P/\sin i$) for stars with spectral 
sub-type dM3. No corrections for metallicity were applied here. One can see 
that (as for the case of dM2 stars) there is a large scatter among the data 
when it is not corrected for metallicity effects.}
\end{figure*}

\begin{figure*} 
\vspace{-0.5cm}
\hspace{-2.5cm}
\includegraphics[width=14cm,angle=-90]{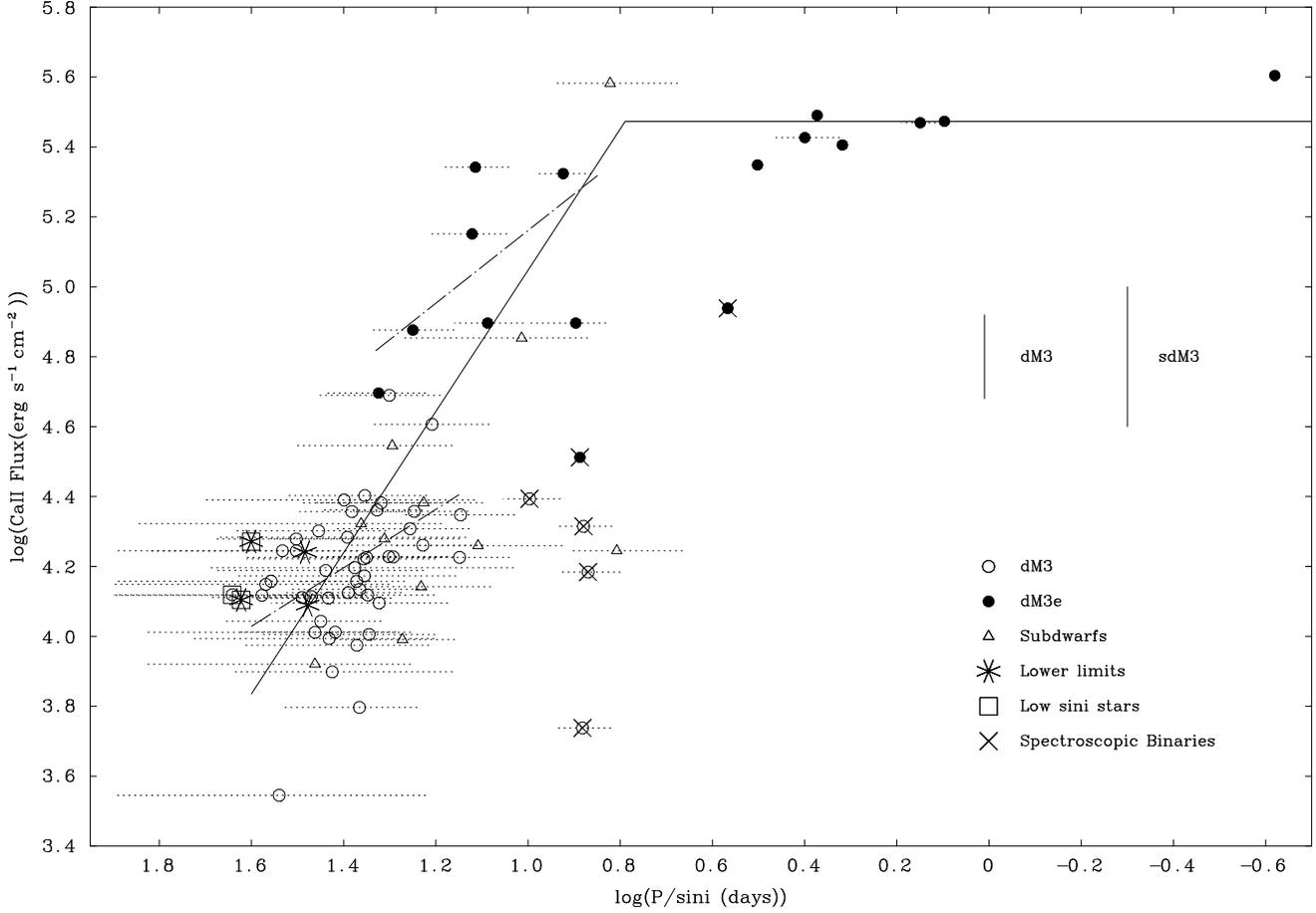}
\vspace{-0.5cm}
\caption[]{Correlation between the logarithm of the mean surface fluxes of the 
Ca\,{\sc ii} resonance doublet corrected for metallicity effects and 
log($P/\sin i$) for stars with spectral sub-type dM3 (see text). Note the 
possible feature of ``saturation" at $P/\sin i < 6$ days. For ``unsaturated" 
stars, i.e. those with $P/\sin i > 7$ days, the solid straight line is the 
heteroscedastic linear LSF to the combined data set of low+high activity stars.
The dot-dashed curves represent the heteroscedastic linear LSFs to the low 
activity star and high activity star sub-samples respectively (see the text). 
We show in the center-right hand of the figure estimates of the uncertainties 
on the Ca\,{\sc ii} mean surface fluxes due to the corrections for metallicity 
effects, for dwarfs and subdwarfs.}
\end{figure*}

We show in Fig.~8 the RAC corrected for metallicity effects for dM3 stars. 
The scatter is reduced compared to the raw data in Fig.~7, but there remains 
a significant scatter among both dM3 and dM3e stars. The corrected correlation 
is not as good as for dM2 stars. We believe this is due partly to the poorer 
statistics we have on the Ca\,{\sc ii} line EWs. Indeed, for dM3 stars we have 
very few measurements of the Ca\,{\sc ii} lines EWs (Table~4), whereas for 
dM2 stars we had several measurements for almost all stars (see Houdebine et 
al. 2012c, Paper XIX). We also observe a larger scatter in the radius-[M/H] 
relationship for dM3 stars (see Paper~I) compared to dM2 stars. This may also 
lead to poorer corrections of the metallicity effects on the RAC.

We observe among our sample of dM3 stars a sub-sample of relatively fast 
rotating-low activity stars. Again, as in the case of dM2 stars, we believe 
that these stars are unresolved spectroscopic binaries. We also found three 
stars with possibly low $v\sin i$ (see Fig.~8). In the corrected RAC, most of 
the subdwarfs now follow the same correlation as normal dwarfs, indicating 
that our metallicity corrections are reasonably correct.

We find that for periods above 7 days or so, the gradient between the dM3 
stars (open circles: the slowest rotators) and the dM3e stars (filled circles: 
the fastest rotators) is very steep. Because we have only a few dM3e stars 
compared to the larger group of dM3 stars, we do not obtain a good linear LSF 
to both the dM3 and dM3e sub-samples. This is due to the fact that the linear 
LSF to the dM3 sub-sample gives a shallower gradient (-0.90, Table~1) compared 
to the dM3+dM3e sample. Therefore, we gave higher weights to our dM3e data 
points: we find that a weight of 7 to dM3e stars and 1 to dM3 stars gives a 
good fit to both data sub-samples. For $P$/sin$i$ $\geq 6.7$ days we obtain 
the following heteroscedastic linear LSF for our dM3+dM3e sample (Table~1):

\begin{equation}
F_{CaII} = 1.169\pm 0.476\times 10^{7} \times (P/sini)^{-2.020\pm 0.11}.
\end{equation}

The gradient in Eq.~(17) (-2.02$\pm$0.11) is significantly steeper than the 
gradients which we determined for our dK4 and dK6 samples (-0.814$\pm$0.059 
and -1.047$\pm$0.042 respectively: see Figs. 2 and 3). But interestingly, the 
mean gradient in Fig.~8 is intermediate between that of dM2 stars 
(-1.575$\pm$0.058) and that of dM4 stars (-2.56$\pm$0.19: Table~1). Once again,
 this is an encouraging sign that the physical parameters we have derived for 
dM3 stars are plausible when compared with stars which are slightly hotter and 
slightly cooler. We plot the linear LSF for stars with periods $P/\sin i > 7$ 
days (we refer to these as unsaturated stars: see below) in Fig.~8 (solid 
line).

For the homoscedastic fit we obtain the following linear LSF for our dM3+dM3e 
sample (Table~1):

\begin{equation}
F_{CaII} = 1.52\pm 1.00\times 10^{7} \times (P/sini)^{-2.075\pm 0.19}.
\end{equation}

The slopes of both the heteroscedastic and the homoscedastic linear fits 
are comparable. Note that the uncertainty on the slope is smaller for the 
heteroscedastic fit than for the homoscedastic fit. Both fits confirm that 
the slope of the RAC is getting steeper as we move to later spectral types.

As we saw in Sect.~1.4, one would expect the RAC to flatten out for rotation 
periods shorter than 2-10 days. We have several objects in our M3 sample that 
have shorter rotation periods. We note that the RAC flattens out (``saturates")
 for the fastest rotating dM3e stars for $P/\sin i<6$~days. In this period 
range, we find 7 fast rotators with about the same surface flux ($\sim 
2.97\times 10^{5}\ erg\ s^{-1}\ cm^{-2}$) which corresponds to the maximum 
observed in the unsaturated portion of the RAC. Although in the present data 
set we do not have enough data to clearly confirm the presence of a saturation 
plateau, one should expect saturation to occur among some of our sample stars 
according to previous investigations. We propose that saturation in our present
 dM3 sample occurs at about $\sim 6$ days, which agrees with the expectations 
of saturation occuring in the range 2-10 days. In Fig.~8 we therefore represent
 the fits of the RAC in the ``unsaturated" portion, and draw a flat plateau in 
the expected ``saturated" portion of the RAC.

The fact that certain stars in our samples probably lie in a saturated 
regime, while others are found to lie in an unsaturated regime, has a 
bearing on a methodological point which was made in Sect. 1.1: a study of 
stellar dynamos is best done (we believe) by focussing on stars in an 
unsaturated regime. Therefore in what follows, the least squares fits which we 
will present refer only to stars in the unsaturated regime. In this regime, 
the observations show that there exist stars of both low (dM) and high (dMe) 
activity, so the sample provides access to a range of ``dynamo strengths". In 
contrast, in the saturated regime, only high-activity stars (dMe) are present.

In the regime of our unsaturated stars, we also performed heteroscedastic and 
homoscedastic linear LSFs to the separate sub-samples of low activity dM3 and 
high activity dM3e stars respectively (see Table~1). The heteroscedastic 
linear LSF to the dM3 low activity stars (all of which are in the unsaturated 
regime) yields:

\begin{equation}
F_{CaII} = 2.33\pm  1.10\times 10^{5} \times (P/sini)^{-0.837\pm 0.20}.
\end{equation}

Therefore, the gradient of the RAC for the low activity stars is again much 
shallower than that of the combined sample of low+high activity stars, and by 
more than $3\sigma$. Therefore, the linear LSF to the combined sample may be  
somehow inadequate. We plot the linear LSF to the low activity sub-sample also 
in Fig.~8 (dot-dashed line).

The homoscedastic linear LSF to the dM3 low activity stars yields:

\begin{equation}
F_{CaII} = 3.09\pm  1.58\times 10^{5} \times (P/sini)^{-0.93\pm 0.22}.
\end{equation}

Both fits yield comparable results within errors. The $\chi^{2}$ for these 
fits are 0.039 and 0.047 respectively. Both fits are highly statistically 
significant at a 99.9\% confidence level. Once again, we see that the slopes 
of the RAC among low-activity dM3 stars (-0.84, -0.93) are significantly 
shallower (by more than 3$\sigma$) than the slopes of the RAC for the combined 
sample of high and low activity dM3 stars (-2.02, -2.08). 

Turning now to the high activity dM3e stars in the unsaturated regime, we find 
that the heteroscedastic linear LSF yields:

\begin{equation}
F_{CaII} = 1.60\pm  1.20\times 10^{6} \times (P/sini)^{-1.041\pm 0.58}.
\end{equation}

In contrast to the results we obtained for the K4, K6, and M2 samples, this 
time for the dM3e sub-sample of unsaturated stars, the slope of the RAC is 
{\em shallower} than that of the combined sample of (unsaturated) low+high 
activity stars. It seems that at the TTCC, the slope of the RAC for 
(unsaturated) high activity stars is falling significantly to a value of about 
-1. Although there remains some uncertainty on our fit for dM3e stars (because 
we have only 8 stars in the unsaturated regime, and with a significant 
scatter), we shall see in the next subsection that this decrease in the 
steepness of the slope is confirmed in our (unsaturated) dM4e sub-sample and 
is significant above the $3\sigma$ level (23 stars). Our M4 sample also 
confirms that, the slope of the RACs for the high activity (unsaturated) 
sub-samples has fallen below the slope of the combined dM+dMe samples.

We have also performed LSFs to the full sub-sample of dM3e stars 
(unsaturated+saturated). We give the results of the fits in Table~1 
(dM3e+sat). The slope for the full sub-sample has fallen to -0.39 which is 
considerably smaller than the value of -1.04 we found above. However, this 
conforts us with the idea that at the TTCC, the slope of the RAC for dM3e 
stars is of the same order than that of the slope for the low activity dM3 
stars (see Sect.~3.7.6).

The relative invariance of the RAC slopes which we have obtained for the 
linear LSF to the {\em low activity stars} in our 5 stellar samples gives an 
important degree of credibility to our results. It is important to note that 
all of our low-activity stars have been found to lie in the unsaturated regime 
of the RAC, i.e. in the regime where (we believe) dynamo theory can best be 
tested. To the extent that the slope of the RAC is determined (in the 
unsaturated regime) by the physical properties of a dynamo, our results 
suggest that there may exist only one dynamo regime for the low activity stars 
from dK4 to dM4 within errors (see Sect.~3.7.2). The parameter 
$\frac{R}{\delta}$ is 18.28 for the high activity star sub-sample
 and 21.77 for the low activity star sub-sample. Therefore, the parameter 
 $\delta =\frac{<error>}{\sqrt{n}}$ which is a normalized estimate of the mean 
error on the measurements is much smaller than the period domains of the RACs 
in both cases. Hence, these LSFs should be relatively well established, which 
corresponds to the high statistical significances we obtain (Table~1). However,
 these results may still preliminary because the domains of the 
RACs are relatively small compared to the typical uncertainties on individual 
measures as hilighted by the referee.

There is the possibility that there exist two different dynamo regimes for dK, 
dM and dKe, dMe respectively throughout the spectral range we investigate. 
Note that, in contrast to the low activity dK and dM stars, the dynamo regimes 
in high activity dKe and dMe stars vary with spectral type, especially at the 
TTCC. Might these changes be an indication that the dynamo mechanisms undergo 
a change at the TTCC, perhaps from an $\alpha-\Omega$ dynamo to an $\alpha^2$ 
type of dynamo ?

There tends to be more and more high activity-relatively slow rotators when we 
move to the spectral sub-types dM3 and dM4, which questions the validity of a 
single RAC for the combined samples at these spectral types. Given the 
consistency of the results we have obtained for the slopes of the RAC for the 
{\it low activity} stars for our 5 spectral sub-types, it is possible that a 
single dynamo mechanism may apply to all low-activity stars in our sample. 
However, we arrive at a different conclusion regarding the dynamo mechanism 
among the high activity stars. {\em At the TTCC and beyond, we propose that 
high-activity stars may switch to a different dynamo mechanism (see 
Sect.~3.7)}.

\subsection{The RAC in M4 dwarfs}

Considering that in our previous study of the RAC in dM4 stars we had only a 
couple of dozen stars (Houdebine 2012a, Paper XVII), we decided to compile 
all available measures of $v\sin i$ published in the literature (see Paper~I). 
With our own measures, we obtained $v\sin i$ measures for 106 dM4 stars 
(Paper~I) from our initial list of 395 dM4 stars. Our dM4 stars in this study 
were selected according to their (R-I)$_{C}$ color, within the range
[1.500:1.700]. This range is somewhat larger than the one used in Paper~XVII 
in order to have more targets. 

In order to obtain the RAC for dM4 stars, we also compiled the Ca\,{\sc ii} 
resonance doublet and the H$_{\alpha}$ line EW and fluxes from the literature. 
We give this compilation of data in Table~6. The data were collected from the 
following authors: Giampapa \& Liebert (1986), Stauffer \& Hartmann (1986), 
Young et al. (1986), Fleming \& Giampapa (1989), Herbst \& Miller (1989), 
Hawley et al. (1996), Delfosse et al. (1998), Gizis et al. (2002), Mohanti \& 
Basri (2003), Wright et al. (2004), Rauscher \& Marcy (2006), Morales et al. 
(2008), Walkowicz \& Hawley (2009), Browning et al. 
(2010), Isaacson \& Fischer (2010), Houdebine (2012b), Reiners et al. (2012), 
Pace (2013). In addition to this data, we also measured the Ca\,{\sc ii} and 
H$_{\alpha}$ EW from FEROS spectra from the ESO archive for 47 dM4 stars. We 
give the results in Table~6. Together with the measurements from Paper XVIII, 
this allows us to derive a correlation between the Ca\,{\sc ii} EW and the 
H$_{\alpha}$ EW for dM4e stars. We show this correlation in Fig.~9. Although 
there is more scatter in this sample than in our dM3 star sample, we have a 
relatively good correlation with a correlation coefficient of 0.877 for 35 
measures (see Table~2). This correlation is significant at a confidence level 
better than 99.8\%. The LSF to this data gives:

\begin{equation}
EW_{CaII} = 1.909\pm 0.18\times EW_{H_{\alpha}} - 1.035\pm 0.58
\end{equation}

\begin{figure*} 
\vspace{-0.5cm}
\hspace{-2.5cm}
\includegraphics[width=14cm,angle=-90]{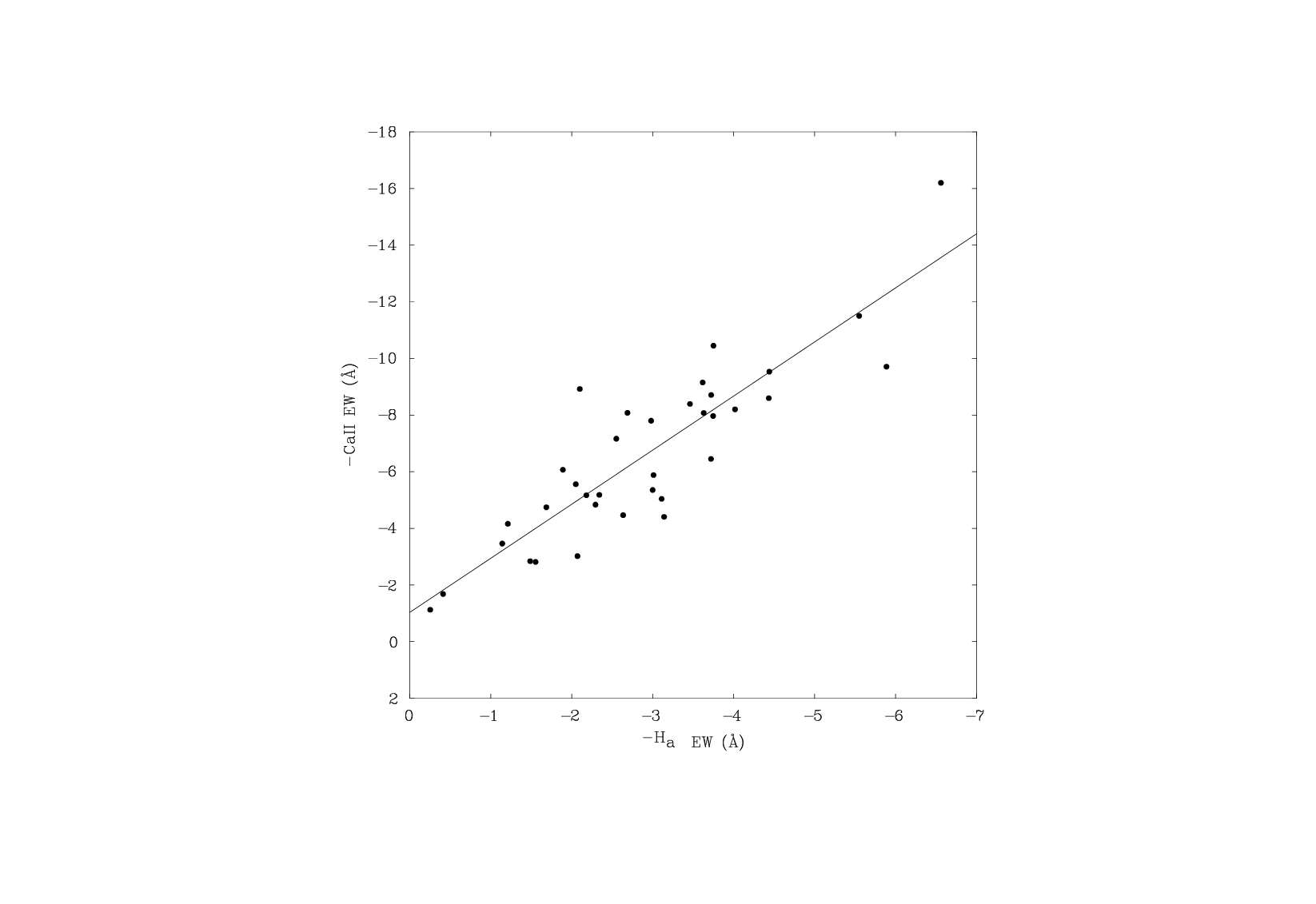}
\vspace{-0.5cm}
\caption[]{Correlation between the mean EW of the Ca\,{\sc ii} resonance 
doublet and the $H_{\alpha}$ EW for stars with spectral sub-type dM4e. The 
solid line is the linear LSF to the data.}
\end{figure*}

\begin{figure*} 
\vspace{-0.5cm}
\hspace{-2.5cm}
\includegraphics[width=14cm,angle=-90]{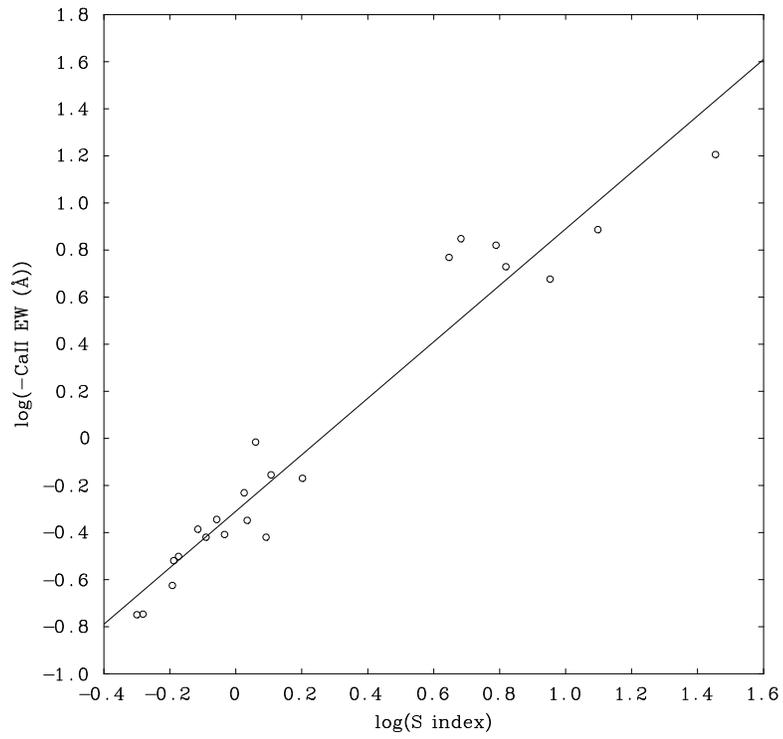}
\vspace{-0.5cm}
\caption[]{Correlation between the mean EW of the Ca\,{\sc ii} resonance 
doublet and the Ca\,{\sc ii} Mount Wilson S index for stars with spectral 
sub-type dM4. The solid line is the linear LSF to the data.}
\end{figure*}

Eq. (22) allows us to obtain an estimate of the Ca\,{\sc ii} EW when we have 
only the H$_{\alpha}$ line EW available for dM4e stars. Similarly, we compiled 
many measures of the Mount Wilson $S$ index (Table~6). We show the empirical 
correlation between the Ca\,{\sc ii} EW and the $S$ index in Fig.~10. Again, 
we obtain a relatively good correlation between the two parameters, with a 
correlation coefficient of 0.970 for 22 data points (see Table~2). The 
relationship between the Ca\,{\sc ii} EW and the $S$ index is:

\begin{equation}
EW_{CaII} = 0.4898\pm 0.047\times S^{1.20\pm 0.07}.
\end{equation}

All these compilations of data now allow us to compute the Ca\,{\sc ii} surface
fluxes (Table~6). We show the Ca\,{\sc ii} surface fluxes as a function of 
$P/\sin i$ for our dM4 stellar sample in Fig.~11. One can see in this diagram 
that the scatter is very large, and is similar to (or even worse than) the raw 
correlation for dM3 stars. It appears that the scatter increases from spectral 
sub-types dM2 to dM4. However, there is a parameter that plays a role in the 
scatter in Fig.~11: the (R-I)$_{C}$ range for our dM4 sample ([1.500:1.700]) 
spans 0.2~dex, whereas for our dM2 and dM3 samples it spans 0.132~dex. This 
larger range in our selected targets contributes to the larger scatter 
observed in Figs.~11 and 12. In Fig.~11, it is noteworthy that there is no 
overlap in Ca\,{\sc ii} fluxes between the low activity dM4 stars (up to 
$log(F_{HK} = 4.4$) and the high activity dM4e stars ($log(F_{HK}\geq 4.6$): 
there is a clear gap in the fluxes, although the two groups overlap in their 
rotation periods. The dM3 stars do not show this clean separation between dM 
and dMe (Fig.~8). The dM2 stars also show some separation (Fig.~5), but it is 
not as clean as for dM4 stars. 

Considering the good correction we had for metallicity in dM2 stars, we 
decided to compile all the metallicities published for our initial selection 
list of 395 dM4 stars, and try to obtain a metallicity-radius correlation for 
these stars (see Paper~I). We found metallicities from the literature for 179 
dM4 stars. Data for M4 stars that have been corrected for metallicity can be 
found in Fig.~12 and Table~7.

\begin{figure*} 
\vspace{-0.5cm}
\hspace{-2.5cm}
\includegraphics[width=14cm,angle=-90]{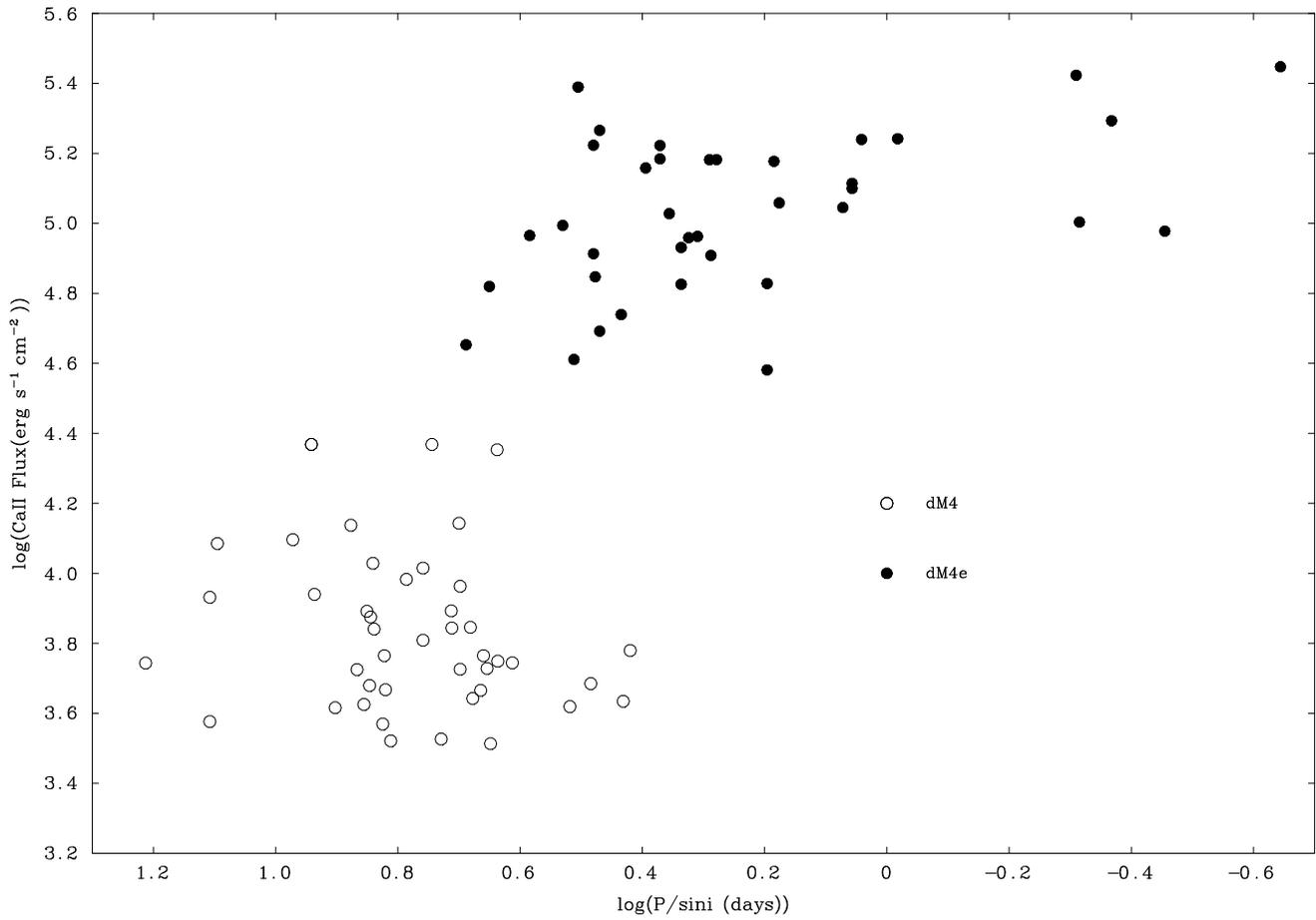}
\vspace{-0.5cm}
\caption[]{Correlation between the mean EW of the Ca\,{\sc ii} resonance 
doublet and log($P/\sin i$) for stars with spectral sub-type dM4. Filled dots: 
dM4e. Hollow circles: dM4 stars.}
\end{figure*}

\begin{figure*} 
\vspace{-0.5cm}
\hspace{-2.5cm}
\includegraphics[width=14cm,angle=-90]{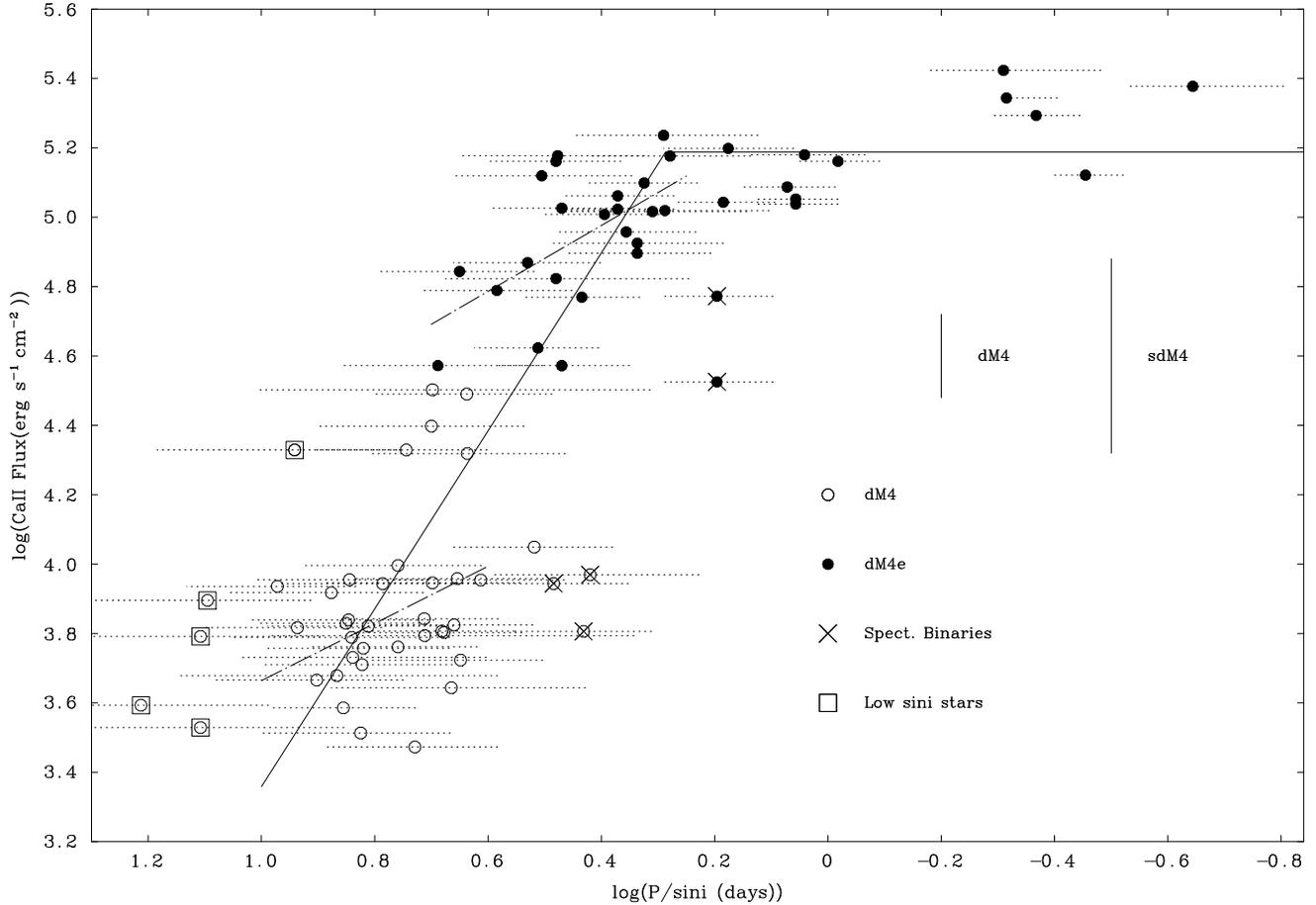}
\vspace{-0.5cm}
\caption[]{Correlation between the logarithm of the mean surface fluxes of the 
Ca\,{\sc ii} resonance doublet corrected for metallicity effects and 
log($P/\sin i$) for stars with spectral sub-type dM4. Notice the onset of 
saturation for stars with $log(P/\sin i) \leq 0.2$. LSFs have been obtained 
for the stars in the unsaturated regime as follows. Solid line: heteroscedastic
 linear LSF; separated dot-dashed lines, the heteroscedastic linear LSF to the 
low activity dM4 and to the high activity dM4e subsamples. We show in the 
center-right hand of the figure estimates of the uncertainties on the Ca\,{\sc 
ii} mean surface fluxes due to the corrections for metallicity effects, for 
dwarfs and subdwarfs.}
\end{figure*}

We show in Fig.~12 the RAC corrected for metallicity effects for dM4 stars. 
The scatter is reduced compared to the raw data in Fig.~11, but there remains 
a significant scatter among both dM4 and dM4e stars. The corrected correlation 
is not as good as for dM2 stars in spite of the relatively good statistics 
we have on the Ca\,{\sc ii} surface fluxes.

We also observe among our sample of dM4 stars a sub-sample of relatively fast 
rotating-low activity stars. Again, as in the case of dM2 and dM3 stars, we 
believe that these stars are unresolved spectroscopic binaries. We also found 
a few stars with possibly low $\sin i$ (see Fig.~12). 

We find that when we apply the linear heteroscedastic and homoscedastic LSFs 
to the combined dM4+dM4e stars in the unsaturated regime, the gradients are 
very steep, steeper even than what we found for our dM2 and dM3 stellar 
samples. We also find that dM4 stars rotate much faster than dM2 and dM3 
stars. The change in Ca\,{\sc ii} flux between dM4 and dM4e stars is very 
abrupt, as already noticed in Fig.~11, a difference of only 0.1 in the 
$log(P/\sin i)$. Our data suggest that saturation in our dM4 sample occurs at 
about $P/\sin i\sim 1.8$ days. This agrees with the values found in previous 
studies. Specifically, for stars in the unsaturated regime, i.e.
for $P/\sin i\geq 1.8$ days we obtain the following heteroscedastic linear 
LSF for the stars in our combined dM4+dM4e sample:

\begin{equation}
F_{CaII} = 8.38\pm 2.05\times 10^{5} \times (P/sini)^{-2.564\pm 0.19}.
\end{equation}

The gradient in Eq.~(24) (-2.564$\pm$0.19) is significantly steeper than the 
gradient which we determined for our dK6 (-1.047$\pm$0.042: Table~1) and dM2 
stars (-1.575$\pm$0.058: Table~1), and is also steeper than the gradient for 
our dM3 stars (-2.020$\pm$0.11: Table~1). This confirms that the slope of the 
linear LSFs increases in absolute value when the spectral type increases from 
dK4 to dM4. This increase is particularly marked at the TTCC (M3) and beyond.

For $P/\sin i\geq 1.8$ days we obtain the following homoscedastic linear 
LSF for the stars in our combined dM4+dM4e sample:

\begin{equation}
F_{CaII} = 7.94\pm 2.05\times 10^{5} \times (P/sini)^{-2.526\pm 0.13}.
\end{equation}

The gradient in Eq.~(25) (-2.526$\pm$0.18) is very close to that determined 
for the heteroscedastic linear fit (-2.564$\pm$0.19: Table~1). As a whole, 
we find comparable results from the heteroscedastic and homoscedastic linear 
fits, suggesting that the errors on the measures do not play a large role in 
the determination of the linear fits (Table~1).

In Fig.~12 we observe a flattening (``saturation") for the fastest rotating 
dM4e stars for $P/\sin i<1.6$~days. This is a range in periods where saturation
 is expected to occur (see Sect.~1.4). Although our data set is not yet 
complete enough to be conclusive of a saturation phenomenon, we assume here 
that for $P/\sin i$ shorter than $\sim 2$~days, saturation occurs. In this 
period range, we observe 12 fast rotators with a mean surface flux of $\sim 
1.54\times 10^{5}\ erg\ s^{-1}\ cm^{-2}$.

We also performed separate linear LSFs to the sub-samples of only the low 
activity dM4  and the sub-set of high activity dM4e stars (all of which we 
assume are in the unsaturated regime) respectively (see Table~1). The 
heteroscedastic linear LSF to the dM4 low activity stars yields:

\begin{equation}
F_{CaII} = 3.08\pm  1.25\times 10^{4} \times (P/sini)^{-0.825\pm 0.35}.
\end{equation}

Therefore, the gradient of the RAC for the low activity stars is again much 
shallower than that of the combined sample of low and high activity stars 
(unsaturated), and by more than $3\sigma$. Therefore, the linear LSF to the 
full sample is somehow inadequate. We note also that the slopes of the 
heteroscedastic linear LSF to the sub-samples of only the low activity stars 
remain approximately constant (within measurement errors) from dK4 to dM4. 
This point will be further developed in Sect.~3.7.

The homoscedastic linear LSF to the dM4 low activity stars yields:

\begin{equation}
F_{CaII} = 3.80\pm  1.89\times 10^{4} \times (P/sini)^{-0.91\pm 0.39}.
\end{equation}

Again, the gradient of the RAC for the low activity stars is shallower than 
that of the combined sample of low and high activity stars, and by more than 
$3\sigma$. We find that for the low activity dM4 stars, both the 
homoscedastic and heteroscedastic models give similar results for the mean 
value of the slope of the RAC. And this mean value (0.825-0.91) is remarkably 
similar (within 1$\sigma$) to the mean values for low-activity stars in the 
other 4 spectral sub-types: 0.624, 0.637, 0.891, 0.93. However, there is one 
aspect in which the dM4 slow rotators differ from the other four sub-types: 
the dM4 stars have significantly larger values of $\sigma$ associated with 
the slope of the RAC in both the homo- and hetero-scedastic LSFs. We will 
return to this topic in Section 3.7.2 below.

We find that the $\chi^{2}$ is much poorer for the heteroscedastic model 
(1.966) than that for the homoscedastic LSF (0.057). Nevertheless, both fits 
are statistically highly significant (better than 96\%). 

The heteroscedastic linear LSF to the 23 unsaturated dM4e high activity stars 
yields:

\begin{equation}
F_{CaII} = 2.27\pm  0.58\times 10^{5} \times (P/sini)^{-0.951\pm 0.31}.
\end{equation}

Therefore, the gradient of the RAC for the high activity stars is also much 
shallower than that of the combined sample of low and high activity stars, and 
by more than $3\sigma$. Therefore, the linear LSF to the full sample 
seems rather inadequate. This point is straightforward if one looks at the 
linear heteroscedastic LSFs in Fig.~12. It is obvious that the LSFs to the 
low activity stars, and to the high activity stars cannot be reproduced by a 
simple linear or quadratic function. We note that the dM4e data confirm the 
decrease in the slope of the LSF fits to unsaturated stars as we pass through 
the TTCC. In addition, we also performed a linear LSF to the full sample of 
dM4e stars (unsaturated+saturated) in order to compare with the previous fit. 
We find a slope of -0.47 for the full sub-sample of dM4e stars (Table~1). This 
result again suggests that the slopes for the sub-samples of low activity 
stars and of high activity stars are of the same order after the TTCC.
The parameter $\frac{R}{\delta}$ is 18.90 for the high activity star sub-sample
 and 19.13 for the low activity star sub-sample. These figures are still very 
high, although that of the low activity sub-sample has decreased compares to 
other spectral types. Therefore, the parameter  $\delta =
\frac{<error>}{\sqrt{n}}$ is still much smaller than the period domains of the 
RACs in both cases. As a consequence, these LSFs should be established with 
a good confidence level, which is in agreement with the high statistical 
significances we obtain for these fits (Table~1). 

We find that the linear fits (to the unsaturated dM4+dM4e samples) are 
incapable of reproducing the slopes among the low activity stars and the 
(unsaturated) high activity stars. We face the same problem as we did in 
dealing with the case of dM3 stars: the problem is that there exists high 
activity-relatively slow rotators and low activity-relatively fast rotators. 
These two types of stars overlap in their range of values of $P/\sin i$.

The dM3 and especially the dM4 datasets lead us to question the validity of 
fitting both the low activity and high activity (unsaturated) sub-samples with 
a single RAC. Instead, it seems to us that it might be better to consider the 
possibility that two distinct RACs are present in the data. To the extent that 
the RACs are determined with dynamo operation, this leads us to wonder: is 
there perhaps a distinct dynamo process operating in low-activity dM4 stars 
from the dynamo process which is at work in (unsaturated) high-activity dM4e 
stars? The slopes of the linear LSF to the low activity stars for our 5 
stellar samples are rather homogeneous, even for the dM4 stars, although the 
latter stars are almost certainly fully convective. This result suggests that 
there may exist a single dynamo regime for the slow rotators from dK4 to dM4 
within errors (see Sect.~3.7.2). This important result should be tested with 
larger samples of low activity stars.

There exist more and more unsaturated high activity-relatively slow rotators 
when we move to the spectral sub-types dM3 and especially dM4. These 
relatively slow rotators overlap in $P/\sin i$ with the low activity stars. 
This poses an important problem: we cannot fit satisfactorily both the low 
activity and (unsaturated) high activity samples with a single RAC. We suggest 
that the dynamo mechanisms behave differently among the high activity stars 
when crossing the TTCC. At the TTCC (dM3) and beyond, the data suggest that 
there exist two different dynamo regimes for and high activity stars (see 
Sect.~3.7). These differences in these dynamo regimes could be related to 
differential rotation: in the young high activity stars differential rotation 
is expected to be larger, and it should vanish with age, i.e. in the low 
activity stars. Large differential rotation would boost the magnetic field 
generation in these fully convective stars, such that we could have high 
activity-relatively slow rotators with large differential rotation.

Now that we have assembled enough data to construct RACs at five different 
spectral sub-types for late K and M dwarfs, it is worthwhile to examine the 
systematics of RACs as a function of spectral sub-type.

\subsection{The slopes of the RACs: variation with spectral sub-type}

\begin{figure*}
\vspace{-1.5cm}
\hspace{-3.5cm}
\includegraphics[width=15cm,angle=-90]{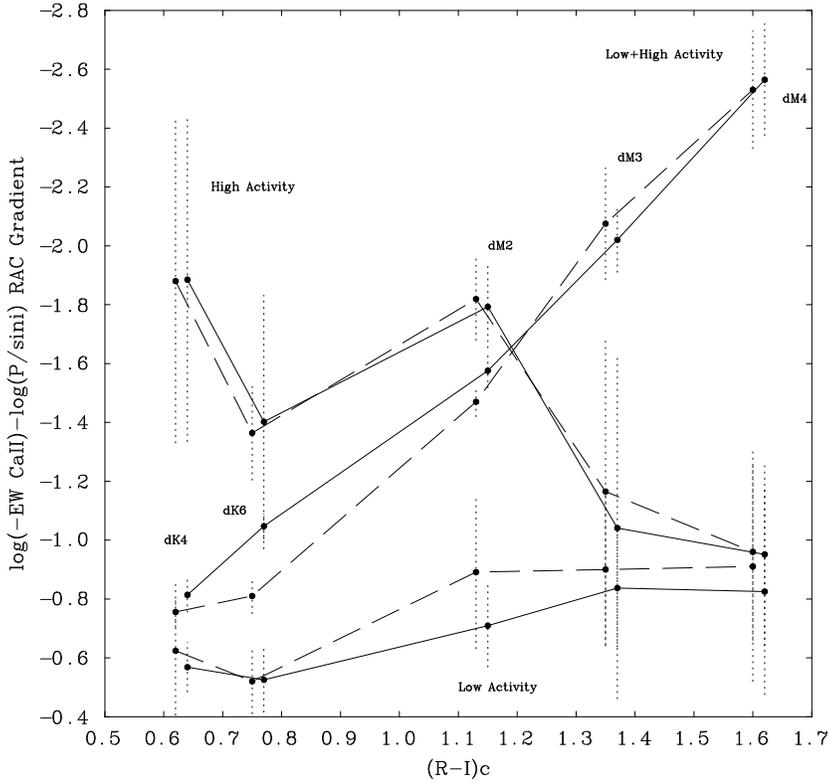}
\vspace{-0.5cm}
\caption[]{Values of the slopes of the heteroscedastic linear LSFs 
(solid lines) and the homoscedastic linear LSFs (dashed lines) to the RACs as 
a function of the infra-red color (R-I)$_{C}$. We indicate the fits to the 
combined samples of high+low activity stars (dK+dKe, dM+dMe) in each spectral 
sub-type (labelled ``Low+High Activity"). The lower curves are the fits to low 
activity stars only (dK, dM, labelled ``Low Activity"). We also indicate the 
fits to the combined samples of high activity star subsamples in the 
unsaturated regime (dKe, dMe, labelled ``High Activity"). The slopes for the 
combined samples increase significantly from dK4 to dM4, with only minor 
differences between heteroscedastic and homoscedastic fits. For the low 
activity stars, as a whole, the slopes for the heteroscedastic and 
homoscedastic models agree well and the results indicate that the slopes remain
 almost constant within the error bars from dK4 to dM4. The magnitudes of the
slopes for the (unsaturated) high activity stars lie significantly above those 
for the low activity stars at spectral types dK4, dK6 and dM2. At the TTCC 
(M3) and beyond (M4), the slopes for the (unsaturated) high activity stars 
significantly diminish.}
\end{figure*}

Now that we have values of the RAC slopes for five different spectral 
sub-types, we provide an overview by plotting the slopes as a function of the 
infra-red color index (R-I)$_{C}$ in Fig.~13. In this Figure, results are 
plotted separately for three distinct samples of stars: the low-activity stars 
only (lower curves), the (unsaturated) high activity stars only (labelled 
``High Activity"), and the combined samples low+high activity stars (labelled 
``Low+High Activity"). In this Figure, the heteroscedastic models are plotted 
as solid lines, and the homoscedastic models are plotted as the dashed lines.

\subsubsection{Linear LSFs: combined samples of low plus high activity stars}

In Fig.~13, the upper curves refer to the combined samples: 
the plotted points indicate the mean values of the RAC gradients which we have 
obtained for each of our 5 sub-samples by means of linear heteroscedastic 
(solid line) and homoscedastic (dashed line) LSFs. Attached to each mean value 
of the gradient is plotted the 1$\sigma$ uncertainty in the mean value of the 
gradient.

We see that the value of the mean gradient becomes monotonically steeper as 
(R-I)$_{C}$ increases. The heteroscedastic (solid line) and homoscedastic 
(dashed line) models give similar results. There are some significant 
differences (above the $3\sigma$ level) between the two models for the dK6 
sample, but this does not affect the overall trend observed for the gradient. 

The fact that the magnitude of the gradient increases monotonically with 
increasing spectral type implies that the magnitude of the effects of rotation 
on the mechanisms of the dynamo increases systematically as we examine our 
combined samples at later spectral types. If this result is connected with the 
physics of the dynamo, and if it is physically permissible to consider low 
activity stars in the same context as (unsaturated) high activity stars, then 
the results in Fig. 13 may provide an important new constraint on the dynamo 
mechanism(s). The magnitude of the gradient increases monotonically as we go 
from the spectral sub-type dK4 to the spectral sub-type dM4. The results in 
Fig.~13 suggest that, if the TTCC occurs at dM3, the effects of rotation on 
chromospheric emission are larger in fully convective stars than in early M 
type and late K dwarfs where a radiative core persists. If this result is 
taken at face value, it does not appear to be consistent with the 
suggestions of Durney et al (1993), mentioned in the first paragraph of the 
Introduction. 

Previous studies have also reported a variation in the RAC gradient of linear 
fits (e.g. Stepie\'n 1989, 1993, 1994) as a function of spectral type: the RAC 
was found to be steepest (i.e. most negative) at dF6 (B-V=0.45) and 
shallowest close to dK6 (B-V=1.41: KH). However, the most complete study 
(Stepien 1989) did not include any stars which are cooler than dK6. Thus, 
the study we report here is complementary to Stepie\'n's work, extending that 
work towards cooler stars as far as dM4 ((B-V)=1.60). However, our results 
indicate that the trend noted by Stepie\'n between dF6 and dK6 (namely, the 
RAC becomes shallower in cool stars) does {\it not} continue at later 
spectral types. On the contrary, we find that the trend in the gradient 
reverses: according to our samples, the RAC for the combined samples becomes 
{\it steeper} as we go towards dM4. 

It is natural to wonder if the pronounced steepening of the RAC gradient is a 
signature of changes in the dynamo mechanism when crossing the TTCC ? As noted 
in the Introduction above, it appears that the TTCC lies between subtypes dM2 
and dM4. In this context, we consider it significant that, among the stars in 
our samples, the RAC gradient for the combined samples steepens rapidly 
between spectral subtypes dM2 and dM4. Moreover, HM report a notable 
lengthening in the mean rotation periods of inactive and active stars also at 
spectral subtype M3. Also at M3, coronal loops in active stars become 
noticeably longer (see HM for a discussion of loop length data reported by 
Mullan et al. [2006]). 

Our data lead us to believe that something interesting takes place in the 
spectral subtype range dM2-dM4 as regards the rotational and activity 
parameters. Is the interesting behaviour between dM2 and dM4 perhaps associated
 with changes in the dynamo regime when crossing the TTCC ? Possibly, although 
we have not yet been successful in identifying a physical explanation which 
accounts for the results reported above. However, whatever the physical 
mechanism which is responsible for the increased steepnesses in Fig. 13, the 
rise in the steepness of the gradient for the combined sample from sub-type K4 
to sub-type M2 indicates that the changes occurring in the RAC slopes (and 
therefore perhaps also in the dynamo mechanism(s)) may be progressive and may 
begin even earlier than sub-type dM2.

\subsubsection{Linear LSFs: low activity stars only}

Now let us consider only the {\it low activity stars}, in order to position 
ourselves as much as possible in the unsaturated regime of the Ca II fluxes, 
and therefore (hopefully) also in the unsaturated regime of the dynamo. The 
continuous line and dashed line in Fig.~13 (lower curves) are a 
plot of the slopes of the heteroscedastic and homoscedastic linear LSFs 
respectively to the RACs for the low activity stars sub-samples only.
The parameter $\frac{R}{\delta}$ lies between 19.13 and 39.68 for these low 
activity star sub-samples. Therefore, the parameter  $\delta = 
\frac{<error>}{\sqrt{n}}$ which is a normalized estimate of the mean 
error on the measurements is much smaller than the period domains of the RACs 
in all cases. Hence, these LSFs should be established to a relatively high 
confidence level, which agrees with the high statistical significances we 
obtain for these fits (Table~1).

In this regard, we note that the process of applying a LSF to a data set 
leads not only to a mean value for the slope: the LSF also yields a standard 
deviation $\sigma$ for the slope. We consider it worthwhile to note that the
values of $\sigma$ contain information about the robustness of the slope.

Fig.~13 shows that the slopes for the slow rotators (dK, dM only)
 do not behave the same as for the combined samples of stars (dK+dKe or 
dM+dMe). The gradient ``$a$" for the low activity stars remains almost constant
 (within the error bars) between dK4 and dM4 at a level of $a$ = -(0.8-0.9). 
More specifically, if we consider not merely the {\it absolute value} of $a$, 
but also the statistical significance of its value {\it compared to its own 
$\sigma$}, we find the following pattern: in dK4 stars, the magnitude of $a$ 
is 5.6 times its own $\sigma$, in dK6 stars, $a$ is  6.1 times its $\sigma$, 
in dM2 stars, $a$ is 7.4 times its $\sigma$, and in dM3 stars, $a$ is 4.3 
times its $\sigma$. All 4 of these cases have slopes which are statistically 
highly significant. The chromospheric emission in these cases does indeed 
depend sensitively on rotation. But in M4 stars, $a$ has a value which is only 
2.3 times its own $\sigma$ for the homoscedastic model and 2.4 times its own 
$\sigma$ for the heteroscedastic model. In this case, at the 3$\sigma$ level 
of significance, the mean slope which we have derived for the RAC of our 
sample of low activity dM4 stars is formally consistent with zero. In a 
situation where the slope of the RAC is zero (or formally consistent with 
zero),the meaning is that the chromospheric emission in low-activity dM4 stars 
does {\it not} depend on rotation at all (or is consistent with zero 
sensitivity to rotation).  

Although our LSFs of the slow rotators seem well established because of the 
large number of measures we have in our sub-samples, we still consider these 
results as preliminary and they should be confirmed with measures from a 
higher resolution spectrograph (such as ESO-ESPRESSO), because our errors on 
the individual measures are still large compared to the period domains of 
the RACs.

\subsubsection{Linear LSFs: High activity stars only}

We also show the slopes of the heteroscedastic and homoscedastic LSFs to the 
sub-samples of only the (unsaturated) high activity stars in Fig.~13. The 
parameter $\frac{R}{\delta}$ lies between 17.20 to 23.45 for these high 
activity star sub-samples (the case of dK4e stars is not considered because we 
have too few measures). Therefore, the parameter $\delta = 
\frac{<error>}{\sqrt{n}}$ which is the normalized estimate of the mean 
error on the measurements is much smaller than the period domains of the RACs 
in all cases. Hence, these LSFs should be established with a rather good 
conficence level, which agrees with the high statistical significances we 
obtain for these correlations ($\geq$90\%, Table~1). However, we consider 
these results as still preliminary because the domains of the RACs are 
relatively small compared to the typical uncertainties on individual measures.
We note that these slopes behave quite differently from the slopes of the low 
activity stars and the slopes of the combined samples of low+high activity 
stars. For the high activity stars, we observe that the slopes are clearly 
steeper than those for the low activity stars at spectral types dK6 and dM2. 
At these spectral types, they are also steeper than the slopes of the combined 
samples of low+high activity stars. These higher slopes suggest the existence 
of two different dynamo regimes for the low and the high activity sub-samples 
for these stars.

However, things change at the TTCC. At the TTCC (dM3), the slopes for the high 
activity stars are found to diminish dramatically, by a factor of about 2. 
Although there remains a large uncertainty on this slope for dM3e stars, the 
fall of the slopes at the TTCC is confirmed by the measure of the slope for 
dM4e stars with a much higher confidence level.
Beyond the TTCC (dM4), the slope continues to diminish. In fact, the slope at 
dM4 reaches a value that is very similar to the slope we have obtained for the 
low activity star sub-samples (although we emphasize that the overall shapes 
of the RACs are very different: see Figs.~8 and 12). This dramatic decrease in 
steepness among (unsaturated) high activity stars at M3 and M4 suggests that 
the dynamo mechanisms operating in dK4, dK6 and dM2 high activity stars 
(perhaps an $\alpha-\Omega$ dynamo?) may be different from those operating in 
the fully convective dM3 and dM4 high activity stars (perhaps an $\alpha^2$ 
dynamo?). In this regard, we detect no major changes in the slopes at the TTCC 
for the low activity star sub-samples apart from a decreased statistical 
significance of the RAC slope. This point is further discussed in the next 
sub-section.

We emphasize that, although our LSFs to the high activity stars seem relatively
 well established, in our sub-samples we have only few measures (from 8 to 23).
 Therefore, these results should be confirmed with larger stellar samples and 
are still preliminary.

\subsubsection{Approaching the TTCC?}

These results lead us to suggest a five-part conclusion. 

(i) The gradients of the RACs which we have obtained in our combined samples 
of low+high activity stars unambiguously increase with increasing spectral 
type with a high level of confidence. But it also appears that the linear LSFs 
only provide a poor description of the full RACs.

(ii) There appears to be a significant dichotomy between the low activity and 
the (unsaturated) high activity sub-samples. This leads us to consider that 
there may exist two distinct dynamo regimes in dKe, dMe and dK, dM stars.

(iii) The gradients of the combined samples suggest that there are changes 
operating in the dynamo mechanisms before, at, and after the TTCC. In addition,
 the gradients to the high activity sub-samples point to important changes in 
the dynamo mechanisms occuring at the TTCC and beyond.

(iv) We can confidently assert that in low activity dK4, dK6, dM2, and dM3 
stars, there is a robust ($>4 \sigma$) increase in Ca\,{\sc ii} flux as the 
period decreases. That is, the dynamo in unsaturated dK4-dM3 stars {\it is} 
clearly sensitive to rotation. 

(v) But in low activity M4 stars, the sensitivity of the dynamo to rotation, 
although still perhaps present, is not as robust, and may even be zero (at the 
3$\sigma$ level of significance). Thus, even in the low activity stars in our 
samples, a rotational dynamo is certainly contributing significantly to 
chromospheric heating in dK4-dM3 stars. But in dM4 low activity stars, the 
signs of rotational control over chromospheric emission are less significant 
(in a statistical sense). 

These two last points, if they can be confirmed, might be relevant in the 
context of Durney et al (1993), who suggested that rotation would be a 
controlling influence in the case of interface dynamos, but should not be as 
effective as a controlling influence in distributed dynamos. On the one hand, 
the robust sensitivity to rotation which we have found in dK4-dM3 low activity 
stars could be evidence for interface dynamos in those stars. On the other 
hand, in dM4 low activity stars, the weaker evidence for statistically 
significant rotational sensitivity might presage the lessening of the effects 
of an interface dynamo. In this context,  M3 might be considered to be the 
latest spectral sub-type to have definitive evidence for the presence of a 
radiative core (so that an interface dynamo is even possible). It is worth 
recalling that HM also concluded that at M3, something unusual happens to the 
rotational braking. Could the HM result, in combination with our results for 
the RAC slope, be a sign that the TTCC occurs between M3 and M4? This will be 
tested in a future study using an even finer grained spectral type sampling
at about the TTCC using new data.

We suggest that in Fig.~13, the striking difference between the 
combined sample (upwardly rising lines) and the low activity stars only (lower 
lines) may be associated with the fact that the combined sample might contain 
two distinct regimes of dynamo operation (high and low), whereas the low 
activity sample contains only a single regime (low). Suppose that the 
separation of dM4e stars from dM4 stars in Fig.~12 is associated with two 
distinct dynamo regimes: if so, the attempt to fit a LSF to the combined 
sample of dM4+dM4e stars may involve an attempt to ``force" two different 
types of dynamos into a single mold. Such an attempt would not be as 
physically meaningful as attempting to fit a truly homogeneous population (e.g.
 the low activity stars alone). Analogous arguments can be made for dM3 stars 
(Fig.~8) and dM2 stars (Fig.~5). According to this argument, it may be more 
profitable to focus separately on the different curves of the low activity 
stars and the high activity stars respectively in Fig.~13 in order to gain 
insight into dynamo theory. 

We conclude that for the low activity stars only, the gradient is definitely 
shallower than that for the samples of the low plus high activity stars. But
we also conclude that the gradient for the low activity stars only does not 
vary significantly with spectral type. We also note that the differences 
between the slopes for the low activity stars and for the low+high activity 
stars increase drastically with increasing spectral type. This result 
emphasizes that {\em the efficiency of the dynamo mechanisms increases 
drastically at short rotation periods, and also increases with decreasing 
stellar mass}. We also conclude that {\em there are no significant changes 
occurring in the efficiency of the dynamo mechanisms for low activity stars
as a function of stellar mass when moving from dK4 to dM4}.

\subsubsection{Different dynamo regimes among low activity and high activity 
stars at the TTCC and beyond ?}

In the plots which we have compiled of each RAC for our five spectral 
sub-types (Figs.~2, 3, 5, 8, 12), we overplot the separate heteroscedastic 
linear LSF to the low activity stars and to the high activity stars for direct 
comparison with the heteroscedastic linear LSF to the 
low+high activity samples. As we saw in the previous Sections, the overplots 
show clearly that the linear LSF to the low+high activity stars fails 
completely to reproduce the slopes of the linear LSF which we have derived 
both from the slow rotators and the fast (unsaturated) rotators. We show that 
the low+high activity samples cannot be described by a single RAC. 

In view of this, it seems to us that, to the extent that the RAC is determined 
by a dynamo mechanism, we need to invoke two different dynamo regimes. The 
first (we suggest) is at work among dK and dM stars (i.e. low-activity stars). 
This first regime seems rather constant from dK4 to dM4 and is therefore 
independent (to a first approximation) of the spectral type. A second dynamo 
regime (we suggest) is at work among dKe and dMe stars. This second regime 
exhibits steeper RAC slopes for the spectral types dK4, dK6 and dM2, and 
shallower slopes for the dM3 and dM4 spectral types. The data suggest that the 
second regime deviates significantly from the first regime at the TTCC and 
beyond. The main problem is that, among the stars in our samples at spectral 
sub-types dM3 and dM4, there are interlopers, namely, a certain number of high 
activity-relatively slow rotators and a certain number of low activity 
relatively-fast rotators that overlap in $P/\sin i$. These interlopers make it 
impossible to reproduce the entire datasets with a single RAC. Our results 
suggest that, whether we are considering low activity stars alone, or the 
combined samples of low+high activity, the properties of the RACs undergo 
changes of some kind, especially at the TTCC and beyond. This conclusion is 
especially true for dM4 stars. 

The findings we present here can be considered as providing evidence in favor 
of the following hypothesis: the dynamo mechanism in fully convective stars is 
different from the dynamo mechanisms in stars where a tachocline exists. We 
propose that differential rotation (probably important in the young dMe stars 
and less important in the older dM stars) plays an important role in defining 
the RACs among dM and dMe stars at the TTCC and beyond.

\subsection{The RACs for five different spectral sub-types}

\begin{figure*} 
\vspace{-0.5cm}
\hspace{-2.5cm}
\includegraphics[width=11cm,angle=-90]{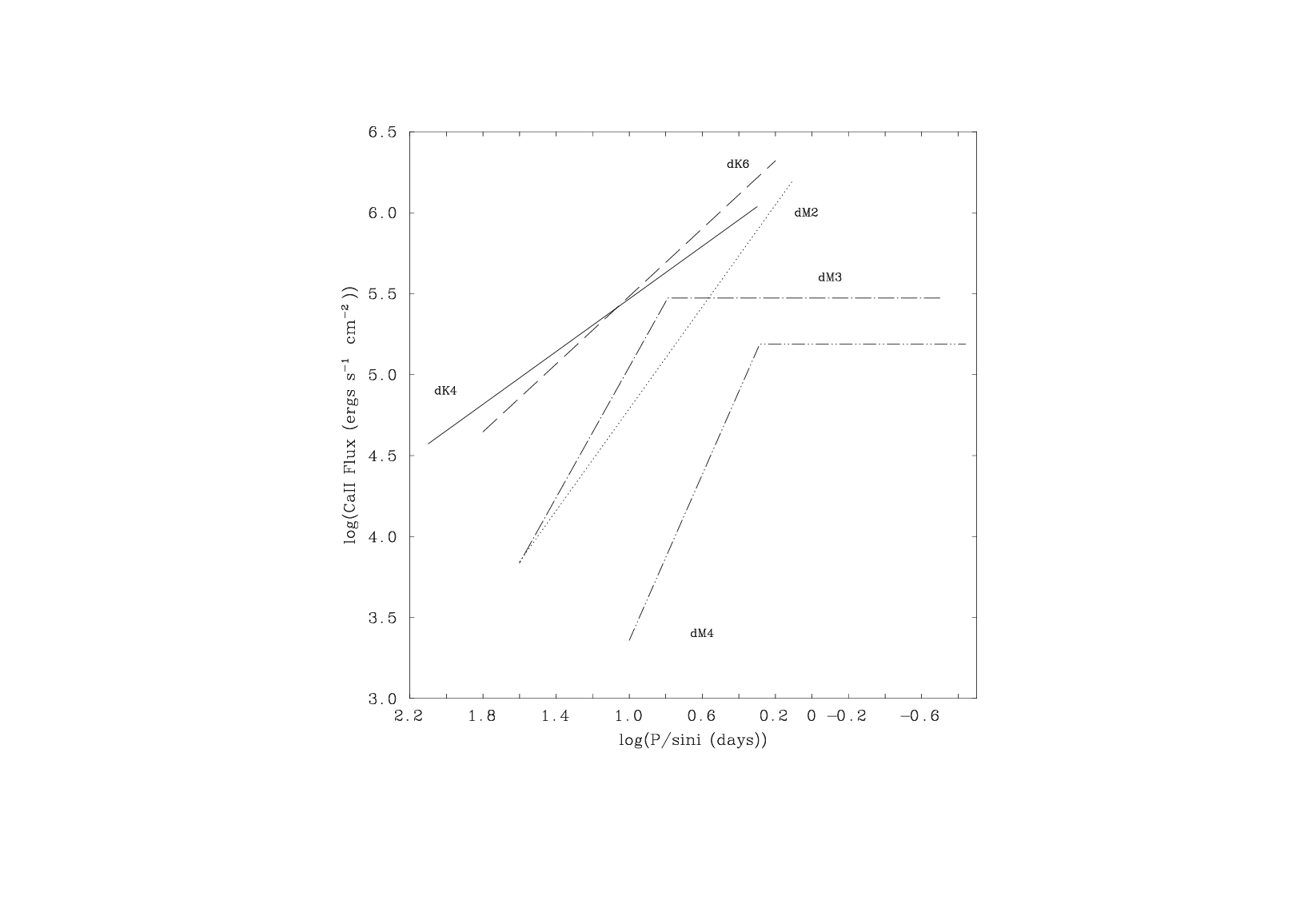}
\vspace{-0.5cm}
\hspace{-2.5cm}
\includegraphics[width=11cm,angle=-90]{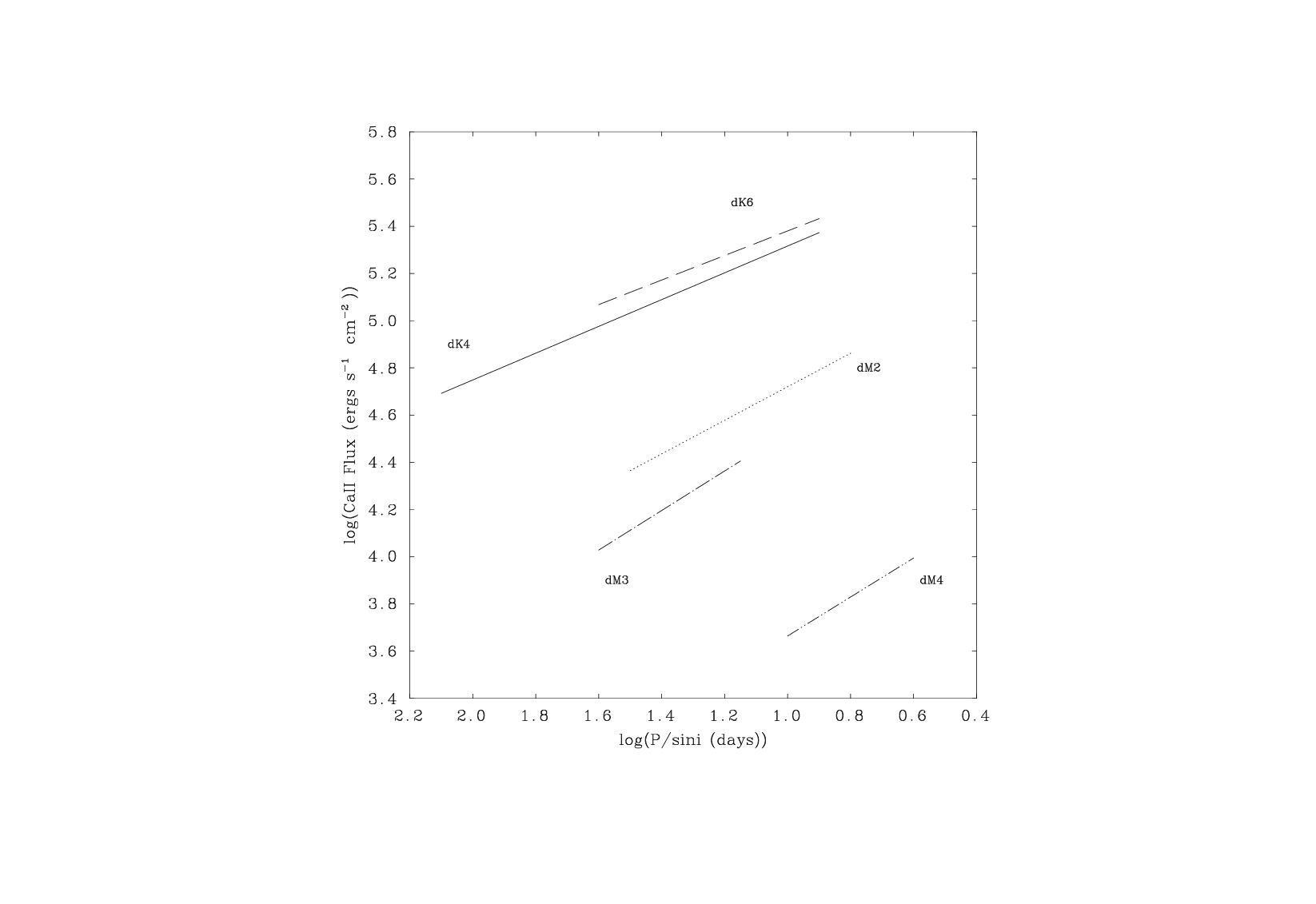}
\vspace{-0.5cm}
\caption[]{Upper panel: RACs for the logarithm of the Ca\,{\sc ii} resonance 
line surface fluxes and $log(P/\sin i)$ for combined samples of low+high 
activity stars in each sub-type dK4, dK6, dM2, dM3 and dM4. Note the major 
differences that exist between different spectral sub-types as regards the 
surface fluxes of the Ca\,{\sc ii} lines and the rotation periods. Lower 
panel: RACs for the same five spectral sub-types, but for the sub-samples of 
low activity stars only (see text).}
\end{figure*}

We bring together the heteroscedastic linear fits to the low+high activity 
RACs (in terms of surface fluxes) for our five spectral sub-types in Fig.~14 
(upper panel). In this section we discuss two key aspects of the figure, and 
how they vary as a function of spectral type: (i) the absolute magnitudes of 
the Ca\,{\sc ii} flux, and (ii) the slopes of the RAC in the unsaturated 
regime.

There are several striking trends in this figure. First, with 
increasing spectral type, the curves tend to shift downward and to the right, 
i.e. the overall level of chromospheric flux decreases, and the periods tend 
to be shorter. Important differences between the RACs of dK4, dM2 and dM4 
stars have already been reported by Houdebine (2012b). Also, variations in the 
RAC properties at different spectral types was previously reported by 
Stepie\'n (1993, 1994). He found that the $log(R'_{HK})-P_{rot}$ relationships 
globally shifted towards shorter rotation periods as (B-V) increases from 0.52 
(dF7.5) to 1.15 (i.e. spectral type dK5, according to KH). On the other hand, 
in an earlier study, the opposite tendency emerged: Stepie\'n (1989) found 
that the $log(\Delta F_{CaII})-P_{rot}$ relationships were globally shifted 
towards longer rotation periods for (B-V) from 0.45 (dF6) to 0.66 (dG5) and 
that they stabilize for (B-V) from 0.75 (dG8.5) to 1.40 (dK6). Also, Patten \& 
Simon (1996) proposed that the projected rotation period becomes longer as one 
move from spectral types G1 to M2. In view of this complicated behaviour, it 
seems that in the $F_{CaII}-P_{rot}$ RACs, first the rotation period becomes 
longer between dF6 and dG7, then remains unchanged between dG7 and dK6, and 
then the period starts to become shorter between dK6 and dM4: the latter 
behaviour persists into later spectral types.

In Fig.~14, in the context of the combined samples (upper panel), one can see 
that the steepness of the RACs (at least in the unsaturated regime)  
increases noticeably as we go towards later spectral type (as already shown in 
Fig.~13). We have not detected any saturation behavior for stars in our dK4, 
dK6 or dM2 stars, whereas we may observe a saturation in dM3 and dM4 stars 
for the most rapid rotators. Might this saturated-unsaturated behavior 
also point towards a different dynamo mechanism in dM3 and dM4 stars from the 
dynamo mechanism in dK4, dK6, and dM2 stars? We also observe that the largest 
Ca\,{\sc ii} surface fluxes are found in the stars with the earliest spectral 
types (dK4, dK6, dM2). For dM4 stars, there is a pronounced decrease in the 
absolute magnitudes of the surface fluxes for the fastest (unsaturated) 
rotators compared to dM2 stars. But even among relatively slow rotators (with 
periods longer than 10 days), where any effects of saturation are expected to 
be minimal, the Ca\,{\sc ii} surface fluxes are observed to decline by more 
than 100 between dK4 and dM4. 

Surveying the results, we see that dK4 and dK6 stars have similar RACs, 
with a slightly larger gradient for the dK6 stars. Then the RAC falls to lower 
surface fluxes for dM2 stars and the RAC steepens. The RAC for dM3 stars is 
distinctly different in shape compared to that of the dM2 stars: dM3e stars 
exhibit saturation whereas the dM2e stars do not. Also, dM3 stars overlap in 
rotation periods with dM2 stars. This is due to the fact that we found (in HM) 
that dM3 stars possess abnormally long rotation periods for both high activity 
and low activity stars compared to the adjoining dM2 and dM4 stars. In HM, it 
was suggested that the anomaly of relatively long rotational periods at type 
M3 might be related to an empirical report (Mullan et al. 2006) that flaring 
loop lengths undergo an increase to larger values at dM3. In dM4 stars, the 
(unsaturated) RAC shape parallels that of dM3.

The trends in surface fluxes of Ca\,{\sc ii} in Fig.~14 suggest that whatever 
is controlling chromospheric emission in our sample stars, the mechanism is 
less effective (by factors of 10 or more), and leads to different RAC shapes, 
in dM3 and dM4 stars than in dM2, dK6, dK4 stars. 

Now, as regards the latter stars, it might at first sight be believed that we 
should be fairly confident that interface dynamos are at work (e.g. Mullan et 
al. 2015). 
(Unfortunately, we cannot be absolutely confident that {\it only} an ID is at 
work in the dM2, dK6, and dK4 stars: the results of Brown et al. (2010) 
indicate that also a DD can operate in the convection zone of partially 
convective stars. If both ID and DD are in fact simultaneously operative in a 
certain star, current dynamo models give no quantitative results for the 
relative strengths of magnetic fields which would be generated by the 
different dynamos. Thus, although we use the words ``fairly confident" in the 
present context, the question of the relative importance of ID and DD in a 
partially convective star remains ambiguous.

Could it be that the different RAC shapes in dM3 and dM4 stars might be due to 
the absence of an interface dynamo? If so, TTCC may be occurring between dM2 
and dM3 sub-types. We also note that the changes in the RACs are progressive 
from dK4 to dM4, and that the fall in the fluxes begins at about dM2 and 
continues from dM3 to dM4. However, whatever the correct physical 
interpretation in terms of different dynamos eventually turns out to be, our 
data show that {\em the efficiency of the dynamo mechanisms falls 
progressively from dK6 to dM4, and that this fall is particularly pronounced 
after the TTCC at spectral sub-type M4}.

\subsubsection{Low activity stars only}

In the lower panel of Fig.~14, we show results of the heteroscedastic models 
for low activity stars only. The most striking difference between the lower 
and upper panels in Fig.~14 is that the low activity stars do not exhibit any 
evidence for saturated behavior. Moreover, a striking characteristic of the 
lower panel is that the five RACs are almost all parallel to one another
 (as already noted in Fig.~13). These features support our claim 
that low activity stars may provide us with a possible ``window" into a more 
homogeneous sample in terms of the dynamo mechanism: it seems likely that only 
unsaturated dynamo operation is at work in the lower panel of Fig.~14.

We note that surface fluxes for unsaturated dM4 stars are again found to be 
10-100 times smaller than those in unsaturated dK4 and dK6 stars. And we also
observe a similar global pattern in both panels (upper and lower): there is a 
trend towards shorter periods with increasing spectral type. 

It is clear that the overall level of Ca\,{\sc ii} flux is diminishing as we 
go from K4 to M4: the $b$ coefficient of the linear LSF to the RAC of dK4 
stars is larger by a factor of 25 than the linear LSF to the RAC of dM4 stars.
This suggests that the dM4 stars at any given period are generating mechanical 
energy flux some 25 times less effectively than dK4 stars with the same 
rotational period. Even in the much narrower gap between dM3 and dM4, the $b$ 
value for the linear LSF to the RAC has decreased by a factor of almost 10 at 
any particular rotational period. These results indicate that dM4 stars really 
are suffering from a weakening of the ability to generate mechanical energy 
(whether in magnetic form or in acoustic form or in a combination of both 
forms) compared to dM3 and earlier sub-types.

We conclude that the overall level of dynamo efficiency (as regards 
chromospheric heating) diminishes markedly as the stellar mass decreases.
However, we also find that, as regards the low activity stars (i.e. unsaturated
dynamos), {\em there is an almost universal dependency of the dynamo efficiency
on the rotation period: i.e. the level of activity varies approximately as 
$\sim P/\sin i^{-0.80}$.} This universal feature is missing when we attempt to 
analyze samples which combine stars of {\it both} low activity and high 
activity. For the combined samples, our results indicate that the dynamo 
efficiency decreases differentially and drastically with decreasing rotation 
period as we move from dK4 dwarfs to M4 dwarfs.

\subsection{$R'_{HK}$ as a function of the Rossby number}

In the spirit of searching for RAC's among a variety of parameters (in case 
the correlations are obscured more in certain cases), we now set aside the 
rotation period which we have used so far, and explore how activity depends on 
Rossby number. Dynamo theory suggests that the dynamo efficiency, and 
therefore the magnetic field strength and its direct diagnostic in the 
chromosphere, the Ca\,{\sc ii} lines, should scale with the dynamo number 
(Montesinos et al. 2001) $N_{D}$ given by:

\begin{equation}
N_{D}\sim \frac{1}{R_{0}^{2}} \frac{\Delta\Omega r_{cz} L}{\nu \Omega d^{2}}
\end{equation}

where $R_{0}=P/\tau_{c}$ is the Rossby number, $\tau_{c}$ the convective 
overturning time, $\Omega$ is a characteristic rotation rate in the lower 
convection zone, $r_{cz}$ the radius at the base of the convection zone, $L$ 
the characteristic length-scale for the differential rotation in the 
overshoot region just below the convection zone, $\nu$ the ratio of the 
diffusivities in the two layers directly below and above the overshoot region 
and, and the quantity $d$ is given by $d\sim \sqrt{\eta \tau_{c}}$ where 
$\eta$ is the turbulent diffusivity in the layer just above the overshoot 
region. Making reasonable assumptions for late-type stars, the dynamo number 
simplifies to (Noyes et al. 1984, hereafter N84):

\begin{equation}
N_{D}\sim (\Omega \tau_{c})^{2} \sim R_{0}^{-2} 
\end{equation}

Therefore, a good evaluation of the efficiency of dynamo mechanisms can be 
done through plotting the chromospheric indices $R'_{HK}=L_{HK}/L_{bol}$ 
(where $L_{HK}$ is the luminosity in the H and K lines of Ca\,{\sc ii}) as a 
function of the Rossby number $R_{0}=P/\tau_{c}$ (e.g. N84, Stepien 1989, 
1994, Hempelmann et al. 1995, Patten \& Simon 1996, Montesinos et al. 2001). 

Montesinos et al. (2001) argue that the dimensionless factor,

\begin{equation}
\frac{\Delta\Omega r_{cz} L}{\nu \Omega d^{2}}
\end{equation}

plays a role in the scatter of the $R'_{HK}$ versus $R_{0}$ diagrams. 
This dimensionless factor depends notably on the internal differential 
rotation which is mostly confined to the overshoot region just below the 
convection zone as well as on the turbulent magnetic diffusivities, neither of 
which are currently reliably known. In view of this lack of information, we 
shall for simplicity here assume this dimensionless factor to be constant for 
our samples of late-type dwarfs. 

\begin{figure*} 
\vspace{-0.5cm}
\hspace{-2.5cm}
\includegraphics[width=11cm,angle=-90]{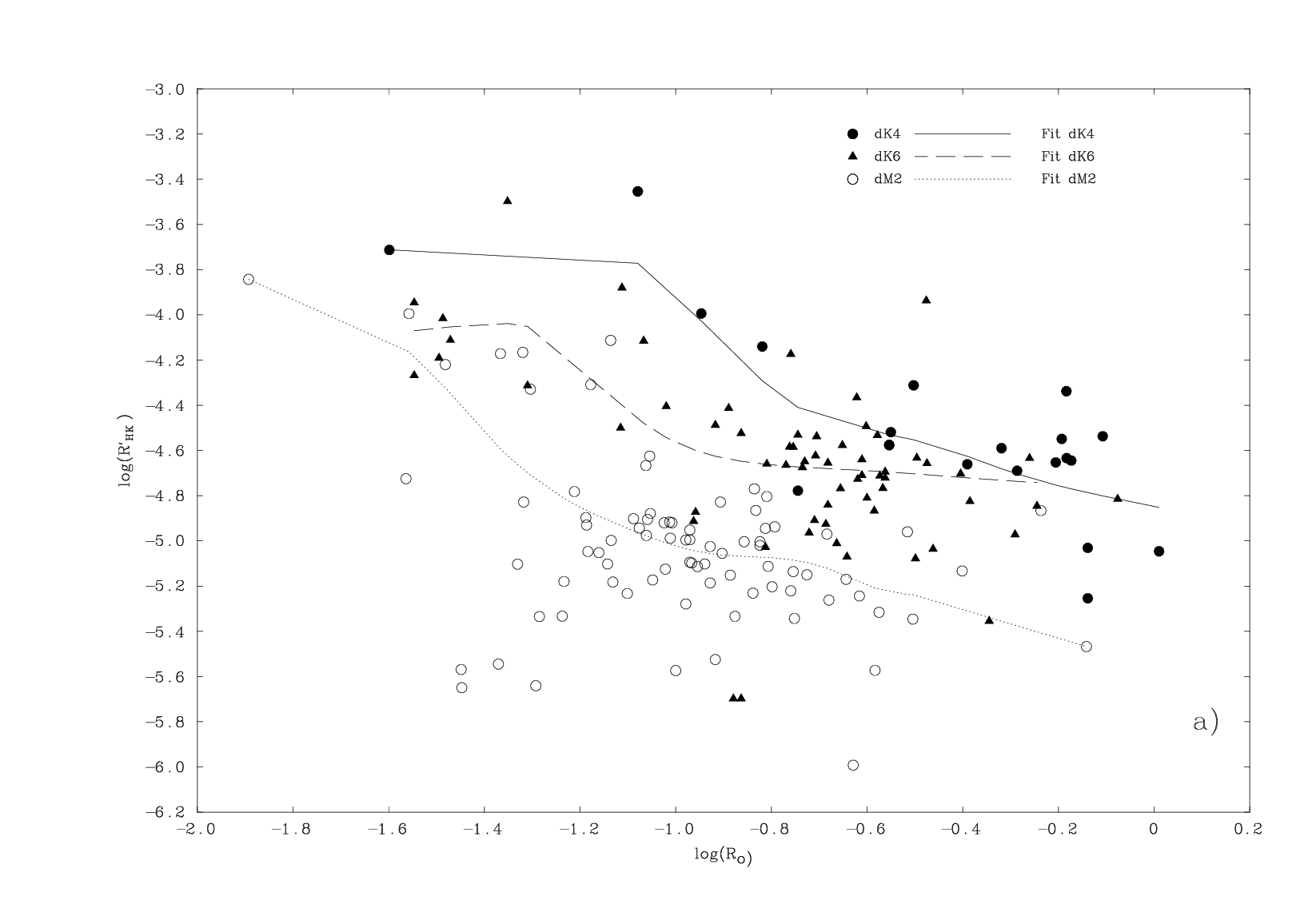}
\vspace{-0.5cm}
\hspace{-2.5cm}
\includegraphics[width=11cm,angle=-90]{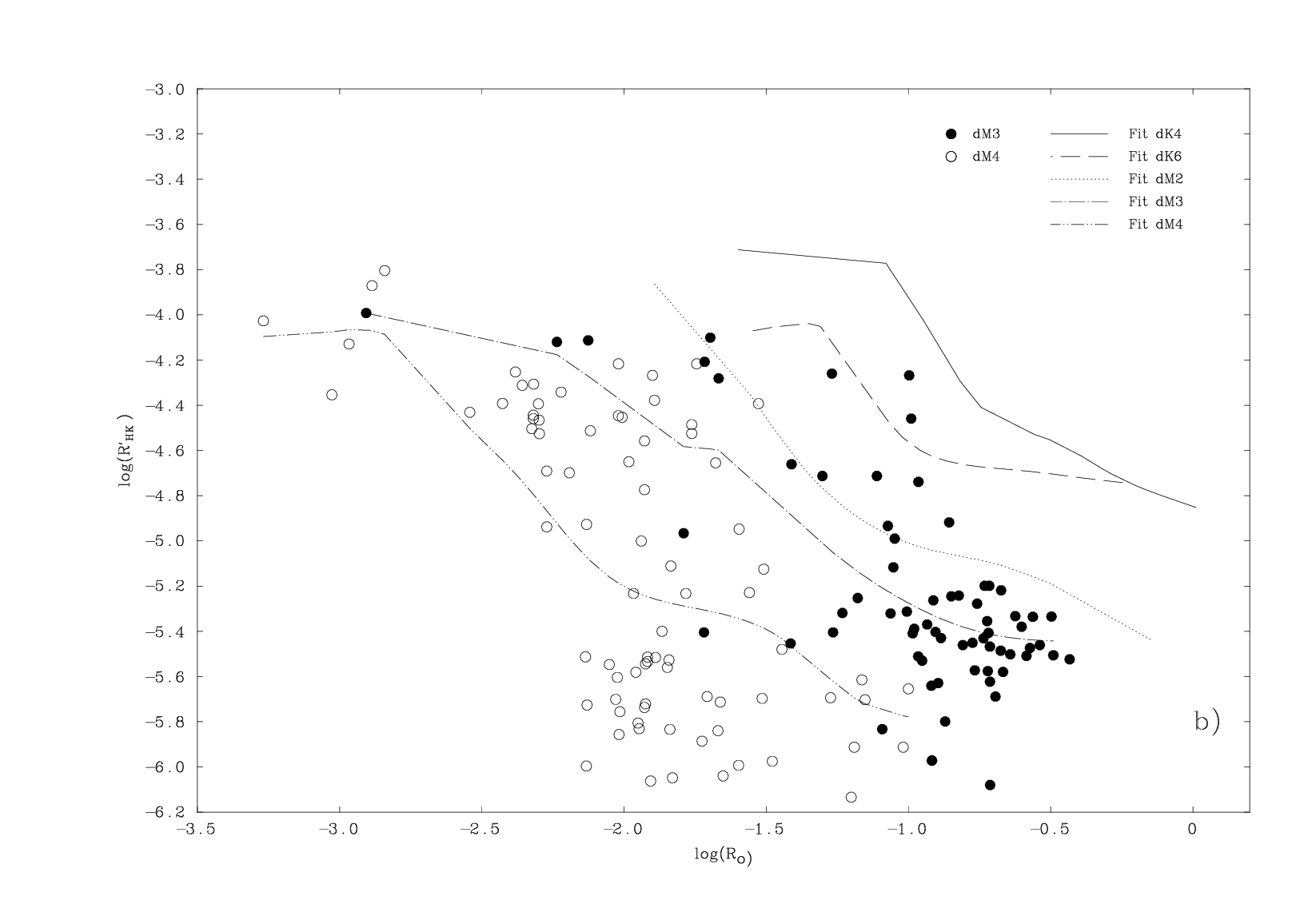}
\vspace{-0.5cm}
\caption[]{Upper panel: $R'_{HK}$ as a function of the Rossby number, $R_{0}$. 
Solid line: the mean empirical relationship for dK4 stars. The data for dK4 
stars is a good representation of the F, G, early K type star correlation. 
For dK6 stars, the mean values of $R'_{HK}$ (long-dashed line) lie on average 
a factor of 3 below the dK4 stars. For dM2 stars, the mean values of $R'_{HK}$ 
(dotted curve) lie a factor of about 10 below the dK4 stars. Lower panel: dM3 
stars (dot-dashed curve) and dM4 stars (double-dot-dashed curve) lie a factor 
of about 20 and 90 below the dK4 stars respectively. Our data indicate a 
gradual decrease in the fluxes in M dwarfs that continues into the fully 
convective M3 and M4 dwarfs, in agreement with our findings on the RACs 
based on the Ca\,{\sc ii} stellar surface fluxes (Sect. 3.6). This suggests 
a progressive change in the dynamo mechanisms from a shell dynamo to a 
distributive dynamo.}
\end{figure*}

In order to evaluate Rossby numbers in the stars of our sub-samples, we 
have adopted values of $\tau_{c}$ from the results of Spada et al. (2013) 
who gives the relationship between stellar mass $M_{\star}$ and $\tau_{c}$. We 
derived a relationship between the stellar radius, $R_{\star}$, and 
$\tau_{c}$ using the mass-radius relationships of Spada et al. (2013) for 
[Fe/H]=0. The results of Spada et al. (2013) on $\tau_{c}$ are in reasonably 
good agreement with other calculations (e.g. Kim \& Demarque 1996). Recent 
values of $\tau_{c}$ agree to some extent with the values of Noyes et al. 
(1984) for solar masses down to $M_{\star}\sim 0.8 M_{\odot}$ but differ 
substantially for lower mass stars. We compare the values of $\tau_{c}$ from 
Spada et al. (2013) and Noyes et al. (1984) for our five spectral sub-types 
in Paper~I. The large increase in the numerical values of $\tau_{c}$ which we 
take from Spada et al (2013) at low masses has important consequences for the 
$R'_{HK}$/$R_{0}$ relationships in M dwarfs as we shall see below.

We use the $\tau_{c}$/$R_{\star}$ tabulation to derive $R_{0}$ for each star 
in our five samples of stars. We also computed the luminosity, $L_{HK}$, in 
the Ca\,{\sc ii} lines according to the continuum surface fluxes derived 
above from the models of de Laverny et al. (2012). We list the Ca\,{\sc ii} 
luminosities $L_{HK}$ and $R_{0}$ for our five samples of stars in Table~8. 
We also computed the bolometric luminosities $L_{bol}$ and the activity 
indice $R'_{HK}=L_{HK}/L_{bol}$ for all our targets. We list the results in 
Table~8. We further computed $L_{HK}$ and $R'_{HK}$ corrected for 
metallicity effects for dM2, dM3 and dM4 stars.

Mullan and MacDonald (2001) suggest that the ratio of the X-ray luminosity 
created by large-scale fields $L_{X}(L)$ to that of the X-ray luminosity 
created by turbulent fields $L_{X}(t)$ may attain factors of 5-10. Although the
 case of chromospheric lines does not necessarily track the coronal X-ray 
emission, one might also expect a decrease in chromospheric emission when the 
dynamo mechanisms change from a shell dynamo to a distributive dynamo. Note, 
however that in the case of the chromosphere some modelling calculations (e.g. 
Ulmschneider et al. 2001, Fawzy et al. 2002, Ulmschneider \& Musielak 2003, 
Ulmschneider et al. 2005) yield one to expect a basal flux in the chromospheric
 lines. Such basal flux has not been observed in our datasets (e.g. this study,
Papers XV and XVIII). However, some previous authors claim to have detected 
such fluxes for various spectral types (e.g. Wilson 1968, Schrijver 1987, 
Rutten et al. 1991, Strassmeier et al. 1994, Fawzy et al. 2002). Note however 
that M dwarfs were usually excluded from those studies. On the other hand, 
there do exist certain data sets for line fluxes of Mg\,{\sc ii} and 
$Ly_{\alpha}$ in inactive M dwarfs which have been found to be consistent with 
{\it ab initio} models of acoustically heated chromospheres (Mullan and Cheng 
1993): thus, there seems to be little reason to exclude the concept of ``basal"
 fluxes from a discussion of dM chromospheres. Here we investigate the 
$R'_{HK}-R_{0}$ relationships from dK4 stars to the fully convective dM3 and 
dM4 stars and search for a signature of the change in the dynamo mechanisms at 
the TTCC.

We plot in Fig.~15 $log(R'_{HK})$ as a function of $log(R_{0})$ for our five 
different spectral types. We found, using the values of $\tau_{c}$ from N84, 
that the mean of the measurements for dK4 stars (the dK4 data smoothed with a 
Gaussian of FWHM=0.1) agree well with the correlation for F, G and K type 
stars found by N84. Thus, our RAC data for dK4 stars are consistent with the 
N84 study and represents well a low-mass extension of the correlation for F, G 
and early K type stars.

In Fig.~15 (upper panel) we can see that for the dK6 stars (filled triangles), 
the mean values (the data smoothed with a Gaussian of FWHM=0.1, dashed line) 
lies slightly below the correlation for dK4 stars, typically by a factor from 
1.5 to 5.0, i.e. a factor of 3 on average. The dK6 curve crosses the dK4 curve 
at $log(R_{0})\sim -0.2$. There is a large scatter in the data for dK4, dK6 
and dM2 stars in spite of the fact that we corrected the data for the effects 
of metallicity in dM2 stars (otherwise, the scatter is even larger): this is 
explained by the variations in radii of our stars at a given spectral 
sub-type, and the large variations of $\tau_{c}$ with radius. This scatter 
increases for the spectral types dM3 and dM4 Fig.~15b) because for these stars 
there is a large increase in $\tau_{c}$ with decreasing stellar radius and 
within a small range in radius (see Spada et al. 2013).

For our dM2 stars, the situation is more striking: the data (dotted line in 
Fig. 15(a)) now lie much lower than the dK4 curve. There is a clear and large 
difference between our dK4 data and our dM2 data. Over the common range in 
$R_{0}$ the curve for dM2 stars is roughly parallel to that of the dK4 stars, 
but is shifted below by a factor from 6 to 16. Our data for the dM2 stars 
cover a larger range in $R_{0}$ than in previous studies, notably for small 
values of $R_{0}$ (i.e. stars with short rotational periods). One can see that 
the correlation suddenly rises for log$R_{0}$ less than -1.3. At small values 
of $R_{0}$, $log(R'_{HK})$ reaches values of about -4.0 or even larger. 

Moving later in spectral sub-type, we show in Fig.~15 (lower panel) the 
analogous data for our dM3 and dM4 stars. We note that in this diagram, the 
mean data for these stars (dash-dotted line) generally lie below the 
correlation of dK4 stars (solid line), and also significantly below our dM2 
data (dotted line). The large variations in $\tau_{c}$ between M2, M3 and M4 
stars imply that the curves for M3 and M4 stars are systematically shifted 
towards lower $R_{0}$. But we emphasize that also, globally, the values of 
$R'_{HK}$ decrease from M2 to M4 (Fig.~15, lower panel). The mean curve for 
our dM3 stars lies significantly below the mean curve for dM2 stars. This 
difference is even more pronounced for the sample of dM4 stars. At 
$log(R_{0})=-1.2$, we find that the dM3 curve lies a factor of 20 below that 
of the dK4 stars, and that the dM4 curve lies a factor of about 100 below the 
dK4 curve (Fig.~15, lower panel). The curves for dM2, dM3 and dM4 stars are 
more or less parallel (except at very low values of $R_{0}$ for dM3 and dM4 
stars, where there are only a few data points). This highlights the consistency
 of the decreases in the $R'_{HK}$-$R_{0}$ curves when going from mid-K type 
stars to the fully convective dM4 stars.

The decrease from dM2 to dM3 stars is typically a factor of 2 (at $log(R_{0})= 
-1.0$, whereas the decrease from dM3 to dM4 stars is typically a factor of 4 
(taken at $log(R_{0})=-1.5$).

Christian et al. (2011) have reported a saturation (or even 
``super-saturation") phenomenon in the chromospheric emission for fast 
rotators in young clusters at about $log(R'_{HK})=-4.08$ but with a significant
 scatter ($\pm$0.5 in the log). In our stellar samples, we do not have enough 
very fast rotators to confirm the existence of super-saturation. However, we 
note that $R'_{HK}$ measures generally lie below -3.8 for all our samples of 
late-K and M dwarfs. We also find that this high level of activity is attained 
for different values of $R_{0}$ that depend on the spectral type. For dK4 
stars, this high level of activity is attained at about $log(R_{0})\sim -1.0$, 
whereas for dM4 stars it is attained at about $log(R_{0})\sim -2.7$. 

Our datasets suggest that M dwarfs (especially M3 and M4) generally do not 
follow the same relationship as F, G, and K dwarfs. This had not been noted 
previously because the M dwarf data sample was too sparse. If this difference 
is due to a difference in dynamo mechanisms, then the change in dynamo 
mechanisms seems to be progressive as is illustrated by our dK6 and dM2 
datasets. Therefore, we cannot conclude from the results in Fig.~15 that there 
is any abrupt change at the TTCC (if this occurs no later than M4): instead, 
there seems to be a gradual decrease in the efficiency of the dynamo mechanisms
 that may start as early as spectral type dK6. This gradual change was 
suggested by Mullan \& MacDonald (2001). The efficiency of an interface (shell)
 dynamo may fall off between dK4 and dM2 as the radiative core occupies an 
increasingly small proportion of the stellar radius (Mullan et al. 2015), and 
the present data suggest that this decrease continues at least as far as M4. 
The latter point is strengthened by our results in Paper~XV which showed a 
{\em progressive} change from high fluxes (large radii M2 dwarfs, partially 
convective) to very low fluxes (M2 subdwarfs, fully convective). The 
$R'_{HK}-R_{0}$ diagram may be one of the most sensitive tests for detecting 
changes occurring in the dynamo mechanisms. However, the results from these 
RAC relationships have to be compared with the other magnetic activity 
indicators we study here. Note that with the present data there is evidence in 
the $R'_{HK}-R_{0}$ diagrams that gradual changes in the dynamo efficiency are 
occurring all the way to M4. Other magnetic activity diagnostics, including 
the mean fluxes (Sect.~3.8), also point to a continued change from K4 to M4.

The results in Fig.~15 strongly suggest that, as far as ${\it chromospheric}$ 
heating is concerned, there may be a gradual decrease in dynamo efficiency in  
M dwarfs compared to F, G and K dwarfs. It is generally believed that an 
$\alpha-\Omega$ type of dynamo dominates in F, G and K dwarfs. In early M 
dwarfs, an interface dynamo is probably still operating (Mullan et al. 2015), 
although eventually, in fully convective M dwarfs (beyond the TTCC), the 
dynamo mechanism is expected to behave differently. The most striking 
conclusion of the present section is that, as regards chromospheric heating 
there are almost two orders of magnitude {\it decrease} in going from dK4 to 
dM4. This further strengthens our previous findings (see Fig. 14) that, in 
the context of chromospheric heating, the  mechanical fluxes decrease by a 
factor of order 100 as we go from dK4 to dM4.

\subsection{The coronal RAC: $L_{X}/L_{bol}$ as a function of the Rossby 
number}

\begin{figure*} 
\vspace{-0.5cm}
\hspace{-2.5cm}
\includegraphics[width=13cm,angle=-90]{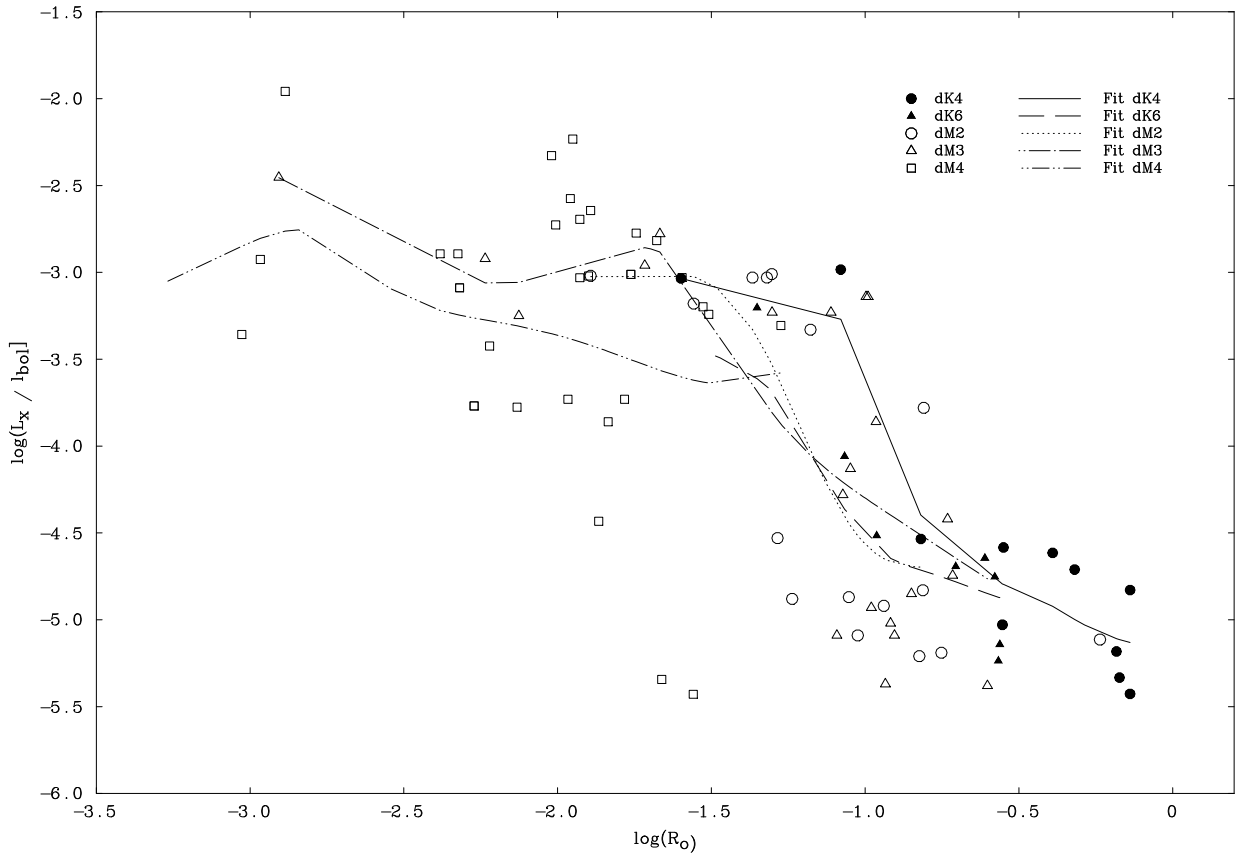}
\vspace{-0.5cm}
\caption[]{$L_{X}/L_{bol}$ as a function of the Rossby number, $R_{0}$, for our
 five different spectral types. We overplot the running means for dK4, dK6, 
dM2, dM3 and dM4 stars. Note the important differences in the behaviour of 
$L_{X}/L_{bol}$ (this figure) compared to that of $R'_{HK}$ (see Fig.~14).}
\end{figure*}

Now we turn to the corona, where deposition of mechanical heating manifests 
itself in the form of X-ray luminosity $L_X$.  Mullan \& MacDonald (2001) 
found that in a sample of some 40 ROSAT measures of dMe stars by Fleming et 
al. (1993), $L_{X}/L_{bol}$ was found to remain essentially invariant over the 
spectral range from early M to at least M7, at least for the most active stars 
(see also Browning et al. 2010). This result, which implies that coronal 
heating efficiency in the most active stars remains unchanged at the TTCC, now 
appears worth re-examining in the context of our findings (see Fig.~14) of 
changes in chromospheric heating before and near the TTCC. Our sample is now 
several times larger than that of Fleming et al. (1993). We therefore 
investigate here the behavior of $L_{X}/L_{bol}$ as a function of $R_{0}$ for 
our five different spectral sub-types, to see if we can identify any signature 
of the TTCC in the coronal data.

In order to illustrate the sensitivity of coronal heating to rotation, we plot 
in Fig.~16 $L_{X}/L_{bol}$ as a function of $R_{0}$ for our five different 
spectral sub-types.  We also plot in this diagram the running mean curves for 
our dK4, dK6, dM2, dM3 and dM4 stars. The differences 
between the behaviours of $R'_{HK}$ (Fig.~15) and $L_{X}/L_{bol}$ (Fig.~16) 
are significant. For dK4, dK6 and dM2 stars, $L_{X}/L_{bol}$ rises slowly as 
$R_{0}$ decreases for the slow rotators ($log(R_{0})$=0 to -1.0), and 
suddenly rises abruptly for the rapid rotators ($log(R_{0})<-1.0$). For the 
fast rotators ($log(R_{0})<-1.5$), $L_{X}/L_{bol}$ seems to saturate at a 
level of about $10^{-3.0}-10^{-2.5}$. Patten \& Simon (1996) also observed a 
similar behavior among their samples of stars in young clusters (IC 2391, 
$\alpha$ Persei, Pleiades, Hyades and main-sequence stars in the field). They 
also observed a saturation at $L_{X}/L_{bol}\sim 10^{-3.0}$ for 
$log(R_{0})<-0.6$. On the other hand, they did not observe a plateau-like of 
behavior from $-3.0<log(R_{0})<-1.3$ such as our dM2, dM3 and dM4 star data 
display in Fig.~16. Instead they observe a strong rise in $L_{X}/L_{bol}$ from 
$10^{-6.5}$to $10^{-3.0}$, more like the behaviour of our data from 
$-1.3<log(R_{0})<0.3$ in Fig.~16. Overall, our impression is that all stars in 
our five sub-samples follow a more or less similar correlation and show 
evidence for saturation for $log(R_{0})<-1.3$.

We should emphasize that our $L_{X}$ data (taken from Hunsch et al. 1999) are 
severely biased towards the most active stars because of the limited 
sensitivity of ROSAT. Therefore, our samples of measures for the least active 
stars ($log~R_{0}>-1.3$) are inevitably biased in a negative sense, and may 
not be dependable when we calculate the means of $L_{X}/L_{bol}$ as a function 
of $R_{0}$.

\subsection{Comparison and contrast between chromospheric and coronal heating 
rates}

In order to facilitate comparison of Fig.~15 (lower panel) (chromosphere) and 
Fig.~16 (corona), we use similar notation for the lines which illustrate 
average values of the quantities at different spectral sub-types. Thus, in 
Fig.~15 (lower), the dM3/dM4 stars (dotted line) lie {\it lowest} in the 
figure, whereas in Fig.~16, the same stars lie at about {\it the same levels} 
as other spectral sub-types in the figure.  For purposes of the present 
discussion, let us suppose that dM3 and dM4 stars may be labelled as at, or 
later than, the TTCC: we refer to these stars by the shorthand notation TTCC+. 
Stars at spectral types dM2 and earlier are referred to as TTCC-. In terms of 
this notation, our results indicate that, in terms of {\it the chromosphere}, 
TTCC+ stars are definitely weaker emitters than TTCC- stars (see Fig.~14 lower 
panel). On the other hand, in terms of the {\it corona}, TTCC+ stars have 
similar emissions to those from TTCC- stars.

Thus, when we compare stars on different sides of the TTCC, the trend in 
{\it coronal} activity does not behave in the same way as the trend in 
{\it chromospheric} activity. 

In order to explore this difference between activity at the chromospheric and 
coronal levels of M dwarf atmospheres  we turn now to investigate how $L_{X}$ 
varies as a function of $L_{HK}$, and also as a function of spectral type.

\subsubsection{$L_{X}$ as a function of $L_{HK}$}

In a previous study Schrijver et al. (1992) reported on the inter-relation 
between $L_{X}$ and $L_{HK}$ with nearly-simultaneous observations of a sample 
of 26 F5-K3 main-sequence stars. Schrijver et al. (1992) obtained a good 
correlation over three orders of magnitude in $L_{HK}$ and over four orders of 
magnitude in $L_{X}$. They found that $L_X$ increased faster than $L_{HK}$: 
$L_{X}\propto L_{HK}^{1.50\pm 0.20}$. In the present study, we compiled 
$L_{HK}$ and $L_{X}$ for our five different spectral subtypes. We find that 
the LSF to our samples give: 

\begin{equation}
L_{X}=4.57\ 10^{-16}\ L_{HK}^{1.54\pm 0.29}\ ergs/s,
\end{equation}

for our dK4 stars,

\begin{equation}
L_{X}=7.94\ 10^{-17}\ L_{HK}^{1.57\pm 0.21}\ ergs/s,
\end{equation}

for our dK6 stars,

\begin{equation}
L_{X}=1.00\ 10^{-15}\ L_{HK}^{1.57\pm 0.17}\ ergs/s,
\end{equation}

for our dM2 stars,

\begin{equation}
L_{X}=7.94\ 10^{-20}\ L_{HK}^{1.74\pm 0.11}\ ergs/s,
\end{equation}

for our dM3 stars, and 

\begin{equation}
L_{X}=2.51\ 10^{-9}\ L_{HK}^{1.37\pm 0.12}\ ergs/s,
\end{equation}

for our dM4 stars.

The correlations are in all cases highly significant (see Table~1). The only 
exception occurs in the dK6 stars: this is because in this sample we have only 
few high activity stars. Since the slope for dK6 stars is the same as for dK4 
and dM2 stars, we believe the dK6 correlation is relevant for comparison to 
other sample correlations. Perhaps the smaller range of $L_X$ values among dK6 
stars also contributes to low significance. We show in Fig.~17 the $L_{X}$ 
versus $L_{HK}$ correlations we found for our dK6, dM3 and dM4 stellar samples.

Including late-K and M dwarfs of all 5 sub-types, we find an average slope 
of 1.56 in our $log(L_{X})-log(L_{HK})$ relationships. This is essentially 
identical to the value reported by Schrijver et al. (1992) for F5-K3 stars. 
Therefore, the same slope seems to prevail for all main-sequence stars 
from F5 to M4. This result emphasizes that in low mass stars, it is a 
general result that {\em the coronal emission grows faster than the 
chromospheric emission, from low activity stars to high activity stars}.
It is interesting to note that when investigating IUE spectra of F to K 
type stars, Ayres et al. (1981) found also a power-law slope of about 
1.5 between the $10^{5}$~K line fluxes and the Mg\,{\sc ii} line fluxes. 
However, in their correlation between the soft X-rays and Mg\,{\sc ii} 
fluxes, they found a larger slope.

\begin{figure*}
\vspace{-0.5cm}
\hspace{-2.5cm}
\includegraphics[width=8cm,angle=-90]{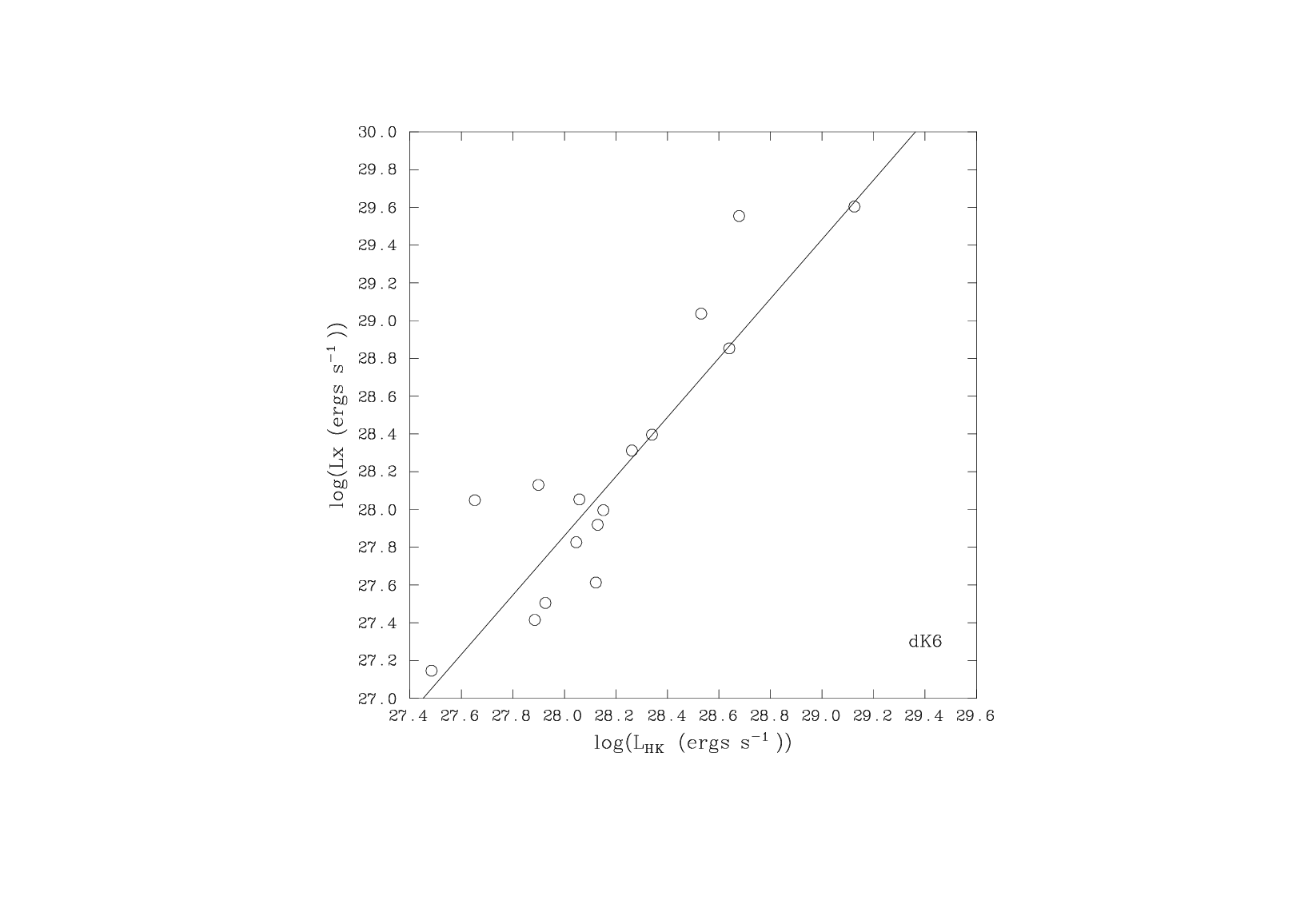}
\vspace{-0.5cm}
\hspace{-4.5cm}
\includegraphics[width=8cm,angle=-90]{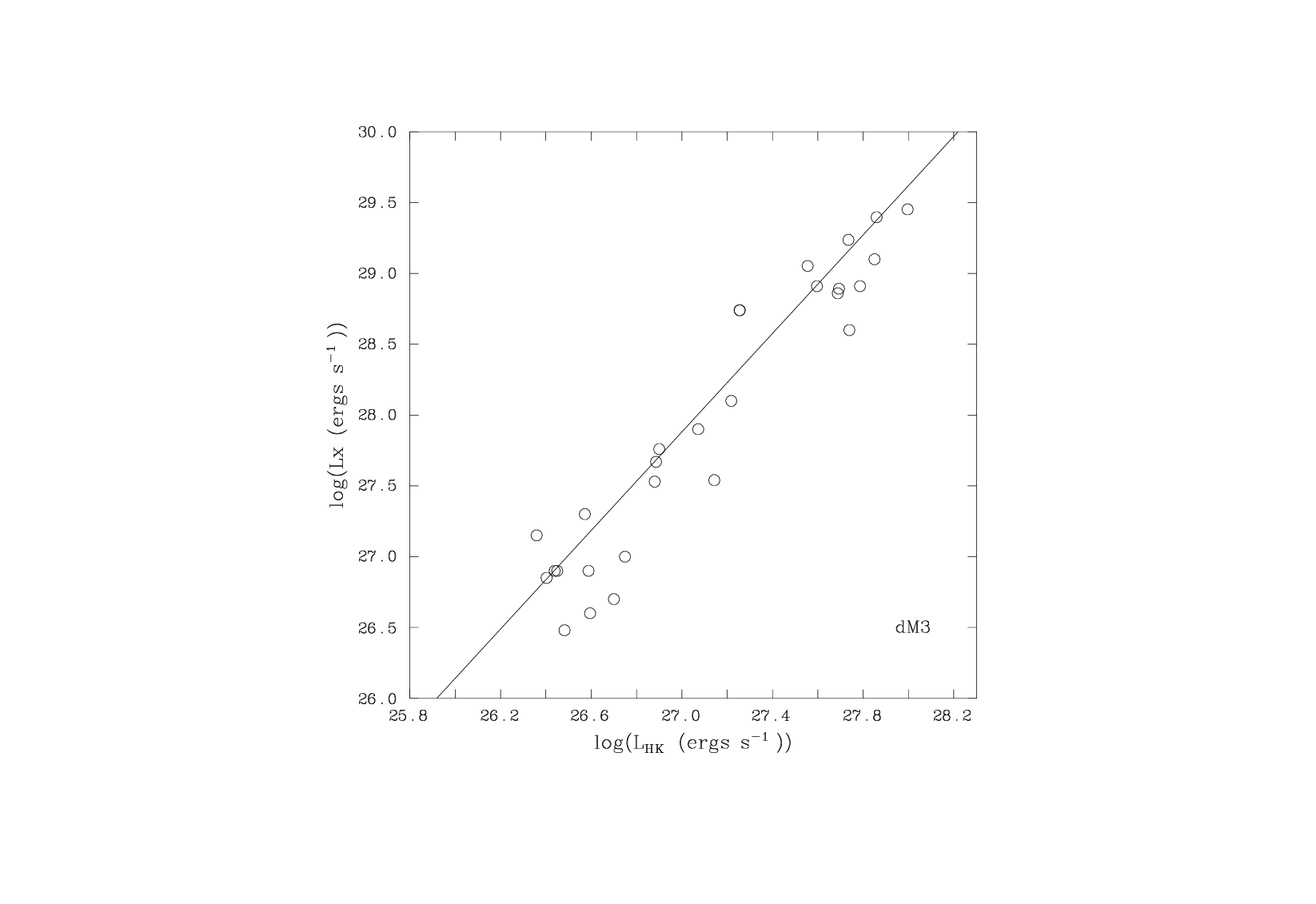}
\includegraphics[width=8cm,angle=-90]{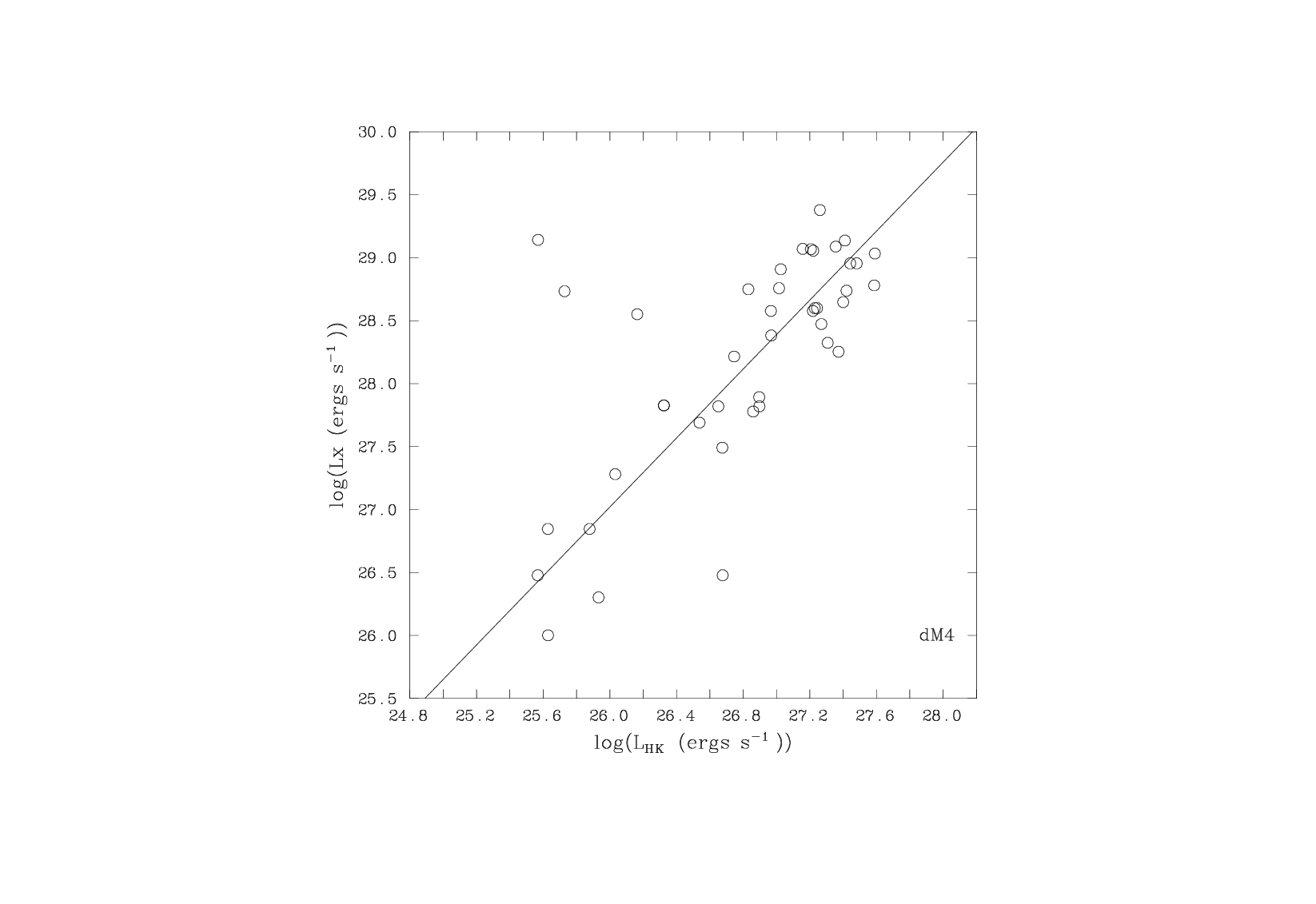}
\vspace{-0.5cm}
\caption[]{$L_{X}$ versus $L_{HK}$ for the stars in our dK6, dM3 and dM4 
samples for which ROSAT X-ray data are available. As regards the empirical 
correlations (solid lines), the {\it slopes} very similar: 1.54, 1.74 and 
1.37 for dK6, dM3 and dM4 stars respectively. Note the differences in the {\it 
scales} on both axes for the different panels: for a given value of {\it 
chromospheric} emission, the dM4 stars have a much larger {\it coronal} 
emission than the dK6 stars.}
\end{figure*}

However, what is more striking than the similarity in {\it gradients} found 
for these correlations, is the {\it multiplicative} coefficients in the 
correlations. We find that this factor changes from $4.47\ 10^{-16}$ for dK4 
stars to $2.51\ 10^{-9}$ for dM4 stars. This large difference indicates that 
the empirical correlations differ greatly from one spectral sub-type to another
 even if the gradients are comparable. To illustrate this,  we plot the LSF of 
$L_{X}$ versus $L_{HK}$ in Fig.~18 for our five spectral sub-types. In all 
cases, it is apparent that the correlations have slopes which do not differ 
greatly. But the absolute values of fluxes from the chromospheres differ 
greatly. E.g., dK4 and dK6 stars have $L_{HK}$ values mainly in the range 
$10^{27.5-29.5} ergs/s$, whereas the chromospheres of dM4 stars emit mainly in 
a much lower range, $10^{25-27.5} ergs/s$. On the other hand, emissions from 
the {\it coronae} of dK4 and dM4 stars overlap in the range $10^{27.5-28} 
ergs/s$. 

As a result, {\em at any given chromospheric emission level, the coronal 
emission strongly increases as we consider later spectral sub-types}. E.g., for
 a chromospheric Ca\,{\sc ii} luminosity of $10^{27.5}~ ergs\ s^{-1}$, the 
X-ray luminosity increases by a factor of at least 110 between dK4 and dM4. 
Therefore, in the case of M dwarfs, the situation is quite different from that 
of earlier F, G and K spectral types (e.g. Schrijver et al. 1992): stars from 
F to G to K all follow a common correlation between $L_X$ and $L_{HK}$. In a 
similar vein, the large differences we observe among M dwarfs explains the 
large scatter observed in the $L_{X}-L_{HK}$ correlation reported by Panagi \& 
Mathioudakis (1993) for K and M dwarfs. 

It is natural to inquire at this point: what makes M dwarfs so different from 
F-G-K dwarfs as regards the ratio of coronal to chromospheric emission? One 
possibility has been suggested by Mullan (1984). Coronal heating depends 
ultimately on tapping into the reservoir of mechanical energy associated with 
convective motions. In the context of electrodynamic coupling, the efficiency 
with which energy contained in that reservoir can be conveyed to the corona 
depends on two time-scales, $\tau_c$ for convection, and $\tau_A$ for the 
coronal loops in which the energy is to be deposited. In the Sun, these time 
scales differ by factors of order 100. This large difference explains why the 
coronal energy flux from the Sun amounts to only 1\% or less of the mechanical 
energy flux in convection. But Mullan (1984) predicted that as one 
goes down the main sequence, there could come a point where the two time 
scales $\tau_c$ and $\tau_A$ could be comparable. This would occur by reduction
 in $\tau_c$ due to smaller convection cells, and increases in $\tau_A$ due to 
longer coronal loops. By making some assumptions, the prediction was made that 
$\tau_c$ $\approx$ $\tau_A$ would occur among stars with $T_{eff}$ 
$\approx$~3400~K (see Mullan 1984: section VI). What spectral type does this 
correspond to? According to Rajpurohit et al (2013), the corresponding I-J 
colour is 1.2-1.3, and the spectral type is M3. Moreover, I-J = 1.2 
corresponds to V-I colour = 2.2, which is also the colour at which coronal 
loops become longer (Mullan et al 2006). 

\begin{figure*} 
\vspace{-0.5cm}
\hspace{-6.5cm}
\includegraphics[width=20cm,angle=-90]{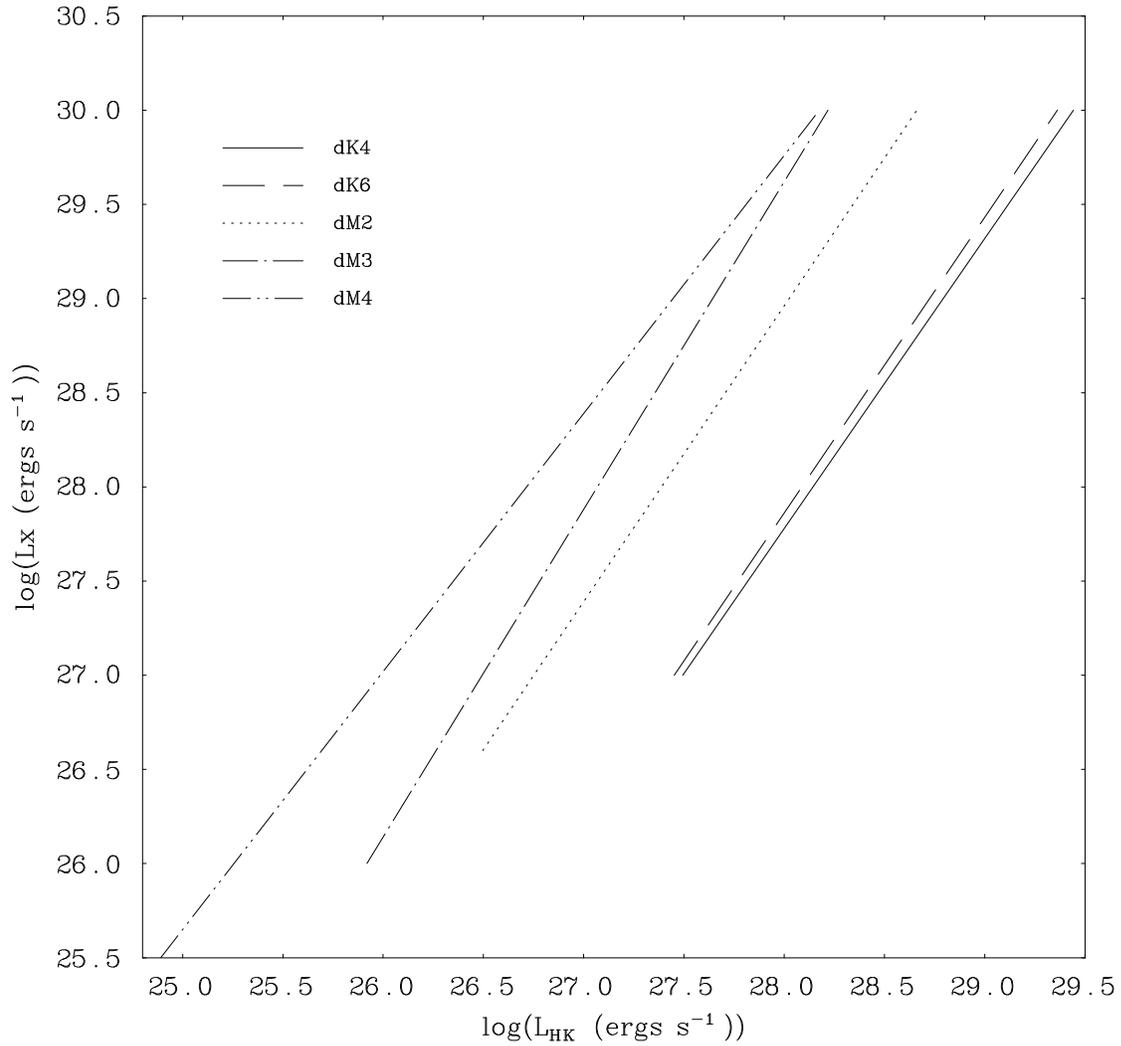}
\vspace{-0.5cm}
\caption[]{$L_{X}$ as a function of $L_{HK}$ for all five of our spectral 
sub-types. The {\it gradients} of the empirical correlations are similar for 
the five spectral types. But the more pronounced aspect of this Figure is 
that, at a given {\it chromospheric} flux, the {\it coronal} flux increases 
towards later spectral types. For instance, for a chromospheric Ca\,{\sc ii} 
luminosity of $10^{27.5} ergs/s$, the X-ray luminosity increases by a factor 
of 110 from spectral type dK4 to dM4.}
\end{figure*}

At the spectral type where $\tau_c$ $\approx$ $\tau_A$, the coronal loops 
would reach a resonance in their efficiency to tap into the convective 
reservoir. As a result, the efficiency of coronal heating in dM3 stars would 
be larger than in the Sun by 80-170. This range overlaps with the excess by 
factors of order 110 reported in Fig.~18 above. Because a resonant process is 
at work in this model, the increase in coronal heating efficiency is expected 
to build up in spectral types as these approach closer to the resonance, i.e. 
approaching M3. Thus, the full increase by factors of 80-170 should be 
realized at M3, but smaller enhancements are expected at (say) M2, K6, and K4. 

It is important to note that the electrodynamic resonance works only for the 
{\it corona}: there is no analogous process occurring in the {\it 
chromosphere}. Heating of the chromosphere, whether in an F star of an M star, 
continues to rely on localized dissipation of acoustic waves (for the ``basal" 
component: Mullan and Cheng 1993) and on localized dissipation of currents, 
via the conductivity tensor, in a partially ionized medium (for the ``magnetic"
 contribution) (e.g. Kazeminezhad and Goodman 2006).

Even if the resonant model (Mullan 1984) turns out to be incorrect, at least 
the empirical results explain the $L_{X}/L_{bol}$-$R_{0}$ relationships 
reported above. We can now interpret the $L_{X}/L_{bol}$-$R_{0}$ relationships 
(in Fig.~16) vis-a-vis the $R'_{HK}$-$R_{0}$ relationships (in Fig.~15). 
Indeed, in late-K and M dwarfs, the 110-fold increase in X-ray coronal 
emission from spectral sub-types K4 and M4 compensates for the 90-fold 
decrease in dynamo/chromospheric efficiency from the same spectral types. As 
a result, in Fig.~16, the M dwarfs have about the same $L_{X}/L_{bol}$ values 
as the dK4 and dK6 stars, whereas in Fig.~15, we found a drop-off for dM2 
stars of about a factor of 10 as regards the chromospheric heating. Note 
however that in Fig.~16, there is some convergence in $L_{X}/L_{bol}$ for 
$R_{0}<-1.3$ for all spectral types. 

This discussion may help us to understand why Mullan and MacDonald (2001) 
failed to identify any signature of the TTCC in $L_X /L_{bol}$ data: their 
failure may result from one aspect of {\it coronal} properties in M dwarfs. As 
a result of these properties, and their variation with spectral type in the 
vicinity of an electrodynamic resonance which (perhaps coincidentally) 
overlaps with the TTCC, the {\it coronal} ratio $L_{X}/L_{bol}$ is not really
suited for diagnosing the changing properties of the dynamo mechanisms before 
and at the TTCC. In view of the data presented in the present paper, we are 
now disposed to believe that the {\it chromospheric} emission (e.g. the 
Ca\,{\sc ii} luminosity) is better suited to respond to a signature of the 
TTCC.   

\subsection{The RACs and the mean rotation periods}

Can the properties of the RAC's discussed above explain the mean $P/\sin i$ 
values obtained by HM for stars ranging in spectral sub-type from dK4 to dM4 
stars? In particular, can they help in understanding the HM results of 
unexpectedly long rotation periods for dM3 stars? To answer this, we note that 
the mean $P/\sin i$ values of the slow rotators tend to decrease towards later 
spectral types. As a matter of fact, this trend is reproduced in the lower end 
of the RACs (Figs.~14). Therefore, the decrease in the mean $P/\sin i$ at 
later spectral types could be due to the fact that the dynamo mechanisms 
become inefficient at a rotation period that decreases towards later spectral 
type (Fig.~14). However, this is not completely the case for the dM3 RAC, 
which extends to somewhat longer periods than the dM2 RAC. Globally, the dM3 
RAC is shifted towards longer periods with respect to the dM2 RAC such that 
the mean $P/\sin i$ for dM3 stars is longer than for dM2 stars (as mentioned 
by HM). For dM4 stars, the whole of the RAC is shifted towards shorter rotation
 periods compared to dM3 stars. This explains why the mean $P/\sin i$ for dM4 
stars is clearly smaller than that for dM3 stars. In fact, the dynamo mechanism
 in dM4 stars apparently become so inefficient at long periods that our sample 
contains not a single dM4 stars with $P/\sin i$ longer than 10 days (see 
Figs.~14). This can be contrasted with the much longer $P/\sin i$ values for 
dM3 stars: as large as 30 days (see Fig.~8). Our discovery in the present 
paper that the level of chromospheric emission in dM4 stars is definitely 
smaller than in dM3 stars (suggesting that dynamo action in dM4 stars is less 
effective than in dM3 stars) could explain why the mean $P/\sin i$ is much 
shorter at dM4 than at dM3: dM4 stars are not as good at generating magnetic 
fields, and as a result they do not have access to as good a ``magnetic brake" 
as dM3 stars. The dM4 stars just keep on spinning fast, whereas the dM3 stars 
are braked. This might supplement (or replace) the hypothesis of long loop 
lengths (cf. HM) as the reason for slow rotation at dM3.

Among dM2 stars also, there are no stars rotating as slowly as the slowest dM3 
stars (see Fig.~5). So dM2 stars do not have access to as good a brake 
as dM3 stars. And yet, their dynamo effectiveness seems to be as high as, or 
higher than, the dM3 stars (see Fig.~14 above). In this case, the onset of 
increased loop lengths at dM3 (cf. HM) would help to explain why dM3 stars 
have access to a better magnetic brake than dM2 stars. 

For the fast rotators, we found (in HM) that the trend for the mean $P/\sin i$ 
is to become in general shorter as the spectral type becomes later, except at 
sub-type dM3 where the mean period is found to be significantly larger. The 
minima of the RACs show the opposite trend, i.e. the minimum $P/\sin i$ 
decreases with increasing spectral type. However, only very few active stars 
are at the extremes of the RACs: most active stars lie at longer periods. In 
the case of dM3 and dM4 stars, most active stars lie in the upper part of the 
rising slope of the RAC (see Figs.~8 and 12). We also refer to Figs.~3 and 5 
for dK6 and dM2 stars respectively. Therefore, the trend observed in the 
active stars can be interpreted by the RACs only if one considers their 
locations on these curves. Nevertheless, we note again that the upper end of 
the RAC for dM3 stars (in its rising part) is shifted again towards periods 
which are clearly longer than those for dM2 and dM4 stars. This again explains 
at least partly the abnormally large mean $P/\sin i$ of the active dM3 stars 
(see HM). 

In summary, we find that the properties of the RACs can explain at least 
partly the observed trends in the mean $P/\sin i$ and also the abnormally long 
$P/\sin i$ values reported for dM3 stars by HM. However, this does not exclude 
another explanation, possibly involving a change in loop lengths (HM), which 
might also contribute to the unusual rise in the mean $P/\sin i$ at dM3. 

\section{Comparison with other studies of rotation and activity}

West \& Basri (2009) investigated the rotation and activity in $H_{\alpha}$ in 
a sample of 14 late-type M dwarfs (M6-M7). They found that many of these 
objects are rotating relatively fast ($> 3.5 km s^{-1}$) but have also 
$H_{\alpha}$ in absorption. These rotational velocities imply rather short 
rotation periods for these small objects. This confirms our observed trend 
(see HM) that in general, among late type dwarfs, rotation periods diminish 
with increasing spectral type. West and Basri (2009) also derived an empirical 
relationship between $L_{H_{\alpha}}/L_{bol}$ and $v\sin i$ from a compilation 
of data: they found that $L_{H_{\alpha}}/L_{bol}$ increases typically by two 
orders of magnitude for a change in $v\sin i$ of only $5\ km\ s^{-1}$. This 
implies a steep gradient in the RAC. Although our spectral range does not go 
to spectral subtypes which are as late as those of West and Basri, our results 
are not inconsistent with theirs in the sense that that the steepest slope 
we have found for the RAC occurs among the latest spectral type in our sample 
(dM4, see Table 2), which is closest to the spectral types of the West-Basri 
sample.

Browning et al. (2010) analyzed the rotational broadening and activity in the 
Ca\,{\sc ii} lines for a sample of 123 M dwarfs. Unfortunately, because of 
limitations on their spectral resolution, they could measure $v\sin i$ for 
only 7 stars, namely, those in which the projected rotational speeds were 
$> 2.5\ km\ s^{-1}$. They found that the rotation was detected 
mostly in stars later than M3 rather than in the range M0-M2.5. This is also 
consistent with our findings. They also found, in agreement with our results, 
that there is a ``gap" in the measures of $L_{CaII}/L_{bol}$ between the 
active stellar group and the low activity stellar group (see Fig.~11 above). 
They found a rough relationship between $L_{CaII}/L_{bol}$ and $v\sin i$. 
However, most of their measures lie in the saturated regime, whereas in the 
present work, we have chosen to perform our RAC analyses on stars which lie in 
the unsaturated regime (so that we may examine stars which probably have 
dynamos also in the unsaturated regime). 

Wright et al. (2011) reported on $L_{X}/L_{bol}$ as a function of $R_{0}$ 
for a sample of (824) stars which is significantly larger than we have 
analyzed here: the Wright et al. sample is 3 times larger than the number we 
used for chromospheric RACs, and about 10 times larger than the number of 
stars we used for the coronal RACs (Fig.~16). However, the sample of Wright et 
al. extends over a much broader range of spectral types than we have studied 
here: the Wright et al. sample extends from spectral type F to M5. They found 
a correlation between $L_{X}/L_{bol}$ and $R_{0}$ although there is a 
significant scatter among the data. Their results reveal a saturated regime
and an unsaturated regime, with a break at $log(R_{0})\approx 
-0.1$. The saturation value of $L_{X}/L_{bol}$ was found to be at a log value 
of $\sim -3$. In the non-saturated regime the power law fit between 
$log(L_{X}/L_{bol})$ and $log(P)$ was found to have a slope of -2.18$\pm$0.16. 
If the unsaturated (coronal) stars in their sample are analogous to our low 
activity (chromospheric) stars, then their coronal slope (-2.2) is much 
steeper than what we found for the chromospheric slope (-0.8) (Table 1). The 
increase in steepness of the coronal RAC relative to the chromospheric slope 
(by an amount of about 1.4 in the slope) is reminiscent of our results in 
Section 3.11 above, Eqs. (42)-(46) where the slopes of $L_X$ are steeper than 
the slope of $L_{HK}$ by values which (within errors) also overlap with 1.4 
(except for dM2 stars). 

Rebassa-Mansergas et al. (2013) studied a sample of white dwarf/M dwarf 
binaries from the Sloan Digital Sky Survey (SDSS DR7). They found indications 
that magnetic braking is less efficient beyond the fully convective boundary.
 This again agrees with our results that M4 dwarfs rotate faster than early M 
type dwarfs. They also studied the $L_{H_{\alpha}}/L_{bol}$-$v\sin i$ 
relationship and found mostly that stars with $v\sin i=5\ km\ s^{-1}$ are 
all in the saturated regime. They also found a rapid increase in 
$L_{H_{\alpha}}/L_{bol}$ (by 2 orders of magnitude) as $v\sin i$ increases 
from $2\ km\ s^{-1}$ to $v\sin i=5\ km\ s^{-1}$. This agrees with the similar 
diagram of West \& Basri (2009), and again suggests a steep gradient in the 
RAC between low activity and high activity late-type M dwarfs (as we have 
found for our samples: see Table 1).

Robertson et al. (2013) investigated the magnetic activity level in 
$H_{\alpha}$ for a sample of 93 K5-M5 dwarfs. They found, in agreement with 
our results on the RACs, that early type M dwarfs (M0-M2) tend to have higher 
levels of activity than later type dwarfs, and that in general, 
$log(L_{H_{\alpha}}/L_{bol})$ continuously decreases from -3.6 to -3.95 as the 
spectral sub-type increases from M0 to M5. This again agrees with our global 
finding that the surface fluxes in the Ca\,{\sc ii} lines decrease from K4 to 
M4.

Recently, an important paper has been published by West et al. (2015: W15)
 reporting on an analysis of chromospheric RACs in a sample of 238 M dwarfs, 
using $H_{\alpha}$ emission as a measure of chromospheric ``activity". The 
sample stars had originally been selected as targets for planet searches, but 
W15 used the photometry to search for rotation periods $P$ for the stars 
themselves. Photometric periods are not subject to the $\sin i$ uncertainty 
which affects the $P/\sin i$ we have obtained in the present study. In 
constructing their RAC, W15 identified 164 stars for which values of $P$ and 
$H_{\alpha}$ data were available. The $P$ values ranged from 0.3 days to 100 
days. The advantage of using photometry (over spectroscopy) to obtain $P$ 
values is clear: there are no observational limits imposed by attempts to 
extract $v\sin i$ values from spectroscopic data. As a result, photometry 
allows W15 to determine rotational periods for slow rotators which are beyond 
the capabilities which we have used here. As a result,  the 
W15 upper limit on $P$ is several times longer than we have been able to 
report on in the present paper. In order to study variations of RAC with 
spectral type, W15 divided their stars into two groups: M1-M4 (64 stars, 
mainly M3 and M4), and M5-M8 (100 stars, mainly M5 and M6). Their plots of RAC 
(in their Figs.~7 and 8) show that the M1-M4 stars have a negative slope: a 
LSF yields a slope of -0.19$\pm$0.036. For the M5-M8 stars, the LSF gives a 
formal slope of essentially zero (-0.016$\pm$0.050). These results suggest 
that since a finite (though small in absolute value) negative slope exists for 
the RAC in M1-M4 stars, rotation might well play a role in the chromospheric 
emission in M1-M4 stars. On the other hand, since the RAC slope was found to 
be essentially zero for M5-M8 stars, rotation may play little or no role in 
determining chromospheric emission in M5-M8 stars. At first sight, this result 
might be considered as evidence for the suggestion of Durney et al (1993) that 
there might be a change in dynamo mode between M1-M4 and 
M5-M8. But we would like to suggest another possible interpretation, as 
follows.

In view of our own results for the RAC's in M4, M3, and M2 stars, we find it a 
matter of interest that the slope of the chromospheric RAC determined by W15 
for M1-M4 stars (-0.19$\pm$0.036) is much shallower than the values we have 
found. For our combined samples of dM+dMe stars, we have found slopes of -1.5 
to -2.5 (see Table~2). And even when we confine attention to the slow rotators,
 we still find slopes for M2, M3, and M4 dwarfs (-0.89 to -0.93) which are 
clearly significantly steeper than W15 report for their M1-M4 sample. Our 
results  indicate that chromospheric emission in M2-M4 stars in our samples 
exhibit much stronger sensitivity to rotation than W15 have reported for 
either of their samples. Is it possible that this dichotomy arises from 
selection effects? We note that the data of W15 include mainly ``active" stars 
with $H_{\alpha}$ {\it in emission} i.e. fast rotators, whereas our samples 
(especially the slow rotator samples) are biased towards slow rotators 
($H_{\alpha}$ {\it not} in emission). Now, fast rotators are more likely to be 
in the saturated regime of the dynamo (see Section 1.1 above): in that regime, 
the RAC flattens out, and takes on a slope which is close to zero. In fact, in 
a discussion of the systematic differences between their M1-M4 sample and 
their M5-M8 sample, W15 point out that ``At similar rotation periods, a much 
larger fraction of the late-type M dwarfs [M5-M8] are active" than is the case 
for the M1-M4 stars. Visual inspection of Fig.~6 in W15 suggests that some 
65\% of their M1-M4 stars are labelled ``active" while almost 90\% of their 
M5-M8 stars are ``active". This is consistent with the W15 statement that a 
``much larger fraction" of the M5-M8 stars are in the saturated regime. This 
leads us to wonder if the presence of saturated behavior is contributing to 
the (essentially) zero slope reported by W15 for M5-M8 stars. And even among 
the M1-M4 stars of W15, where the fraction of active stars is admittedly 
smaller than in the M5-M8 sample, that fraction is by no means small: some 
65\% of the M1-M4 stars are ``active", and therefore may, lie in the saturated 
regime. Suppose that 65\% of the M1-M4 stars in the W15 sample lie in the 
saturated regime (with a slope of zero), while the remaining 35\% lie in the 
unsaturated regime, where the slope is finite (and negative), with a value of 
-a. Then the complete W15 sample of M1-M4 stars would have an RAC with an 
average slope a(W15) = (0.65$\times$0) + (0.35$\times$(-a)). For slow rotators 
with spectral sub-types extending from dK4 to dM4, we have found that -a can 
take on values in the range from -0.6 to -0.9. This leads us to predict that 
a(W15) could range from -0.21 to -0.32. In fact, the empirical slope reported 
by W15 (-0.19$\pm$0.036) contains, within a 3$\sigma$ range, values of a(W15) 
extending from -0.08 to -0.30. This overlaps extensively with our predicted 
range of slopes. 

Thus, it is possible that: (i) the existence of a small but finite negative 
slope obtained by W15 for M1-M4 stars is due to the presence of a minority 
(35\%) of stars in the unsaturated regime, and, (ii) the change in slope in 
going from M1-M4 to M5-M8 is due to the larger number of stars in the 
saturated regime among the M5-M8 sample. In order to test these possibilities, 
and in order to get a more definitive test of a change in dynamo mode at the 
TTCC, we suggest that it would be best to concentrate on stars in the 
unsaturated regime. So far, our own studies have not yet reached into M5-M8 
stars: it will be a matter of great interest to determine if the W15 findings 
of significant change in RAC slope between M4 and M5 can be replicated using 
unsaturated stars.
  

\section{Conclusion}

In this section, we first (Section 5.1) summarize the approach we have adopted 
in order to study dynamos in low-mass stars. Then we go on (Sections 5.2-5.5) 
to describe how we have characterized quantitatively the observational data in 
terms of chromospheric observations. A discussion of coronal data follows 
(Section 5.6). In Section 5.7 we present the major conclusions of our study in 
terms of three Hypotheses about dynamo action in low-mass dwarf stars. Lastly, 
we discuss evidence which has a bearing on crossing the TTCC (Section 5.8).
 
\subsection{Dynamos: saturated and unsaturated}

In order to account for the presence of active chromospheres and coronae in 
low-mass stars, dynamos are believed to be operative. Dynamos may rely on 
rotation plus turbulence, or on turbulence alone, to generate magnetic fields 
which then, as a result of ``magnetic activity" (of some kind) lead to heating 
of the chromospheric and coronal gas. It is possible that different types of 
dynamos are at work in stars of different masses. In this paper, we bring 
together observational evidence to see if it is possible to identify signatures
 of different types of dynamos. Our study is based on the assumption that 
different types of dynamos are expected to lead to different answers to the 
following question: how does the amount of ``activity" behave as a function of 
``rotation"? In order to address this, we need to compile data which provide 
us with quantitative measures of ``activity" and ``rotation".

An important aspect of the present paper is that we measure rotational speeds 
in stars which are rotating more slowly than has been reported in previous 
studies. Why are we interested in slow rotators? The reason has to do with the 
physics of dynamos: theoretical work suggests that rotational dynamo operation 
can saturate when the rotational speed reaches a certain limit. It seems to us 
that it would be difficult to extract information from a saturated dynamo, 
where the activity level no longer depends on how fast the star is rotating. 
For this reason, we prefer in this paper to deal with dynamos in an unsaturated
 condition, where an increase in rotation leads to a clear and measurable 
increase in activity. This is what has driven us in the present paper to 
undertake a concerted effort towards identifying stars with the slowest 
possible rotations: to do this, we rely on spectroscopic data which were 
obtained with the highest possible resolution.  

\subsection{Data being used in this paper}

Data have been compiled on chromospheric emission and (projected) rotational 
periods for a sample of 418 stars (Paper~I) ranging in spectral sub-type from 
dK4 to dM4 (42 dK4, 118 dK6, 94 dM2, 81 dM3, 83 dM4). 

The analysis which we apply to our data has the goal to derive a 
rotation-activity correlation (RAC) for stars in each spectral sub-type, and 
then see if there are any systematic variations in the RACs as the spectral 
sub-type approaches the limit where main sequence stars make a Transition To 
Complete Convection (TTCC). The location of the TTCC is a matter of some 
dispute, possibly as early as dM2, possibly as late as dM4. Our choice of 
spectral sub-types is meant to overlap with TTCC.

\subsection{Constructing RACs at various spectral sub-types: chromospheric 
data}

As a quantitative measure of ``activity", we use the mean surface flux 
$F_{CaII}$ of emission in the Ca\,{\sc ii} H and K lines. As a quantitative 
measure of ``rotation", we combine the projected rotational speeds $v\sin i$ 
with stellar radii to obtain a ``projected rotation period" $P/\sin i$. We 
construct an RAC for stars in each spectral sub-type by plotting (in log-log 
format) the surface flux versus $P/\sin i$. The RACs which we have obtained for
 stars in each of our 5 spectral sub-types can be found in Figs. 2, 3, 5, 8, 
and 12.

\subsection{Chromospheric data: the slopes of the RACs}

A general feature of many RACs is that as rotational periods become shorter, 
the Ca\,{\sc ii} surface flux becomes larger, up to a point. Beyond that point,
 shorter periods do not lead to any increase in Ca\,{\sc ii} flux: at these 
shortest periods, the RAC is probably ``saturated". A key aspect of the 
present paper is that we discuss only the unsaturated part of the RAC. In this 
case, we obtain least squares fits of the RAC to a function of the form 
$F_{CaII}= b (P/\sin i)^a$. In a log-log plot, $a$ is the slope of the RAC, 
and is a negative number. The coefficient $b$ is a measure of the amplitude of 
chromospheric heating.

Numerical values of the slope $a$ which we have obtained from least squares 
fitting to our data for the various spectral sub-types are shown in Fig.~13. 
The different curves show the results we have obtained when 
we group our target stars in three different ways. (i) The curve which rises 
monotonically from lower left to upper right refers to the combined sample of 
all ``unsaturated" stars, both those with low activity (dK, dM) and those with 
high activity (dKe, dMe). (ii) The curve which runs (almost) horizontally near 
the lower boundary of the figure refers to stars which are confined to the low 
activity sub-samples (dK, dM). (iii) The ``jagged" curve extending from middle 
left to lower right refers to stars which are confined to the high activity 
sub-samples (dKe, dMe). The results in Fig. 13 will guide our discussion of 
dynamos in the final sub-section below.

\subsection{Chromospheric data: the amplitude of chromospheric heating}

As a second step in obtaining RACs, we consider not the slopes, but the 
overall level of chromospheric heating in terms of the ratio of $L_{HK}$ to 
$L_{bol}$. Based on this ratio, we have obtained RACs based on a more 
physically relevant parameter (the Rossby number $R_o$). These RACs (see 
Fig.~15) show that, at the longest rotation periods ($log(R_o)$ = -1.2) in our 
data set, chromospheric heating in dM3 stars is {\it less effective by a 
factor of about 20} than in dK4 stars. Our results suggest that the {\it 
chromospheric} heating efficiency in dM4 stars is {\it less effective} by a 
factor of about 100 than in F-G-K type stars. We also observe a progressive 
decline in the location of the RACs as we go from dK4 to dM4: dK6 stars lie 
slightly below (a factor of 3) the RAC of F, G and K type stars, while dM2 
stars lie a factor of 10 below the RAC of F, G, and K type stars. The data 
point to a conclusion which we consider reliable: the efficiency of 
chromospheric heating decreases progressively between dK4 and dM4. The 
amplitude of the overall decrease is 20-90. This also implies that the overall 
efficiency of the dynamo mechanisms also decreases by the same factors when 
moving from K4 to M4 dwarfs.

\subsection{coronal data}

Turning now to the RAC associated with {\it coronal} emission, we find a very 
different behavior from the chromospheric data (Fig.~16). The coronal emission 
$L_{X}/L_{bol}$ {\it does not decrease} significantly as we go from dK4 to dM4.
The lack of decrease is especially marked when we compare the coronal emission
to the chromospheric emission (see Fig.~18): for a given value of $L_{HK}$ 
(say $10^{27.5}$ ergs/s), the value of $L_X$ in dM4 stars is larger by a 
factor of order 100 compared to $L_X$ in dK4/dK6 stars. Thus, while the {\it 
chromospheric} heating efficiency is {\it decreasing} as we go from dK4 to dM4 
(by a factor of up to 100), the {\it coronal} heating efficiency is 
simultaneously {\it increasing} (by a factor of 100 or so).

In terms of a dynamo interpretation, this raises the question: which part of a 
stellar atmosphere should we study in order to obtain more reliable 
information about dynamo efficiency in M dwarfs, the chromosphere or the 
corona? The behaviors are so different that it is not clear that the same 
information about the dynamo will emerge from both data sets. We have argued 
(Section 3.11) that it may be preferable to concentrate on 
the chromosphere. 

In view of this, we now present some conclusions which, in our opinion, help to
 bring order to the chromospheric data which have been analyzed in the present 
paper.

\subsection{Hypotheses about two distinct dynamos in low-mass stars}

Inspection of Fig. 13 leads us to offer the following hypotheses.

(i) In the case of low-activity stars, we see that the RAC slope $a$ is 
essentially unchanged as we go from dK4 to dM4. To the extent that a particular
 RAC slope is associated with a particular dynamo mechanism, our results 
indicate that low-activity stars have essentially the {\it same} dynamo at work
 in stars which range in spectral sub-type from dK4 to dM4. Such a dynamo must 
have something to do with a physical property which is present in {\it all} low
 mass stars from dK4 to dM4, i.e. on both sides of the TTCC. What property is 
common to all such stars? The answer is: all of them have a deep convective 
envelope in which turbulence provides a ready supply of energy to drive a 
distributed dynamo (DD) i.e. a dynamo which could be described by either an 
$\alpha ^2$ model or an $\alpha^2$ $\Omega$ model. This leads us to {\bf 
Hypothesis (A)}: low-activity stars from dK4 to dM4 may be dominated by a 
turbulent (DD) dynamo. 

(ii) In the case of the RAC for high-activity stars, the aspect of Fig. 13 
 that is most likely to catch the eye is probably the ``jagged" 
behavior at early spectral types. But we would like to draw attention at first 
to a different aspect of the RACs: namely, the values of the slopes $a$ at the 
{\it latest} spectral types (dM3e and dM4e). In these cases, our results 
indicate that the RAC slopes for high-activity stars are essentially the {\it 
same} as for the low-activity stars. Once again relying on a putative 
association between an RAC slope and a dynamo mechanism, these results suggest 
that at spectral types dM3 and dM4, the dynamo mechanism in low- {\it and} 
high-activity stars are actually the {\it same}. It has already been suggested 
(see (i)) that a DD is at work in all low-activity stars. This leads us to 
{\bf Hypothesis (B)}: in the high activity stars at the latest sub-types in our
 samples (dM3e and dM4e), a DD is at work. This is not a surprising conclusion:
 dM3e and dM4e stars are completely convective (CC), so they also have access 
to DD operation in the form of either an $\alpha ^2$ dynamo model or an 
$\alpha^2$-$\Omega$ dynamo model.

(iii) Moving now to high-activity stars at earlier spectral types (dK4e, dK6e, 
and dM2e), we see in Fig. 13 that the RAC slopes {\it are very 
different from the slopes of the low-activity stars}. This suggests that 
high-activity stars in the range dK4e-dM2e have access to a {\it different kind
 of dynamo} from the type (DD) that may dominate in low-activity stars. What 
might give rise to a non-DD type of dynamo in stars with spectral types in the 
range dK4-dM2? The answer is surely related to the fact that such stars have 
interfaces between the outer convective envelope and an inner radiative core. 
This leads to {\bf Hypothesis (C)}: in high-activity stars at the earliest 
spectral types in our samples, an interface dynamo (ID) is at work. It seems 
probable that such a dynamo could be described by an $\alpha$-$\Omega$ dynamo 
model. To be sure, dK4e-dM2e stars also have deep convective envelopes: 
therefore a DD is probably also at work. But the clear difference in RAC slopes
 between low-activity stars in the range dK4-dM2 and high-activity stars in the
 range dK4e-dM2e indicates that the DD (with its shallow RAC slope) is not 
playing a dominant role in dK4e-dM2e stars. 

(iv) If our hypotheses have any validity, we can conclude that the primary 
quantitative difference between ID and DD is this: the RAC slopes are steeper 
(by up to 1.5 units) among the ID stars than the DD stars. Thus, the ID stars 
are much more sensitive to rotation than the DD stars. Specifically, if we 
compare two stars which differ in rotation by a factor of (say) 10, two stars 
where ID operates will differ in activity level by a factor which is 30 times 
larger than the difference in activity level of two stars where DD operates. 

(v) The conclusion in item (iv) is reminiscent of a suggestion that was made by
 Durney et al (1993): the RAC in a star with ID should depend sensitively on 
period, but the RAC in a star with DD should not. The results in Fig. 13 are 
at least partially consistent with Durney et al.: ID stars (as we 
identify them) are definitely more sensitive to rotation than DD stars (as we 
identify them). Admittedly, we have not found that the DD has {\it zero} slope 
for its RAC: but we have found that the slope is at least smaller than in the 
ID stars. 

The fact that the DD stars have an RAC with a non-zero slope suggests that 
rotation $\Omega$ does have $\it some$ effect on the activity level in these 
stars. In terms of the two options which we have proposed in Hypotheses A and B
 above, it might be preferable to conclude that the dominant dynamo model in DD
 stars may be the $\alpha ^2$-$\Omega$ model.

Finally, we make a point about the overall methodology which has been 
ultimately responsible for this paper. The opening sentence in Item (i) above, 
which provides a start to our 3 hypotheses, would not have been possible if we 
did not have access to a large sample of {\it low-activity} stars. Such stars 
are slow rotators. Therefore, if the lead author of this study (ERH) had not 
paid attention to extracting the lowest possible values of $v/\sin i$ (of order
 1 $km ~sec^{-1}$) in as many stars as possible, our sample of low-activity 
stars might have been so small as to prevent us from drawing statistically 
significant conclusions. 

\subsection{Any evidence for crossing the TTCC?}

The theoretical concept that main sequence stars undergo a transition to 
complete convection (TTCC) at a particular mass (in the vicinity of spectral 
types M2-M4) has been in the literature for 50 years or more. Nevertheless, the
 search for an empirical signature which might support this concept has yielded
 no definitive evidence. For example, Mullan and MacDonald (2001) sought such 
evidence in X-ray data, but were unable to identify any signature of the TTCC. 
Also in our own coronal data (Fig.~16), we see no definitive sign of a 
transition. Neither could we find evidence for the TTCC in the RACs of 
chromospheric data when we combined low- and high-activity stars (Fig.~13, 
monotonically rising line from lower left to upper right). Nor could we
 find any evidence of TTCC when we confined our attention to the chromospheric 
RACs of low-activity stars (lowest lying line in Fig.~13.

However, in one particular sample, our study has led to what we believe is a 
potentially valuable signature. Specifically, in our sample of high-activity 
stars between dK4e and dM4e, the slopes of the chromospheric RAC (Fig.~13, 
``jagged" line) consists of 2 distinct regimes. In one regime, the 
slopes overlap with the (shallow) slopes of low-activity stars. In the other 
regime, the slopes are found to be much steeper. We interpret the transition 
between the two regimes as evidence for a transition between dynamo modes. The 
transition occurs between dM2e and dM3e. We suggest that this cross-over from 
steep to shallow RAC slopes may provide empirical evidence that the TTCC has 
been crossed. 

We have shown how empirical information about rotation and chromospheric 
emission may contribute to understanding some aspects of dynamo mechanisms in 
cool dwarfs. Data for M dwarfs with spectral types later than M4, and also for 
L and T dwarfs would be of great interest to complete our view on the dynamo 
mechanisms in low-mass stars. For example, is there any evidence that the 
efficiency of coronal heating actually decreases after we pass though the 
electrodynamic resonance which is expected to occur around dM3-dM4? It would 
also be of interest to investigate the variations of rotational and 
chromospheric properties by using finer grained samples of stars near the 
TTCC, i.e. using samples of stars all of which are confined to spectral types 
of dM2.5, dM3.5, and dM4.5 stars. This could contribute to better 
understanding of the behavior of possible changes in the dynamo mechanism(s) 
at the TTCC.

\section*{acknowledgements}
This research has made use of the SIMBAD database, operated at CDS, Strasbourg,
France. DJM is supported in part by the NASA Space Grant program. This study 
was based on data obtained from the ESO Science Archive Facility and the 
Observatoire de Haute Provence SOPHIE database. This research made use of 
Astropy\footnote{http://www.astropy.org/ and 
http://astroquery.readthedocs.org/en/latest/}, a community-developed core 
Python package for Astronomy (Astropy Collaboration, 2013). We also used the 
tutorial developed by Paletou \& Zolotukhin (2014)\footnote{http://www.astropy.org/}. This research was achieved using the POLLUX database 
(http://pollux.graal.univ-montp2.fr) operated at LUPM (Universit\'e 
Montpellier- CNRS, France) with the support of the PNPS and INSU. This 
research has made use of the VizieR catalogue access tool, CDS, Strasbourg, 
France. The original description of the VizieR service was published in A\&AS 
143, 23.

\begin{appendix}

\begin{center}
{\bf APPENDIX: Quadratic heteroscedastic regression models}
\end{center}

\renewcommand{\thesection}{\Alph{section}} 
\renewcommand{\theequation}
{\thesection.\arabic{equation}} \setcounter{section}{1}  
\setcounter{equation}{0}


This appendix is devoted to classical mathematical results on stochastic 
linear regression models. Our purpose is to show how to better take into 
account the measurement errors of $P/\sin i$ in order to explain the 
variability of Ca\,{\sc ii} surfaces fluxes. In the previous literature, only 
linear or quadratic homoscedastic LSF were investigated in order to explain 
this variability. Our strategy is slightly different as we propose to make use 
of linear or quadratic heteroscedastic LSF in order to better explain this 
variability. 

More precisely, consider one of our 5 groups of dK4, dK6, dM2, dM3, or dM4 
stars, for example dM2. Let $n>3$ be the total number of the dM2 stars 
under study. For each dM2 star, denote by $x_k$ the $\log(P/\sin i)$ 
associated measure and by $\sigma_k>0$ the measurement error of $\log(P/
\sin i)$. Moreover, let $Y_k$ be the $\log(\text{CaII} \ \text{Flux})$ measure 
of this dM2 star. Then, we shall deal with the linear or quadratic 
heteroscedastic regression models, respectively given, for $k=1,\ldots,n$, by

\begin{equation}
\label{Linreg}
Y_k =  a + bx_k + \sigma_k \veps_k, 
\end{equation}

and, 

\begin{equation}
Y_k =  a + bx_k + cx_k^2 + \sigma_k \veps_k
\label{Quadreg}
\end{equation}

\noindent
where $a,b$ and $c$ are unknown parameters and the random noise $(\veps_k)$ is 
a standard Gaussian white noise with mean zero and unknown variance $\tau^2$. 
The important point is the crucial role played by the error term $\sigma_k$.
The variance of the additive noise clearly depends on the explanatory variable 
$x_k$ as it is given by $\sigma_k^2 \tau^2$ where $\sigma_k$ is the 
measurement error associated with $x_k$. We shall only focus our attention on 
the quadratic heteroscedastic regression model given by (A.2) inasmuch as 
the linear regression model (A.1) is a particular case of (A.2).

The quadratic heteroscedastic regression model (A.2) can be rewritten into 
the matrix form

\begin{equation}
\label{Matrixreg} 
Y=X\theta +\Gamma^{1/2}\veps 
\end{equation} 

\noindent
where the vector of observations $Y$, the vector of unknown parameters 
$\theta$, and thevector containing the random noises $\veps$, are respectively 
given by 

\begin{equation*} 
  Y= 
  \left(\begin{matrix}
    Y_1 \\ 
    \vdots\\ 
    Y_n  
  \end{matrix}\right), \hspace{1cm} 
  \theta= 
  \left(\begin{matrix}
    a \\ 
    b \\ 
    c  
  \end{matrix}\right), \hspace{1cm} 
  \veps= 
  \left(\begin{matrix} 
    \veps_1\\  
    \vdots\\ 
    \veps_n  
   \end{matrix}\right). 
\end{equation*} 

In addition, the heteroscedastic matrix $\Gamma=\text{diag}( \sigma_1^2, 
\sigma_2^2, \cdots, \sigma_n^2)$ and the design matrix $X$ is given by

\begin{equation*} 
  X= 
  \left( 
  \begin{matrix} 
    1  & x_{1} &  x_{1}^2\\ 
    1  & x_{2} &  x_{2}^2\\ 
    \vdots  & \vdots &  \vdots \\ 
    1 &  x_{n}  & x_{n}^2 
   \end{matrix} 
   \right).
\end{equation*}

On the one hand, the least squares estimator (LSE) of $\theta$ is the value 
$\wt{\theta}$ which minimizes the strictly convex function $\wt{\Delta}(\theta)
=||Y-X\theta||^2$. Straightforward calculation leads to

\begin{equation*}
\wt{\theta}=(X^t X)^{-1}X^tY.
\end{equation*} 

However, this estimator does not taken into account the heteroscedasticity of 
model (A.3). On the other hand, the weighted least squares estimator (WLSE)
of $\theta$ is the value $\wh{\theta}$ which minimizes the strictly convex 
function 

$$\wh{\Delta}(\theta)=(Y-X\theta)^t\Gamma^{-1}(Y-X\theta).$$ 

It is not hard to see that

\begin{equation}
\label{WLSE}
\wh{\theta}=(X^t \Gamma^{-1}X)^{-1}X^t\Gamma^{-1}Y.
\end{equation} 

We immediately deduce from (A.3) and (A.4) that

$$
\wh{\theta}  = \theta + (X^t \Gamma^{-1}X)^{-1} X^t\Gamma^{-1/2} \veps.
$$

Consequently, as $\veps$ is a $n$-dimensional Gaussian vector $\cN(0, \tau^2 
I)$ where $I$ stands for the identity matrix of order $n$, we obtain that 
$\wh{\theta}$ is a $3$-dimensional Gaussian vector,

\begin{equation}
\label{DISTWLSE}
\wh{\theta} = \left(\begin{matrix}
    \wh{a} \\ 
    \wh{b} \\ 
    \wh{c}  
  \end{matrix}\right) \sim \cN(\theta, \tau^2 (X^t \Gamma^{-1}X)^{-1}).
\end{equation}

Hereafter, the linear least squares fit (LSF) is the line

\begin{equation}
\label{LSF}
y=\wh{a} + \wh{b}x,
\end{equation}

\noindent
while the quadratic LSF is the curve

\begin{equation}
\label{QLSF}
y=\wh{a} + \wh{b}x + \wh{c}x^2.
\end{equation}

Furthermore, denote by $H$ the hat matrix

$$
H=X(X^t \Gamma^{-1}X)^{-1}X^t\Gamma^{-1}
\hspace{1cm}\text{and}\hspace{1cm}
L=I-H.
$$

It follows from (A.3) and (A.4) that $\wh{Y}=HY=X\wh{\theta}$ which implies 
that $HY$ has an $\cN(X\theta, \tau^2 H\Gamma)$ distribution and $LY=Y-\wh{Y}$ 
has an $\cN(0, \tau^2 L\Gamma)$ distribution. Hence, as $\text{rank}(L)=n-3$, 
we obtain that the sum of squared errors (SSE) has a chi-squared distribution,

\begin{equation}
\label{CHIS}
||\Gamma^{-1/2}(Y - \wh{Y})||^2 \sim \tau^2 \chi^2(n-3).
\end{equation}

The SSE is a way to evaluate the discrepancy between the data $Y$ and its 
estimate $\wh{Y}$ and a small value of the SSE indicates a tight fit of our 
model to the data. Let us recall the celebrated decomposition of total sum of 
squares which is in fact a direct application of Pythagoras's theorem 

\begin{equation*}
|| \Gamma^{-1/2} Y ||^2 = || \Gamma^{-1/2}\wh{Y} ||^2 + || \Gamma^{-1/2}(Y-\wh{Y})||^2.
\end{equation*}

Finally, a natural estimator of the variance $\tau^2$ is

\begin{equation}
\label{ESTVAR}
\wh{\tau}^{\,2}=\frac{||\Gamma^{-1/2}(Y - \wh{Y})||^2}{n-3}.
\end{equation}

It is not hard to see that the random vector $\wh{\theta}$ and 
$\wh{\tau}^{\,2}$ are independent. Consequently, we deduce from (A.5) that for 
any real number $x$,

$$
 (\wh{a} - a) + (\wh{b} - b)x + (\wh{c} - c)x^2 \sim \cN(0, \tau^2 \xi(x))
$$

\noindent
where 

$$
\xi(x)= \left(
\right).
$$

Dividing on both sides by $\wh{\tau}$, the ratio has a Student distribution,

\begin{equation}
\label{STUDENT}
\frac{(\wh{a} - a) + (\wh{b} - b)x + (\wh{c} - c)x^2}{\wh{\tau} \sqrt{\xi(x)}} \sim t(n-3).
\end{equation}

\end{appendix}

\onecolumn

\small
\begin{center}
\begin{longtable}{|lccccccc|}
\caption[ ]{In this table, we give the parameters of our homoscedastic and 
heteroscedastic regression models.} \\

\hline     &                   &                         &                   &             &         &             &             \\
             \multicolumn{8}{c}{Homoscedastic Least Square Fits}                                                                \\
           &                   &                         &                   &             &         &             &             \\

\hline     &                   &                         &                   &             &         &             &             \\
           & \multicolumn{7}{c}{$F_{CaII} = b\times (P/\sin i)^{a}$}                                                             \\
Spect. Type&         a         &            b            &                   & Corr. Coef. &$\chi^{2}$&Nb of Stars & Stat. Sign. \\
           &                   &                         &                   &             &         &             &             \\
\hline 
\end{longtable}
\end{center}
\normalsize

\tiny
\begin{landscape}
\begin{longtable}{|lcccccccccccccc|}
  \caption[ ]{Equivalent width (EW, in units of \AA ) of the Ca\,{\sc ii} 
resonance doublet and H$_{\alpha}$ line for our sample of dK6 stars. Results 
were obtained using spectra from HARPS, SOPHIE and FEROS. For each star, we 
give the $S/N$ ratio for the sum of all available spectra, as well as the 
number of spectra $N_b (meas.)$ which we used to compute the equivalent 
widths.} \\

\hline  &       &              &      HARPS   &               &              &   SOPHIE     &               &           &          &       &       &      &       &        \\
        &       &   CaII H     &   CaII K     &  H$_{\alpha}$ &   CaII H     &   CaII K     &  H$_{\alpha}$ & K/H ratio &$<$CaII$>$&Ca\,{\sc ii}&Ca\,{\sc ii}&  Nb. &  S/N  &  S/N   \\
Star    & No. of&     EW       &     EW       &     EW        &     EW       &     EW       &     EW        &           &    EW    &  Flux & Flux\footnotemark[1]&meas.& HARPS & SOPHIE \\
        & meas. &   (\AA)      &    (\AA)     &   (\AA)       &   (\AA)      &    (\AA)     &    (\AA)      &           &  (\AA)   &$(10^{5}\ ergs/s/cm^{2})$ &$(10^{5}\ ergs/s/cm^{2}$)& &@5000&@5000\\

\endfirsthead
\caption{continued.}\\

\hline  &       &              &      HARPS   &               &              &   SOPHIE     &               &           &          &       &       &      &       &        \\
        &       &   CaII H     &   CaII K     &  H$_{\alpha}$ &   CaII H     &   CaII K     &  H$_{\alpha}$ & K/H ratio &$<$CaII$>$&Ca\,{\sc ii}&Ca\,{\sc ii}&  Nb. &  S/N  &  S/N   \\
Star    & No. of&     EW       &     EW       &     EW        &     EW       &     EW       &     EW        &           &    EW    &  Flux & Flux\footnotemark[1]&meas.& HARPS & SOPHIE   \\
        & meas. &   (\AA)      &    (\AA)     &   (\AA)       &   (\AA)      &    (\AA)     &    (\AA)      &           &  (\AA)   &$(10^{5}\ ergs/s/cm^{2})$ &$(10^{5}\ ergs/s/cm^{2})$& &@5000&@5000 \\
\endhead 
\hline
\endfoot

        &       &              &              &               &              &              &               &           &          &       &        &      &       &         \\
GJ 1056 &   9   &-0.61$\pm$0.01&-0.68$\pm$0.01&0.643$\pm$0.020&      -       &      -       &      -        &   1.11    &  -0.645  & 2.052 &        &  9   & 151   &    -   \\
GJ 1066 &   3   &-0.91$\pm$0.10&-1.01$\pm$0.10&0.605$\pm$0.020&      -       &      -       &      -        &   1.11    &  -0.96   & 3.295 &        &  3   &  64   &    -   \\
GJ 1067 &   3   &      -       &       -      &       -       &-0.79$\pm$0.10&-1.11$\pm$0.10&0.554$\pm$0.020&   1.41    &  -0.95   & 3.171 &        &  3   &   -   &  119   \\
Gl 1177A&   -   &      -       &      -       &      -        &      -       &      -       &      -        &    -      &  -1.919  & 4.525 &        &  -   &   -   &     -  \\
GJ 1248 &   3   &      -       &       -      &0.239$\pm$0.020&      -       &      -       &      -        &    -      &     -    &   -   &        &  3   &  29   &    -   \\
GJ 1267 &   1   &-1.01$\pm$0.15&-1.08$\pm$0.15&0.518$\pm$0.020&      -       &      -       &      -        &   1.07    &  -1.045  & 1.755 & 2.944  &  1   &  38   &    -   \\
GJ 1279 &   2   &-0.79$\pm$0.01&-0.87$\pm$0.01&0.638$\pm$0.020&      -       &      -       &      -        &   1.10    &  -0.83   & 2.849 &        &  2   & 177   &    -   \\
\hline 
\end{longtable}
\footnotetext[1]{Corrected from metallicity effects.}
\footnotetext[2]{FEROS spectra.}
\end{landscape}
\normalsize

\small
\begin{center}
\begin{longtable}{|lcccccc|}
  \caption[ ]{We give the  values of $v\sin i$, $P/\sin i$ as well as 
[M/H] from Paper~I for our dM2 stars. We also give the Ca\,{\sc ii} 
EW (from the compilation of Paper~XVIII), surface fluxes $F_{HK}$ and 
the surface fluxes corrected for the metallicity effects.} \\

\hline Star&$v\sin i$& $P/\sin i$ & [M/H]   &Ca\,{\sc ii} EW&$F_{HK}$&$F_{HK}$\footnotemark[1] \\
        & $km/s$  &    (days)     & (dex)   & (\AA)  &$10^{4}erg/s/cm^{2}$&$10^{4}erg/s/cm^{2}$\\
        &         &               &         &        &        &      \\
\endfirsthead
\caption{continued.}\\

\hline Star&$v\sin i$& $P/\sin i$ & [M/H]   &Ca\,{\sc ii} EW&$F_{HK}$&$F_{HK}$\footnotemark[1] \\
        & $km/s$  &    (days)     & (dex)   & (\AA)  &$10^{4}erg/s/cm^{2}$&$10^{4}erg/s/cm^{2}$\\
        &         &               &         &        &        &      \\
\endhead 
\hline
\endfoot

GJ 1010A&    -    &      -        & -0.098  & -0.605 &  2.621 &10.555 \\
GJ 1062 &   9.8   & 1.92-0.53+0.65& -0.584  &    -   &    -   &   -   \\
GJ 1114 &   3.8   & 5.73-1.50+2.07& -0.339  & -0.117 &  0.507 & 0.664 \\
GJ 1264 &   6.46  & 6.82-1.51+2.06& +0.378  & -4.45  & 20.85  & 8.732 \\
GJ 2085 &   2.46  &10.80-3.23+5.31& +0.000  & -0.77  &  4.071 & 4.071 \\
GJ 3084 &   3.16  & 9.74-4.07+5.98& +0.023  & -0.841 &  4.373 & 4.147 \\
GJ 3098 &   2.60  & 8.97-2.49+3.98& -0.200  & -0.86  &  4.215 & 6.680 \\
GJ 3207 &    -    &      -        & -0.346  & -0.19  &  0.801 & 1.777 \\
\hline 
\end{longtable}
\footnotetext[1]{Surface Flux corrected from metallicity effects}
\end{center}

\tiny
\begin{table*}
  \caption[ ]{Equivalent width (EW, in units of \AA ) of the Ca\,{\sc ii} 
resonance doublet and H$_{\alpha}$ line for our sample of dM3 stars 
(see HM). Results were obtained using spectra from HARPS, SOPHIE and 
FEROS.} 

\tiny
\begin{tabular}{|lccccccccc|}

\hline
        &       &               &       HARPS   &               &              &   SOPHIE     &               &           &            \\
        &       &    CaII H     &    CaII K     &  H$_{\alpha}$ &   CaII H     &   CaII K     &  H$_{\alpha}$ & K/H ratio & $<$CaII$>$ \\
Star    & No. of&      EW       &      EW       &     EW        &     EW       &     EW       &     EW        &           &            \\
        & meas. &     (A)       &      (A)      &    (A)        &    (A)       &     (A)      &    (A)        &           &            \\
        &       &               &               &               &              &              &               &           &            \\
GJ 1046 &   6   &-0.370$\pm$0.01&-0.470$\pm$0.01&0.329$\pm$0.020&      -       &      -       &      -        &   1.27    &  -0.420    \\
GJ 1050 &   9   &-0.101$\pm$0.01&-0.125$\pm$0.01&0.261$\pm$0.020&      -       &      -       &      -        &   1.24    &  -0.113    \\
GJ 1097 &   7   &-0.431$\pm$0.01&-0.506$\pm$0.01&0.306$\pm$0.020&      -       &      -       &      -        &   1.17    &  -0.469    \\
\hline 
\end{tabular}
\footnotetext[1]{FEROS spectrum}
\end{table*}
\normalsize

\small
\begin{center}
\begin{longtable}{|lcccccc|} 
  \caption[ ]{The values of $v\sin i$, $P/\sin i$ and [M/H] 
from Paper I for our targets in our list of dM3 stars. We also 
give the Ca\,{\sc ii} EWs, surface fluxes $F_{HK}$ and the 
surface fluxes corrected for the metallicity effects.} \\

\hline Star&$v\sin i$& $P/\sin i$ & [M/H] &Ca\,{\sc ii} EW&$F_{HK}$&$F_{HK}$\footnotemark[1] \\
        &($km/s$) &    (days)     & (dex) & (\AA)  &$10^{4}erg\ s^{-1}\ cm^{-2}$&$10^{4}erg\ s^{-1}\ cm^{-2}$\\
        &         &               &       &        &      &      \\
\endfirsthead
\caption{continued.}\\

\hline Star&$v\sin i$& $P/\sin i$  & [M/H] &Ca\,{\sc ii} EW&$F_{HK}$&$F_{HK}$\footnotemark[1] \\
        &($km/s$) &     (days)     & (dex) & (\AA)  &$10^{4}erg\ s^{-1}\ cm^{-2}$&$10^{4}erg\ s^{-1}\ cm^{-2}$\\
        &         &                &       &        &      &      \\
\endhead 
\hline
\endfoot

GJ 1046 &  2.63   &  7.41-1.1+1.2  &-0.078 & -0.420 & 1.276 & 1.527\\
GJ 1050 &  0.59   & 34.4-18+310    &-0.035 & -0.113 & 0.324 & 0.351\\
GJ 1054B& 32.13   & 0.45-0.07+0.08 &-0.323 &    -   &   -   &  -   \\ 
GJ 1097 & $<$0.5  &  $>$41.91      &-0.033 & -0.469 & 1.178 & 1.271\\
GJ 1125 &  0.99   & 10.3-2.9+8.3   &-0.691 & -0.319 & 1.454 & 7.138\\
GJ 1203 &  0.82   & 23.5-7.1+18    &-0.087 & -0.283 & 0.772 & 0.943\\
GJ 1212A&  0.69   & 25.3-9.6+57    &-0.170 &    -   &   -   &  -   \\
GJ 1212B&  0.90   & 19.3-7.1+13    &-0.170 &    -   &   -   &  -   \\
\hline 
\end{longtable}
\footnotetext[1]{Surface Flux corrected for metallicity effects}
\end{center}

\small
\begin{landscape}
\begin{longtable}{||lccccccccc|}
  \caption[ ]{Equivalent width (EW, in units of  \AA ) of the Ca\,{\sc ii} 
resonance doublet and H$_{\alpha}$ line for our sample of dM4 stars. 
Results were obtained using FEROS spectra as well as a compilation of data 
published in the literature. When the H$_{\alpha}$ EW is known but with no 
observations of the Ca\,{\sc ii} EW, we derived the Ca\,{\sc ii} EW 
according to the H$_{\alpha}$ EW-Ca\,{\sc ii} EW correlation (see Fig.~8).} \\

\hline  &       &              &      FEROS   &               &              &              &              &               &                  \\
        &       &Ca\,{\sc ii} H&Ca\,{\sc ii} K&  H$_{\alpha}$ &Mean Ca\,{\sc ii}&$S_{HK}$   &Mean Ca\,{\sc ii}&  H$_{\alpha}$ & Mean Ca\,{\sc ii}\\
Star    & No. of&     EW       &     EW       &     EW        &     EW       &              &EW\footnotemark[1]&   EW      &        EW        \\
        & meas. &   (\AA)      &    (\AA)     &   (\AA)       &   (\AA)      &              &    (\AA)     &    (\AA)      &      (\AA)       \\

\endfirsthead
\caption{continued.}\\

\hline  &       &              &      FEROS   &               &              &              &              &               &                  \\
        &       &Ca\,{\sc ii} H&Ca\,{\sc ii} K&  H$_{\alpha}$ &Mean Ca\,{\sc ii}&$S_{HK}$   &Mean Ca\,{\sc ii}&  H$_{\alpha}$ & Mean Ca\,{\sc ii}\\
Star    & No. of&     EW       &     EW       &     EW        &     EW       &              &EW\footnotemark[1]&   EW      &        EW        \\
        & meas. &   (\AA)      &    (\AA)     &   (\AA)       &   (\AA)      &              &    (\AA)     &    (\AA)      &      (\AA)       \\
\endhead 
\hline
\endfoot

        &       &              &              &               &              &              &              &               &           \\
GJ 1001A&   1   &-0.20$\pm$0.04&-0.09$\pm$0.07&0.215$\pm$0.02 &    -0.235(2) &      -       &      -       &    0.218(15)  & -0.235(2) \\
GJ 1005 &   -   &      -       &      -       &      -        &      -       &      -       &      -       &    0.224      &     -     \\
GJ 1006A&   2   &-4.09$\pm$0.04&-4.72$\pm$0.04&-3.14$\pm$0.20 &    -4.405(2) &      -       &      -       &   -3.14(2)    & -4.405(2) \\
GJ 1006B&   2   &-7.05$\pm$0.05&-8.55$\pm$0.05&-2.98$\pm$0.20 &    -7.80(2)  &      -       &      -       &   -2.98(2)    & -7.80(2)  \\
GJ 1065 &   3   &-0.26$\pm$0.08&-0.24$\pm$0.08&0.167$\pm$0.02 &    -0.25(3)  &      -       &      -       &    0.189(4)   & -0.25(3)  \\
GJ 1105 &   -   &      -       &      -       &      -        &    -0.62(1)  &      -       &      -       &    0.203(3)   & -0.62(1)  \\
GJ 1129 &   3   &-0.302$\pm$0.03&-0.353$\pm$0.03&0.190$\pm$0.02&   -0.328(3) &      -       &      -       &    0.178(4)   & -0.328(3) \\
\hline 
\end{longtable}
\footnotetext[1]{Inferred from the H$_{\alpha}$ EW-Ca\,{\sc ii} EW correlation}
\end{landscape}
\normalsize

\tiny
\begin{center}
\begin{longtable}{|lccccccc|} 
  \caption[ ]{The values of $v\sin i$, $P/\sin i$ and [M/H] compiled 
from Paper I for our targets in our list of dM4 stars. We also give the 
Ca\,{\sc ii} EW (from the compilation of Table~6), surface fluxes 
$F_{HK}$ and the surface fluxes corrected for the metallicity effects.} \\

\hline Star&$v\sin i$& $P/\sin i$& [M/H]& [M/H]\footnotemark[1] &Ca\,{\sc ii} EW&$F_{HK}$&$F_{HK}$\footnotemark[2] \\
        &$(km/s)$&   (days)   & (dex) & (dex) & (\AA)  &$10^{4}erg/s/cm^{2}$&$10^{4}erg/s/cm^{2}$\\
        &      &                &       &       &        &       &       \\
\endfirsthead
\caption{continued.}\\

\hline Star&$v\sin i$& $P/\sin i$& [M/H]& [M/H]\footnotemark[1] &Ca\,{\sc ii} EW&$F_{HK}$&$F_{HK}$\footnotemark[2] \\
        &$(km/s)$ &  (days)     & (dex) & (dex) & (\AA)  &$10^{4}erg/s/cm^{2}$&$10^{4}erg/s/cm^{2}$\\
        &      &                &       &       &        &       &       \\
\endhead 
\hline
\endfoot

GJ 1001A& 1.15 &12.81-5.66+18.02&   -   &+0.047 & -0.235 & 0.377 & 0.338 \\
GJ 1005AB& 3.00& 3.41-0.96+1.44 & -0.47 &-0.210 &   -    &  -    &   -   \\
GJ 1034 & 4.50 & 2.05-0.61+0.79 & -0.38 &-0.275 &   -    &  -    &   -   \\
GJ 1065 & 4.00 & 2.70-0.65+0.88 &  0.20 &-0.172 & -0.25  & 0.431 & 0.640 \\
GJ 1105 & 2.00 & 6.93-2.28+4.24 &  0.24 &+0.022 & -0.62  & 1.068 & 0.615 \\
GJ 1129 & 3.00 & 4.62-1.94+2.91 &  0.05 &+0.022 & -0.328 & 0.463 & 0.440 \\
GJ 1134 & 4.10 & 3.11-0.74+1.00 & -0.16 &-0.036 &   -    &  -    &   -   \\
GJ 1207 & 10.70& 1.18-0.21+0.25 & -0.15 &-0.042 & -7.098 & 11.10 & 12.23 \\
\hline  
\end{longtable}
\footnotetext[1]{Metalicity from the radius-metallicity relation}
\footnotetext[2]{Surface Flux corrected from metallicity effects}
\end{center}

\small
\begin{center}
\begin{longtable}{|lcccccccc|}
\caption[ ]{In this table, we list $log(L_{HK})$, $log(L_{bol})$, $R'_{HK}$ 
and $R_{0}$ our samples of dK4, dK6, dM2, dM3 and dM4 stars.} \\

\hline Star&$log(L_{HK})$&$log(L_{HK})$\footnotemark[1]&$log(L_{bol})$&$R'_{HK}$&$R'_{HK}$\footnotemark[1]&$log(L_{X})$&$log(L_{X}/L_{bol})$&$R_{0}$\\
Name      &($ergs\ s^{-1}$)&($ergs\ s^{-1}$)&($ergs\ s^{-1}$)&      &        &($ergs\ s^{-1}$)&      &       \\
\hline
          &       &        &        &       &        &      &        &        \\
\endfirsthead
\caption{continued.}\\

\hline Star&$log(L_{HK})$&$log(L_{HK})$\footnotemark[1]&$log(L_{bol})$&$R'_{HK}$&$R'_{HK}$\footnotemark[1]&$log(L_{X})$&$log(L_{X}/L_{bol})$&$R_{0}$\\
Name      &($ergs\ s^{-1}$)&($ergs\ s^{-1}$)&($ergs\ s^{-1}$)&      &        &($ergs\ s^{-1}$)&      &       \\
\hline
          &       &        &        &       &        &       &       &        \\
\endhead 
\hline
\endfoot

          &       &        &        &       &        &       &       &        \\
          &       &        &   K4   &       &        &       &       &        \\
          &       &        &        &       &        &       &       &        \\
GJ 4061   & 28.062&   -    & 32.840 &-4.778 &   -    &   -   &   -   & 0.1801 \\
GJ 4322B  & 27.739&   -    & 32.993 &-5.254 &   -    & 28.164&-4.829 & 0.7267 \\
Gl 106    & 28.338&   -    & 32.999 &-4.661 &   -    & 28.384&-4.615 & 0.4069 \\
Gl 131    & 28.322&   -    & 32.661 &-4.338 &   -    &   -   &   -   & 0.6554 \\
Gl 160.2  & 27.948&   -    & 32.638 &-4.690 &   -    & 27.342&-5.296 & 0.5172 \\
\hline 
\end{longtable}
\end{center}
\normalsize
\footnotetext[1]{Corrected from metallicity effects}

\end{document}